\documentclass[12pt]{iopart}
\usepackage{iopams}
\expandafter\let\csname equation*\endcsname\relax
\expandafter\let\csname endequation*\endcsname\relax
\usepackage{amsmath}
\usepackage{amsfonts}
\usepackage{amsthm}
\usepackage{graphics}
\usepackage{graphicx}
\usepackage{subfigure}
\usepackage{amssymb}
\usepackage{color}
\usepackage{url}
\usepackage{cite}
\usepackage{hyperref}

\def\kmax{k_{\rm max}}

\begin{document}

\title[Turbulence in the 2D Fourier-truncated GP equation]{Turbulence in 
the two-dimensional Fourier-truncated Gross-Pitaevskii equation}

\author{Vishwanath Shukla$^1$, Marc Brachet$^2$ and Rahul Pandit$^{1}$
\footnote{Also at Jawaharlal Nehru Centre For Advanced Scientific Research,
 Jakkur, Bangalore, India}
}
\address{$^1$ Centre for Condensed Matter Theory, Department of Physics,
Indian Institute of Science, Bangalore 560012, India}
\address{$^2$ Laboratoire de Physique Statistique de l'Ecole Normale 
Sup{\'e}rieure, associ{\'e} au CNRS et aux Universit{\'e}s Paris VI et VII,
24 Rue Lhomond, 75231 Paris, France}
\ead{\mailto{vishwanath@physics.iisc.ernet.in},
\mailto{brachet@physique.ens.fr},
\mailto{rahul@physics.iisc.ernet.in}}

\begin{abstract}
We undertake a systematic, direct numerical simulation (DNS) of the two-dimensional,
Fourier-truncated, Gross-Pitaevskii equation to study the turbulent evolutions of its 
solutions for a variety of initial conditions and a wide range of parameters. We find that
the time evolution of this system can be classified into four regimes with qualitatively 
different statistical properties. First, there are transients that depend on the initial conditions. 
In the second regime, power-law scaling regions, in the energy and the occupation-number
spectra, appear and start to develop; the exponents of these power-laws and the extents 
of the scaling regions change with time and depended on the initial condition. 
In the third regime, the spectra drop rapidly for modes with wave numbers 
$k > k_c$ and partial thermalization takes place for modes with $k < k_c$; the 
self-truncation wave-number $k_c(t)$ depends on the initial conditions and it
grows either as a power of $t$ or as $\log t$. Finally, in the fourth regime, 
complete-thermalization is achieved and, if we account for finite-size effects carefully, 
correlation functions and spectra are consistent with their nontrivial 
Berezinskii-Kosterlitz-Thouless forms.
\end{abstract}

\pacs{47.27.Ak, 47.27.Gs, 47.37.+q, 67.25.dk, 67.25.dj}





\section{Introduction}
\label{section:introduction}
The elucidation of the nature of superfluid turbulence, which
began with the pioneering studies of Feynman~\cite{Feynman1955}
and of Vinen and
Hall~\cite{Vinen1957MFI,Vinen1957MFII,Vinen1957MFIII,
Tilley1990,Donnellybook1991}, has continued to engage the attention of
experimentalists, theoreticians, and numerical simulators 
\cite{Barenghicollection2001,Vinenniemela2002,Niemelaviewpoint2008,Procacciasreenireview2008,Tsubotareview2008,Paolettireview2011,Skrbeksreeni2012}.
Experimental systems, in which such turbulence is studied,
include the bosonic superfluid $^4$He, its fermionic counterpart
$^3$He, and Bose-Einstein condensates (BECs) of cold atoms in
traps and their optical analogues; for representative studies we refer the reader
to~\cite{Maurer1998epl,Smith1993vinen,
Stalp1999donnelly,Skrbek2000,skrbek2000decayhit,Stalp2002disspgridturb,
Fisherhe3prl,Henn2009bec1st,NeelyarXiv2012,Sunexptnatphys2012}. 
Theoretical and numerical studies have used a variety of models to 
study superfluid turbulence; these include the two-fluid
model~\cite{Roche2009qt2fluidcascade,Salort20112fluid},
Biot-Savart-type models with~\cite{Schwarz1985,Schwarz1988} or
without~\cite{Barenghi1999biotsavart,Adachi2010withoutLIA} the
local-induction approximation, and the Gross-Pitaevskii (GP) or
nonlinear Schr\"odinger (NLS)
equations~\cite{Kagan1992,Zakharov2005}. These models have been
studied by a combination of theoretical methods, such as
wave-turbulence
theory~\cite{Svistunov1991,Kagan1992,YLvov2003wtbecs,Zakharov2005},
and numerical
simulations~\cite{Koplik1993prl,Nore1997prl,Berloff2002pra,Kobayashi2005prl,
Krstulovic2011prl,White2010prlvpdf,Connaughton2005prl}.  Most of these studies have
been carried out in three dimensions (3D); numerical simulations
of two-dimensional (2D) models for superfluid turbulence have
been increasing steadily over the past few
years~\cite{Nazarenko2007freedecay2d,Numasato2Dgp2010,Nowak2011prb,BradleyPhysRevX2012}.
Here we undertake a systematic direct numerical simulation (DNS)
of the dissipationless, unforced, Fourier-truncated, 2D, GP
equation with a view to identifying what, if any, features of the
turbulent evolution of the solutions of this equation are
universal, i.e., they do not depend on initial conditions.  Some,
though not all, parts of our results are contained in earlier
simulations~\cite{Nazarenko2007freedecay2d,Numasato2Dgp2010,Nowak2011prb,
Nowak2012pra,BradleyPhysRevX2012,BradleyPRAtrapQT2012,Small2011bktphotonlat}.
The perspective of our study is different from earlier studies of
the 2D GP equation; in particular, we elucidate in detail the 
dynamical evolution of this system and examine the various stages
of its thermalization; in this sense our work is akin to recent
studies of thermalization in Euler and other hydrodynamical
equations~\cite{Cichowlas2005prl,Frischprl2008,Ssaytyger2011}; a similar study
for the 3D GP equation has been carried out by Krstulovic and
Brachet~\cite{Krstulovic2011prl,Krstulovic2011pre}.

It is useful to begin with a qualitative overview of our
principal results.  We find that the dynamical evolution of the
dissipationless, unforced, 2D, Fourier-truncated GP equation can
be classified, roughly, into the following four regimes, which
have qualitatively different statistical properties: (1) The
first is the region of initial transients; this depends on the
initial conditions. (2) This is followed by the second regime, in
which we see the onset of thermalization; here the energy and
occupation-number spectra begin to show power-law-scaling
behaviours, but the power-law exponent and the extents of the
scaling regions change with time and depend on the initial
conditions. (3) In the third regime, which we call the region of
partial thermalization, these spectra show clear, power-law,
scaling behaviours, with a power that is independent of the
initial conditions, and, at large wave vectors, an
initial-condition-dependent, self-truncation regime, where spectra
drop rapidly; (4) finally, in the fourth regime, the system
thermalizes completely and exhibits correlation functions that
are consistent with the predictions of the Berezinskii-Kosterlitz-Thouless (BKT)
theory~\cite{Kogut1979rmp,Chaikin1995book,Foster2010pra,Small2011bktphotonlat},
if the simulation domain and simulation time are large enough.  Although some of these
regimes have been seen in some earlier numerical studies of the
2D GP equation, we are not aware of any study that has
systematized the study of these four dynamical regimes.  In
particular, regime 3, which shows partial thermalization and
self-truncation in spectra, has not been identified in the 2D,
Fourier-truncated, GP equation, even though its analogue has been
investigated in the 3D
case~\cite{Svistunov1991,Krstulovic2011prl,Krstulovic2011pre}.

The remaining part of this paper is organised as follows.  In
section~\ref{section:theoryandnumerics}, we describe the
2D, GP equation and the different statistical measures we use 
to characterize turbulence in the Fourier-truncated, 2D, GP
equation (section~\ref{subsection:gptheory}); the details of
our numerical methods and initial conditions are given in 
section~\ref{subsection:numericsandics}.  In
section~\ref{section:results}, we present our results; these are
described in the four
subsections~\ref{subsection:temporalevolution}-\ref{subsection:completethermalization}
that are devoted, respectively, to the following:
(a) the temporal evolution of
the energy components, velocity-component probability
distribution functions (PDFs), and the population $N_0$ in the
zero-wave-number mode; (b) the statistical characterization of
the first two regimes of the dynamical evolution (by using various
energy and the occupation-number spectra for different initial
conditions); (c) a similar statistical characterization, as in
subsection~\ref{subsection:initialtransientsandonset}, but for the 
regime with partial thermalization, and the study of the nature of 
the growth of the self-truncation region; (d) the final, completely 
thermalized state of the Fourier-truncated, 2D, GP equation.
Section~\ref{section:conclusions} contains our
conclusions. A note on the units used for the GP equation and the 
details of some analytical calculations are presented in 
~\ref{app:one} and~\ref{app:two}, respectively.

\section{Model, Initial Conditions, and Numerical Methods} 
\label{section:theoryandnumerics}

In this Section, we describe the 2D, GP equation. We define all 
the statistical measures that we use to
characterize the time evolution of this equation, given the three
types of initial conditions that we describe below. We also
describe the numerical methods, and computational procedures
that we use to solve this equation.

\subsection{The Gross-Pitaevskii Equation}
\label{subsection:gptheory}

The GP equation, which describes the dynamical evolution of the
wave function $\psi$ of a weakly interacting 2D Bose gas at
low temperatures, is
\begin{equation} \label{eq:2dgpe}
i\frac{\partial\psi(\mathbf{x},t)}{\partial t} =
-\nabla^2\psi(\mathbf{x},t) + g|\psi|^2\psi(\mathbf{x},t);
\end{equation}
$\psi(\mathbf{x},t)$ is a complex, classical field and $g$ is the 
effective interaction strength~\cite{Pethickbook2001,Posazhennikova2006rmp}. This equation conserves the
total energy 
\begin{equation} \label{eq:totalenergy}
E = \int_{\mathcal{A}} 
\left [ |\nabla \psi|^2 + \frac{1}{2}g|\psi|^4 \right]d^2x 
\end{equation}
and the total number of particles 
\begin{equation} \label{eq:particleN}
N = \int_{\mathcal{A}} |\psi|^2d^2x ,
\end{equation}
where $\mathcal{A}=L^2$ is the area of our 2D, periodic, computational domain
of side $L$. From \eqref{eq:2dgpe} 
we obtain the continuity equation
\begin{equation} \label{eq:eqconti}
\frac{\partial \rho}{\partial t} + 
\nabla \cdot \left(\rho \mathbf{v} \right) =0,
\end{equation}
where $\rho=|\psi|^2$ is interpreted as the particle density and 
the velocity is
\begin{equation} \label{eq:velocity}
\mathbf{v}(\mathbf{x},t)=\frac{\psi^*\nabla\psi-\psi\nabla\psi^*}{i|\psi|^2}.
\end{equation}
We can use the Madelung transformation 
$\psi(\mathbf{x},t)=\sqrt{\rho}e^{i\theta(\mathbf{x},t)}$, where 
$\theta(\mathbf{x},t)$ is the phase of $\psi(\mathbf{x},t)$, to
write $\mathbf{v}(\mathbf{x},t)=2\nabla \theta(\mathbf{x},t)$,
whence we get~\cite{Nore1997prl}
\begin{equation}
E = \int_{\mathcal{A}} 
\left [ \frac{1}{4}\rho v^2 + \frac{1}{2}g|\psi|^4 + [ \nabla\rho^{1/2}]^2
\right]d^2x
= E_{kin}+E_{int}+E_{q},
\end{equation}
where the kinetic, interaction, and quantum-pressure
energies are defined, respectively, as
\begin{subequations}
\begin{align} 
E_{kin}& = \frac{1}{4}\int_\mathcal{A} |\sqrt{\rho}v|^2 d^2x, \\
E_{int}& = \frac{1}{2} \int_\mathcal{A} g|\psi|^4 d^2x,  \\
E_{q}&= \int_\mathcal{A} \nabla |\rho^{1/2}|^2 d^2x.
\end{align}
\end{subequations}
We separate the compressible (supercript $c$) and the incompressible 
(superscript $i$) parts of the kinetic energy by making use of the decomposition
\begin{equation} \label{eq:sqrhov}
\rho^{1/2}\mathbf{v} = (\rho^{1/2}\mathbf{v})^i + (\rho^{1/2}\mathbf{v})^c,
\end{equation}
where $\nabla \cdot (\rho^{1/2}\mathbf{v})^i =0$ and
$\nabla\times(\rho^{1/2}\mathbf{v})^c= 0 $, whence we obtain the following:
\begin{subequations} \label{eq:partkin}
\begin{align}
E^i_{kin}& = \frac{1}{4}\int_\mathcal{A} |(\sqrt{\rho}v)^i|^2 d^2x; \\
E^c_{kin}& = \frac{1}{4}\int_\mathcal{A} |(\sqrt{\rho}v)^c|^2 d^2x.
\end{align}
\end{subequations}
The spectra for these energies are defined as follows:
\begin{equation} \label{eq:ikes}
E^i_{kin} = \frac{1}{4}\int |\widehat{(\rho^{1/2}\mathbf{v})^i}|^2 d^2k 
 \equiv \int E^i_{kin}(k) dk;
\end{equation}
\begin{equation} \label{eq:ckes}
E^c_{kin} = \frac{1}{4}\int |\widehat{(\rho^{1/2}\mathbf{v})^c}|^2 d^2k 
 \equiv \int E^c_{kin}(k) dk;
\end{equation}
\begin{equation} \label{eq:intes}
E_{int} = \int |\widehat{\sqrt{g/2}|\psi|^2}|^2 d^2k 
 \equiv \int E_{int}(k) dk;
\end{equation}
and
\begin{equation} \label{eq:qpes}
E_q = \int |\widehat{\nabla\rho^{1/2}}|^2 d^2k 
 \equiv \int E_{q}(k) dk;
\end{equation}
furthermore, we define an occupation-number spectrum $n(k)$ via
\begin{equation} 
N = \int |\widehat{\psi}|^2 d^2k \equiv \int n(k) dk;
\end{equation}
here we denote the Fourier transform of $A(\mathbf{x})$ by
$\widehat{A}$; and, for notational convenience, we do not show
explicitly the dependence of these spectra on time $t$. In any
computational study, we must limit the number of Fourier modes
that we use in our study of the GP equation; we refer to such a
GP equation as a Fourier-truncated GP equation
(cf.~\cite{Cichowlas2005prl,Frischprl2008} for studies of the Fourier- or
Galerkin-truncated Euler equation).

The Bogoluibov dispersion relation $\omega_{\rm B}(k)$
is obtained by linearizing \eref{eq:2dgpe} around a constant $\psi$. For a total
number of particles \eref{eq:particleN}  $N=1$, it is
\begin{equation}
\omega_{\rm B}(k)=k c\sqrt{1+ \frac{\xi^2k^2}{2}}, \label{Eq:bogu}
\end{equation}
where the sound velocity is $c=\frac {\sqrt{2 g} }{L }$
and the coherence length is
\begin{equation}
\xi =\frac {L} {\sqrt{ g} }.\label{Eq:cohlength}
\end{equation}

We investigate thermalization in the 2D GP equation, so it is 
useful to recall that a uniform, interacting, 2D Bose gas has a high-temperature
disordered phase and a low-temperature, Berezenskii-Kosterlitz-Thouless
(BKT) phase~\cite{Mermin1966,Hohenberg1967,Berezinskii1971,Kosterlitz1973}, 
which shows quasi-long-range order with an algebraic
decay of the spatial correlation function~\cite{Kogut1979rmp}
\begin{equation} \label{eq:phasecorr}
c(r) = \langle \left[ e^{-i\theta(\mathbf{x})} -\langle e^{-i\theta(\mathbf{x})}\rangle
\right]\left[e^{i\theta(\mathbf{x}+\mathbf{r})}-\langle
e^{i\theta(\mathbf{x}+\mathbf{r})}\rangle\right)]\rangle;
\end{equation}
for temperatures $T$ below the transition temperature $T_{BKT}$ (or energy
$E_{BKT}$ in the microcanonical ensemble), 
\begin{equation} \label{eq:corrpowerlaw}
c(r) \sim r^{-\eta},
\end{equation}
where $r \equiv |\mathbf{r}|$ and the critical exponent $\eta<0.25$ for $T<T_{BKT}$; and 
$\eta=0.25$ at $T=T_{BKT}$~\cite{Chaikin1995book}. The BKT phase shows bound
vortex-antivortex pairs; these unbind above $T_{BKT}$, so
\begin{equation} \label{eq:correxpo}
c(r) \sim e^{-r/\ell},
\end{equation}
in the disordered phase, with $\ell$ the correlation length.
\subsection{Numerical Methods and Initial Conditions}
\label{subsection:numericsandics}

To perform a systematic, pseudospectral, direct numerical
simulation (DNS) of the spatiotemporal evolution of the 2D, 
Fourier-truncated, GP equation, we have developed a parallel, MPI
code in which we discretize $\psi(\mathbf{x},t)$ on a square
simulation domain of side $L=32$ with $N_c^2$ collocation points.
We use periodic boundary conditions in both spatial directions, 
because we study homogeneous, isotropic turbulence in this 2D system, 
and a fourth-order, Runge-Kutta scheme, with time step $\Delta t$, for 
time marching. We evaluate the linear term in \eqref{eq:2dgpe} 
in Fourier space and the nonlinear term in physical space; for 
the Fourier-transform operations we use the FFTW
library~\cite{fftw}. Thus, the maximum wave number $k_{max} =
(N_c/2)\Delta k$, where $\Delta k = 2\pi/L$, and 
\begin{equation}\label{eq:xikmax} 
\xi k_{max}=\frac{\pi N_c}{\sqrt{g}}.  
\end{equation}
We have checked that, for the 
quantities we calculate, dealiasing of our pseudospectral code does 
not change our results substantially; here we present the results from our
pseudospectral simulations that do not use dealiasing.

To initiate turbulence in the 2D GP equation we use three types
of initial conditions $\tt IC1$~\cite{Nazarenko2007freedecay2d}, 
$\tt IC2$, and $\tt IC3$~\cite{Krstulovic2011pre}, always normalized to correspond 
to a total number of particles \eref{eq:particleN}  $N=1$. The first 
of these is best represented in Fourier space as follows:
\begin{equation} \label{eq:ic1inconfig04}
 \widehat{\psi}(\mathbf{k},t=0) = \frac{1}{\sqrt{\pi^{1/2}\sigma}}
\exp{\biggl(-\frac{(k-k_{0})^2}{2\sigma^2}\biggr)}\exp{(i\Theta(k_x,k_y))}, 
\end{equation} 
where $k=\sqrt{k^2_{x}+k^2_{y}}$, $\Theta(k_x,k_y)$ are random
numbers distributed uniformly on the interval
$\bigl[0,2\pi\bigr]$; $k_0= \mathcal{N}_0\Delta k$ and $\sigma =
\mathcal{B} \Delta k$, where the integer $\mathcal{N}_0$ controls
the spatial scale at which energy is injected into the system,
and the real number $\mathcal{B}$ specifies the Fourier-space
width of $\widehat{\psi}$ at time $t=0$.  The initial condition
$\tt IC2$ is like $\tt IC1$ but, in addition, it has a finite
initial condensate population $N^i_0 = \mid
\widehat{\psi}(\mathbf{k}=0,t)\mid^2(\Delta k)^2$ at time $t=0$.

We obtain the initial condition $\tt IC3$ by solving the 2D,
stochastic, Ginzburg-Landau equation (SGLE), which follows 
from the free-energy functional 
\begin{equation} \label{eq:freeenergy}
\mathcal{F} = \int_\mathcal{A} d^2x 
\left(|\nabla \psi|^2 -\mu|\psi|^2 + \frac{1}{2}g|\psi|^4
\right),
\end{equation}
where $\mu$ is the chemical potential\footnote{Recall that the SGLE can be
thought of as an imaginary-time GP equation with external, additive noise 
(see, {\it e.g.} reference~\cite{Krstulovic2011prl})}.
The SGLE is 
\begin{equation} \label{eq:sglfunctional}
\frac{\partial\psi}{\partial t} = -\frac{\delta \mathcal{F}}{\delta\psi^*} +
\zeta(\mathbf{x},t),
\end{equation}
where $\zeta$ is a zero-mean, Gaussian white noise with
\begin{equation} \label{eq:corrwn}
\langle \zeta(\mathbf{x},t)\zeta^*(\mathbf{x'},t')\rangle =
D\delta(\mathbf{x}-\mathbf{x'})\delta(t-t'),
\end{equation}
where $D = 2T$, in accordance with the fluctuation-dissipation
theorem~\cite{Kuboreviewflucdiss1966}, $T$ is the temperature, and $\delta$ the Dirac delta 
function. Finally, the SGLE \eqref{eq:sglfunctional} becomes
\begin{equation} \label{eq:sgle}
\frac{\partial \psi}{\partial t} = \nabla^2\psi - 
\mu \psi + g|\psi|^2\psi + \zeta,
\end{equation}
which we solve along with the following, ad-hoc equation
\begin{equation} \label{eq:cheml}
\frac{d\mu}{dt} = -\frac{\nu_N}{\mathcal{A}}\left(N-N_{\rm av} \right),
\end{equation}
to control the number of particles $N$; the parameter $N_{\rm av}$ 
controls the mean value of $N$; and $\nu_N$ governs the rate at
which the SGLE equilibrates.  We solve the SGLE by using a
pseudospectral method, similar to the one described above for the
2D, GP equation, with periodic boundary conditions in space, an
implicit-Euler scheme, with time step $\Delta t$, for time
marching and the method of reference~\cite{Toral2000} 
(see page 25 of this reference).

\begin{table}
\small
   \begin{tabular}{@{\extracolsep{\fill}} c c c c c c c c c c c c c }
    \hline
    $ $ & $N_c$ & $k_0 (\times \Delta k)$ & $\sigma (\times \Delta k)$ & $g$ & $N^i_0$ &
$\sqrt{D} (\times 10^{-3})$ & $k_c^{in}$ & $E$\\ 
   \hline \hline
    {\tt A1}  &  $1024$ &  $5$	& $2$ &  $1000$ & $-$ & $-$ & $-$ & $2.120$ \\
    {\tt A2}  &  $1024$ &  $5$	& $2$ &  $2000$ & $-$ & $-$ & $-$ & $3.045$ \\
    {\tt A3}  &  $1024$ &  $5$	& $2$ &  $5000$ & $-$ & $-$ & $-$ & $5.82$ \\
    {\tt A4}  &  $1024$ &  $35$	& $5$ &  $1000$ & $-$ & $-$ & $-$ & $49.69$ \\
    {\tt A5}  &  $512$ &  $5$	& $2$ &  $1000$ & $-$ & $-$ & $-$ & $2.15$ \\
    {\tt A6}  &  $256$ &  $5$	& $2$ &  $1000$ & $-$ & $-$ & $-$ & $2.07$ \\
    {\tt A7}  &  $128$ &  $5$	& $2$ &  $1000$ & $-$ & $-$ & $-$ & $2.1$ \\
    {\tt A8}  &  $64$ &  $5$	& $2$ &  $1000$ & $-$ & $-$ & $-$ & $2.2$ \\
    {\tt A9}  &  $256$ &  $5$	& $2$ &  $2000$ & $-$ & $-$ & $-$ & $2.94$ \\
    {\tt A10}  &  $256$ &  $5$	& $2$ &  $5000$ & $-$ & $-$ & $-$ & $5.57$ \\
    {\tt A11}  &  $256$ &  $15$	& $2$ &  $1000$ & $-$ & $-$ & $-$ & $9.86$ \\
    {\tt A12}  &  $256$ &  $15$	& $2$ &  $2000$ & $-$ & $-$ & $-$ & $10.82$ \\
    {\tt A13}  &  $256$ &  $15$	& $2$ &  $5000$ & $-$ & $-$ & $-$ & $13.68$ \\
    {\tt B1}  &  $128$  &  $5$	& $1$ &  $10000$ & $0.95$ & $-$ & $-$ & $5.44$ \\
    {\tt B2}  &  $128$  &  $5$	& $1$ &  $1000$ & $0.95$ & $-$ & $-$ & $0.59$ \\
    {\tt C1}&  $256$  &  $-$	& $-$ &  $5000$ & $-$ & $8$ & $6$ & $2.536$ \\
    {\tt C2}&  $256$  &  $-$	& $-$ &  $1000$ & $-$ & $8$ & $6$ & $0.583$ \\
    {\tt C3}&  $256$  &  $-$	& $-$ &  $1000$ & $-$ & $10$ & $6$ & $0.637$ \\
    {\tt C4}&  $256$  &  $-$	& $-$ &  $1000$ & $-$ & $8$ & $9$ & $0.7$ \\
    {\tt C5}&  $256$ &  $-$	& $-$ &  $1000$ & $-$ & $8$ & $15$ & $1.085$ \\
    {\tt C6}&  $256$ &  $-$	& $-$ &  $1000$ & $-$ & $8$ & $20$ & $1.557$ \\
\hline
\end{tabular}
\caption{\small Parameters for our DNS runs $\tt A1$-$\tt A13$,
$\tt B1$-$\tt B2$, and $\tt C1$-$\tt C6$: $N^2_c$ is the
number of collocation points, $k_0$ is the energy-injection 
scale, $\sigma$ is the Fourier-space width of
$\widehat{\psi}$ at $t=0$; $g$ is the effective
interaction strength; $N^i_0$ is the initial condensate
population; $D$ and $k^{in}_c$ are respectively, the variance of the
white-noise and the initial value of the truncation wave number,
which we use in the initial conditions of type $\tt IC3$;
$E$ is the total energy; we use a square simulation
domain of area $\mathcal{A}=L^2$; we choose $L=32$.
}
\label{table:param}
\end{table}
\section{Results}
\label{section:results}

We first present the time evolution of the different energies,
the probability distribution functions (PDFs) of the velocity
components, and the population $N_0$ in the zero-wave-number
mode. We then give a detailed statistical characterization of the
temporal evolution of the Fourier-truncated, 2D, GP equation in
the four regimes mentioned in the Introduction (section~\ref{section:introduction}).

\subsection{Evolution of energies, 
velocity PDFs, and the zero-wave-number population
}
\label{subsection:temporalevolution}

\begin{figure*}
\includegraphics[height=4.cm]{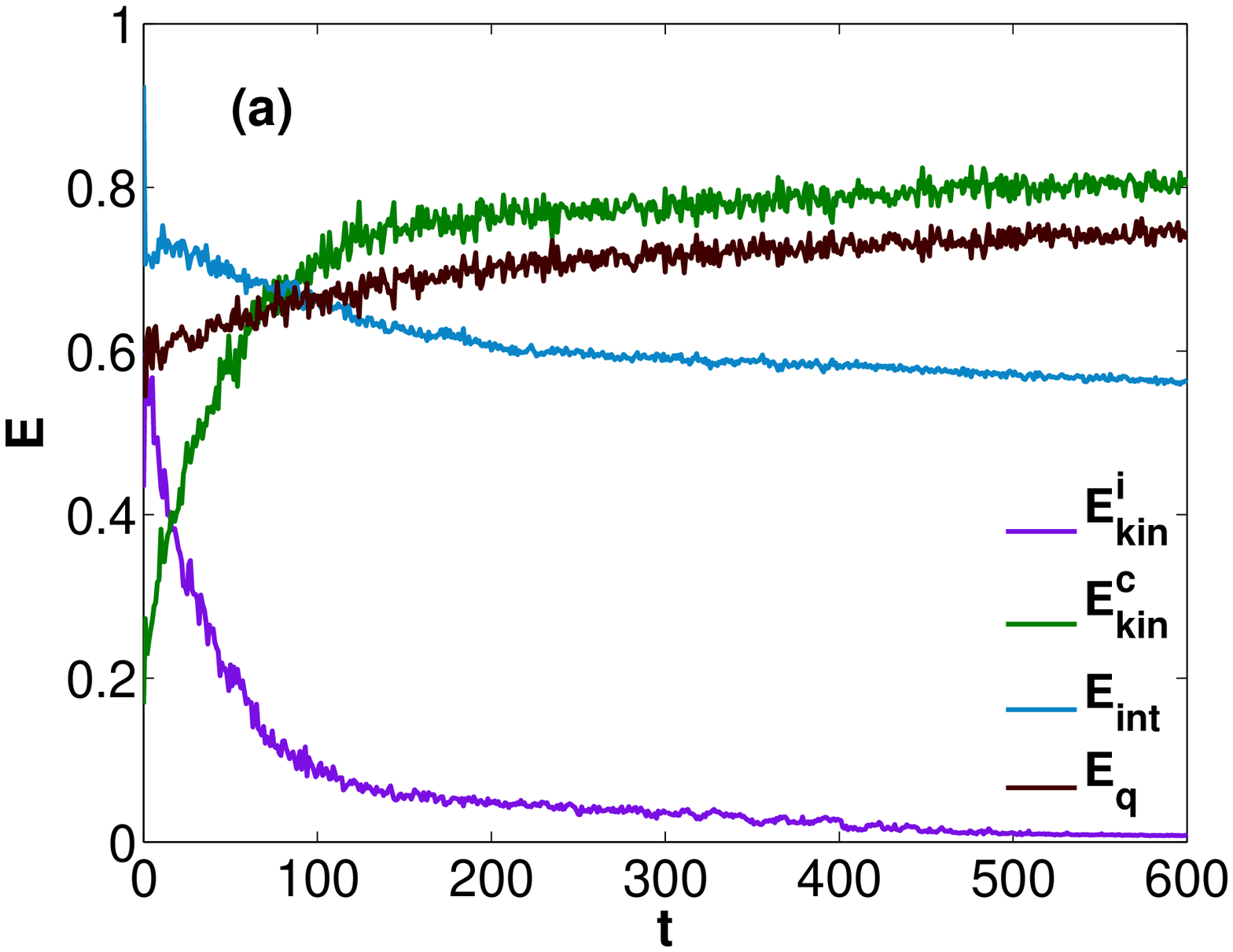}
\includegraphics[height=4.cm]{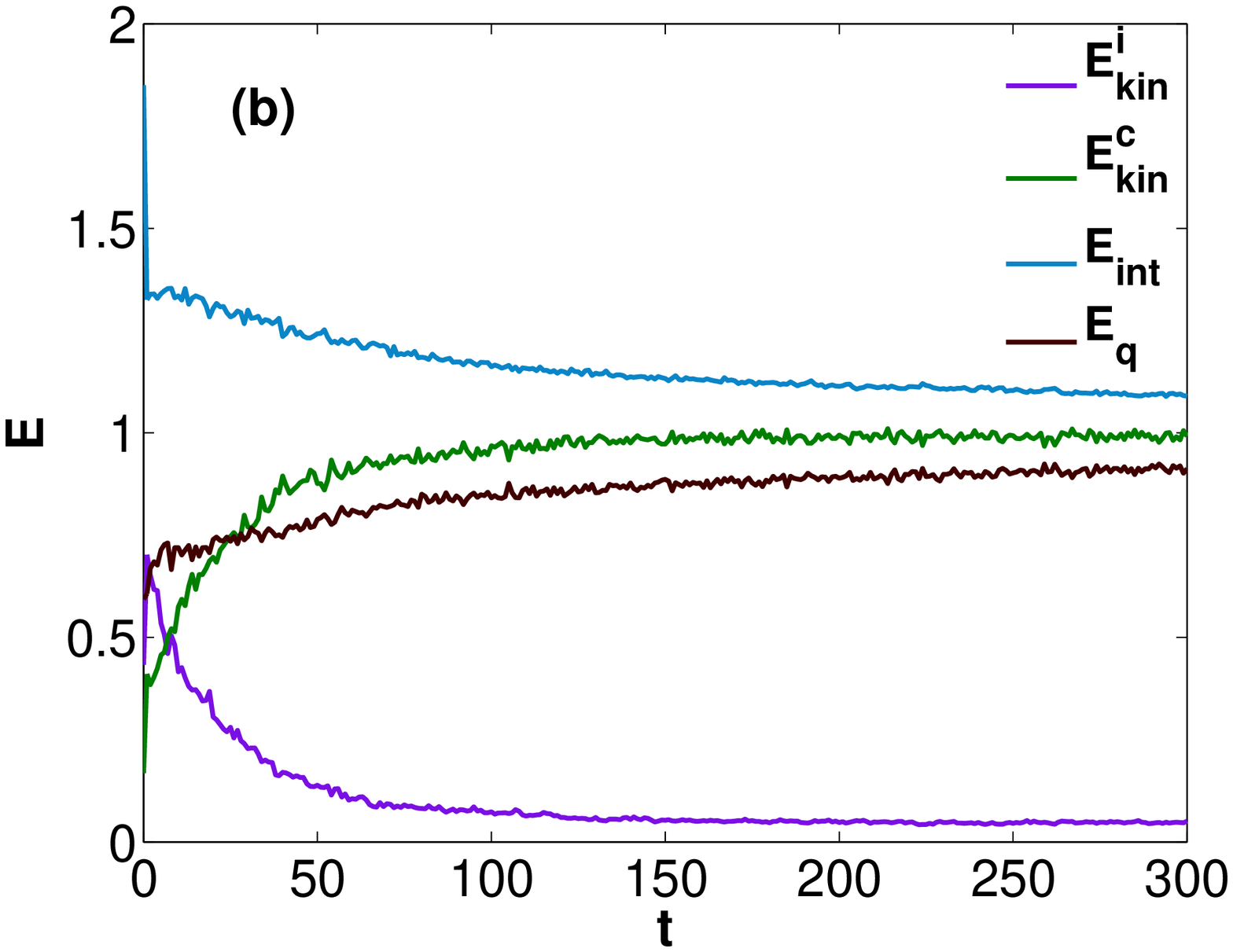}
\includegraphics[height=4.cm]{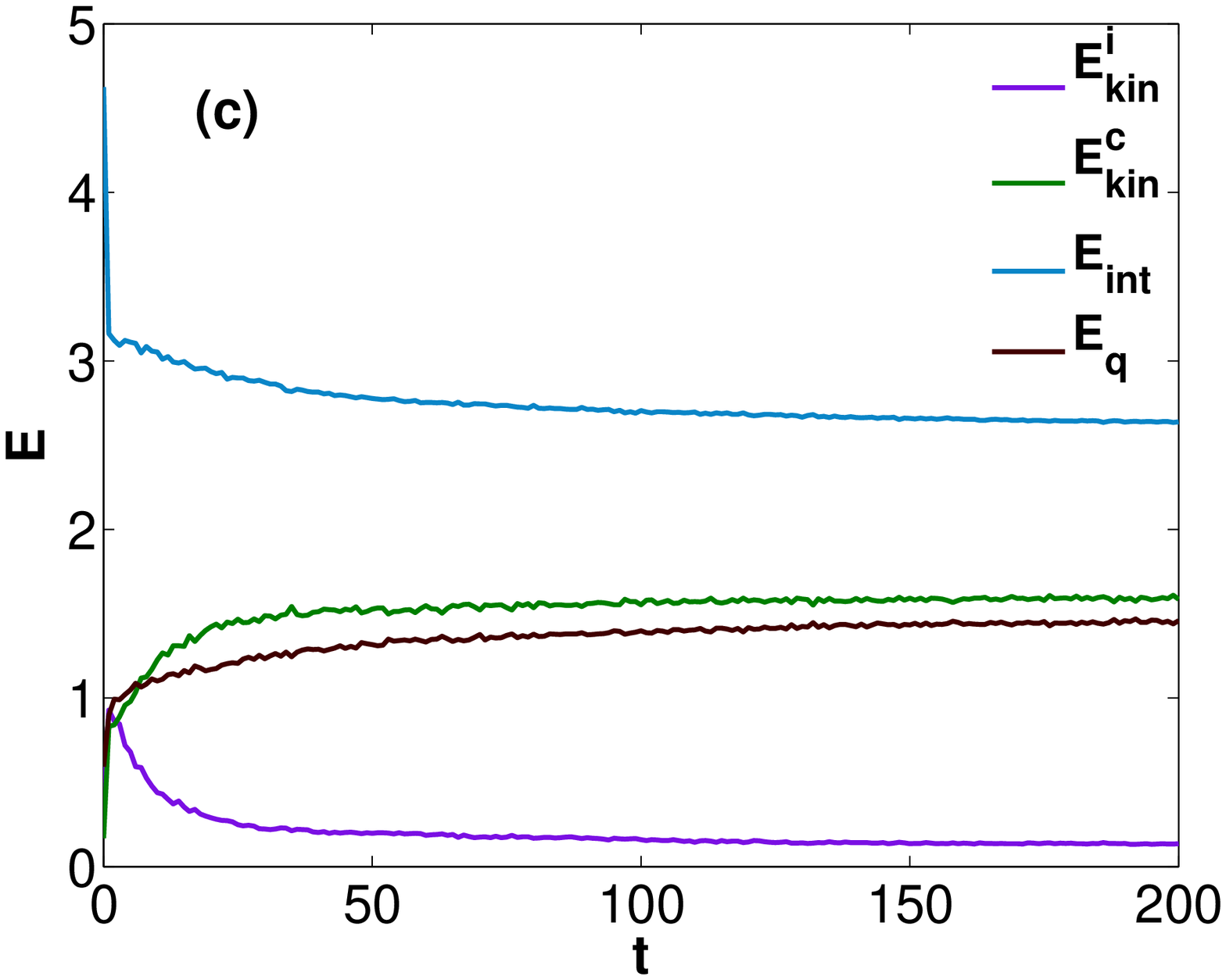}
\includegraphics[height=4.cm]{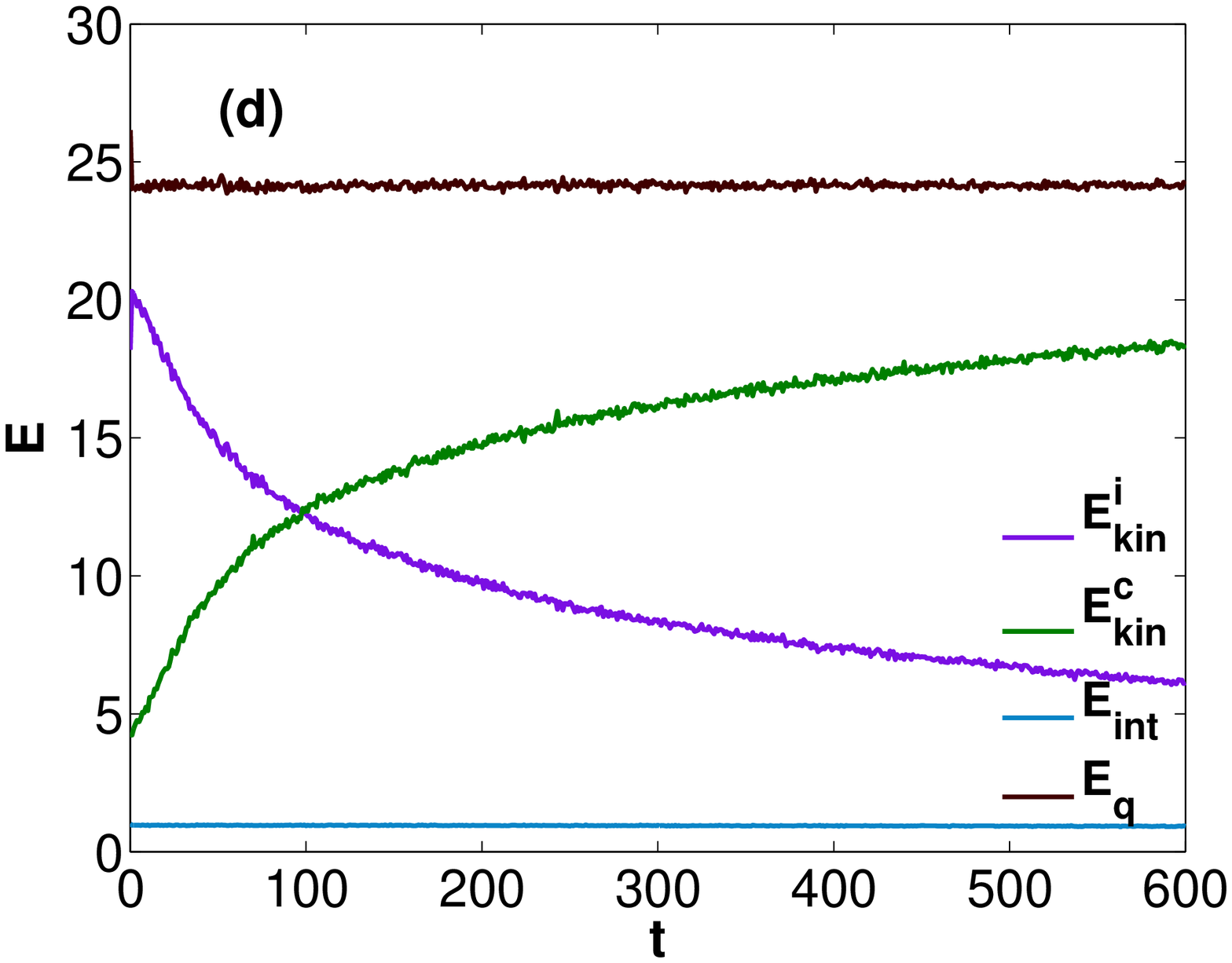}
\includegraphics[height=4.cm]{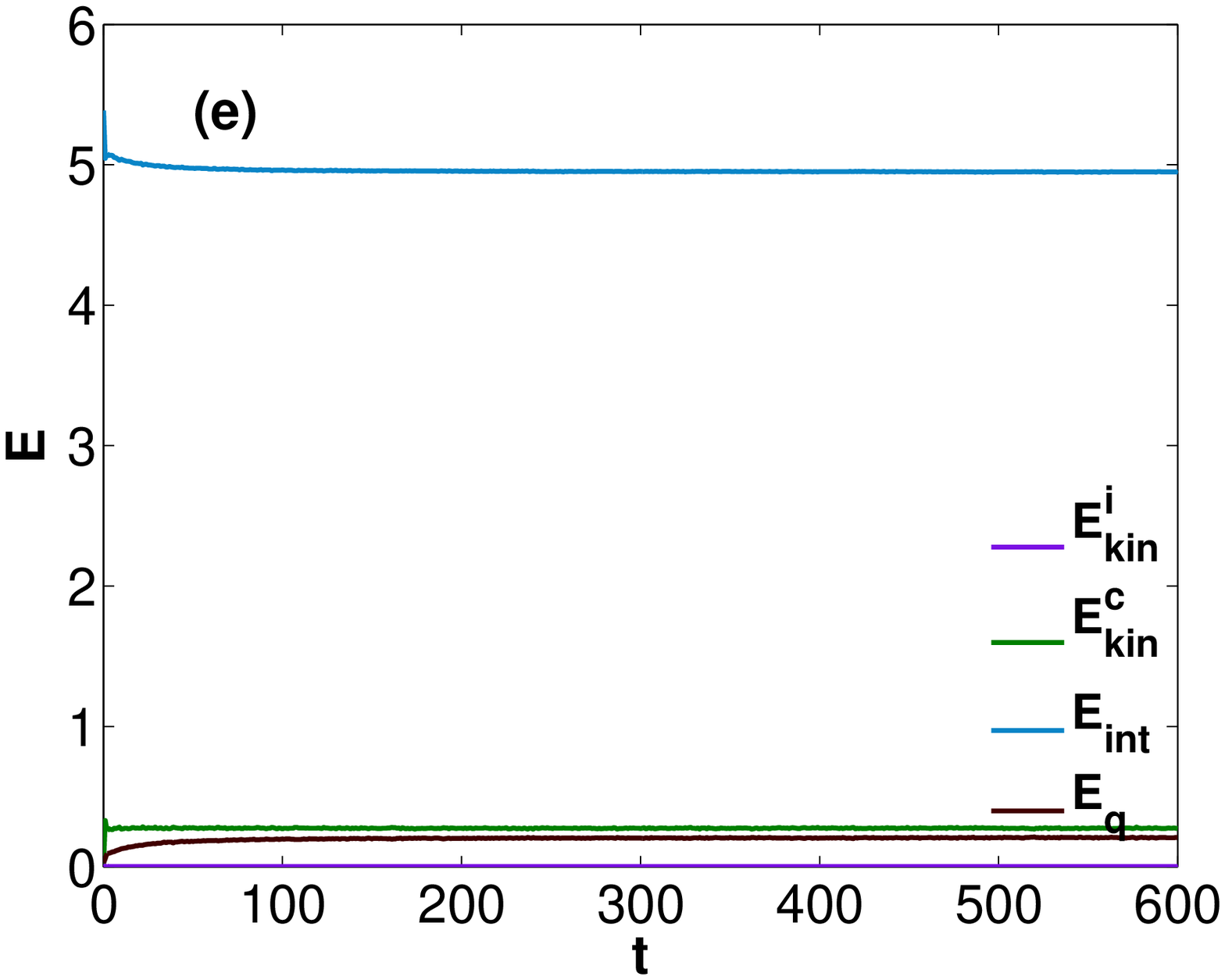}
\includegraphics[height=4.cm]{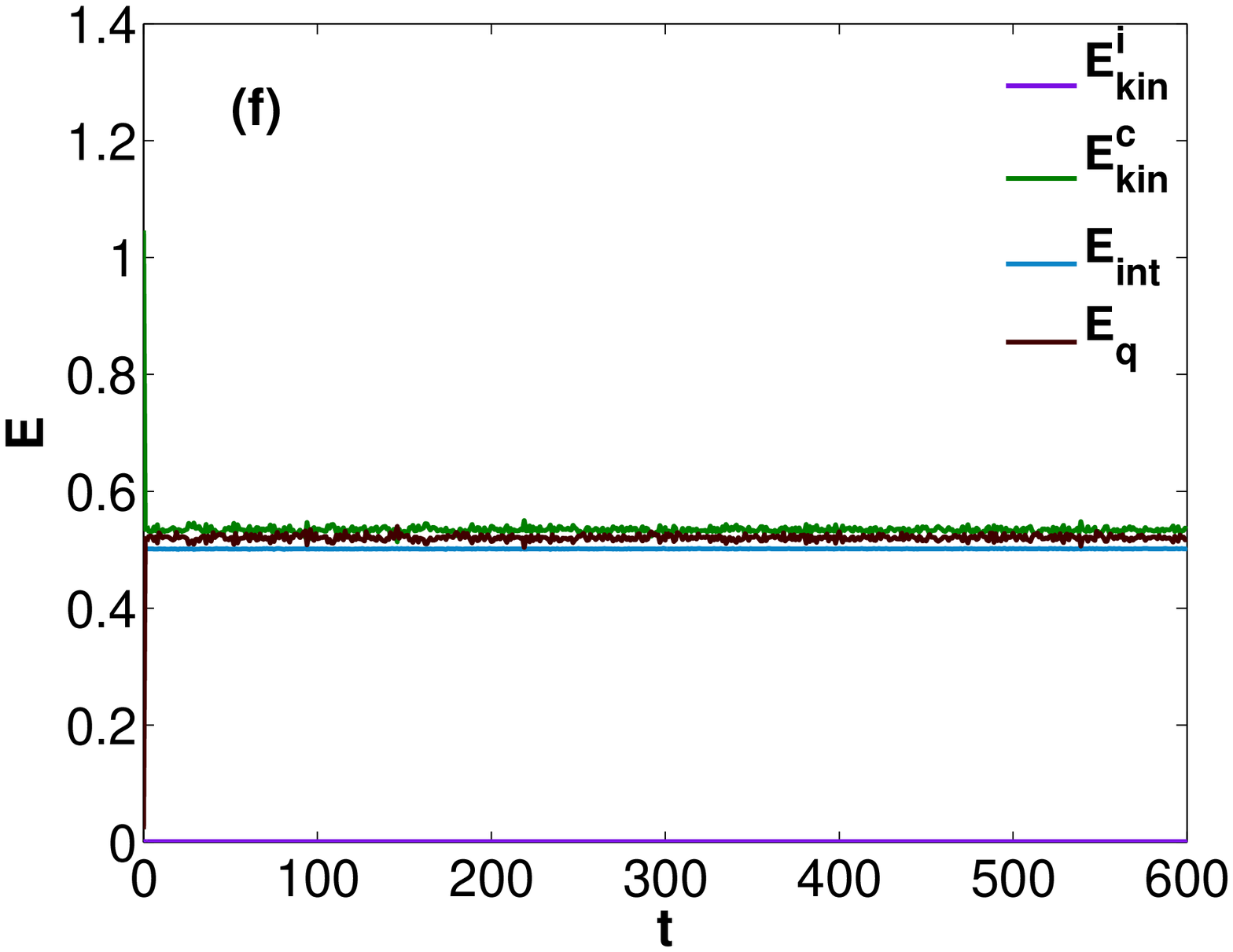}
\caption{\small Plots versus time $t$ of the four components of the 
total energy  $E^i_{kin}$,
$E^c_{kin}$, $E_{int}$, and $E_q$, during the initial stages of evolution, 
from our DNS runs (a) $\tt A1$,
(b) $\tt A2$ (c) $\tt A3$, (d) $\tt A4$, (e) $\tt B1$, and 
(f) $\tt C6$ (see table~\ref{table:param}).} 
\label{fig:energycomp} 
\end{figure*}

We show the early stages of the time evolution of the energies
$E^i_{kin}$, $E^c_{kin}$, $E_{int}$, and $E_q$, from our DNS runs
$\tt A1$-$\tt A4$, $\tt B1$, and $\tt C6$ in figure
\ref{fig:energycomp}.  The runs $\tt A1$-$\tt A4$ use initial
conditions of type $\tt IC1$, in which $E^i_{kin}$ is a significant 
fraction of the total initial energy; the runs $\tt B1$ and $\tt C6$ start with
initial configurations of type $\tt IC2$ and $\tt IC3$,
respectively, in which $E^i_{kin}$ is negligibly small at $t=0$.
The transient nature of the early stages of the dynamical
evolution of the dissipationless, unforced, 2D, GP equation is
evident from figure~\ref{fig:energycomp}, in which we observe a
rapid conversion of $E^i_{kin}$ into the other three components,
with a significant fraction being transferred to
$E^c_{kin}$; moreover, the transient stage depends on the initial
conditions, as we describe below.  Figures~\ref{fig:energycomp}
(a)-(c), show comparisons of the temporal evolution of the
energies, from the runs $\tt A1$-$\tt A3$; we observe, in
particular, that the conversion of $E^i_{kin}$ into the other energy
components is accelerated as $g$ increases from $1000$ to $5000$
(cf.~\cite{Numasato2Dgp2010}); and there is a corresponding
acceleration in the approach to thermalization. Moreover, the
larger the value of  $E^i_{kin}$ the larger is the time required
for thermalization, as we can see by comparing
figures~\ref{fig:energycomp} (a) and (d), for the runs $\tt A1$
and $\tt A4$, respectively; the run $\tt A4$ starts with a high
value of $E^i_{kin}(t=0)$ because of a large number of vortices
and anti-vortices, so it takes a long time to thermalize; indeed,
if the spatial resolution of our DNS is very high, the
computational cost of achieving a statistically steady state is
prohibitively high for intial conditions $\tt A1$-$\tt A4$. In
contrast, the runs $\tt B1$ and $\tt C6$ have negligibly small
values of $E^i_{kin}(t=0)$ to begin with
(figures~\ref{fig:energycomp} (e) and (f), respectively); and
$E^i_{kin}(t)$ remains close to zero throughout the dynamical
evolution here.  For run $\tt B1$, both $E^c_{kin}$ and $E_q$
start from values close to zero, grow at the cost of $E_{int}$,
and finally saturate to small, statistically steady values. For
run $\tt C6$, there are hardly any vortices in the initial
configuration, so the energies start fluctuating about their
statistically steady values very rapidly.

\begin{figure*}
\begin{center}
\includegraphics[height=4.cm]{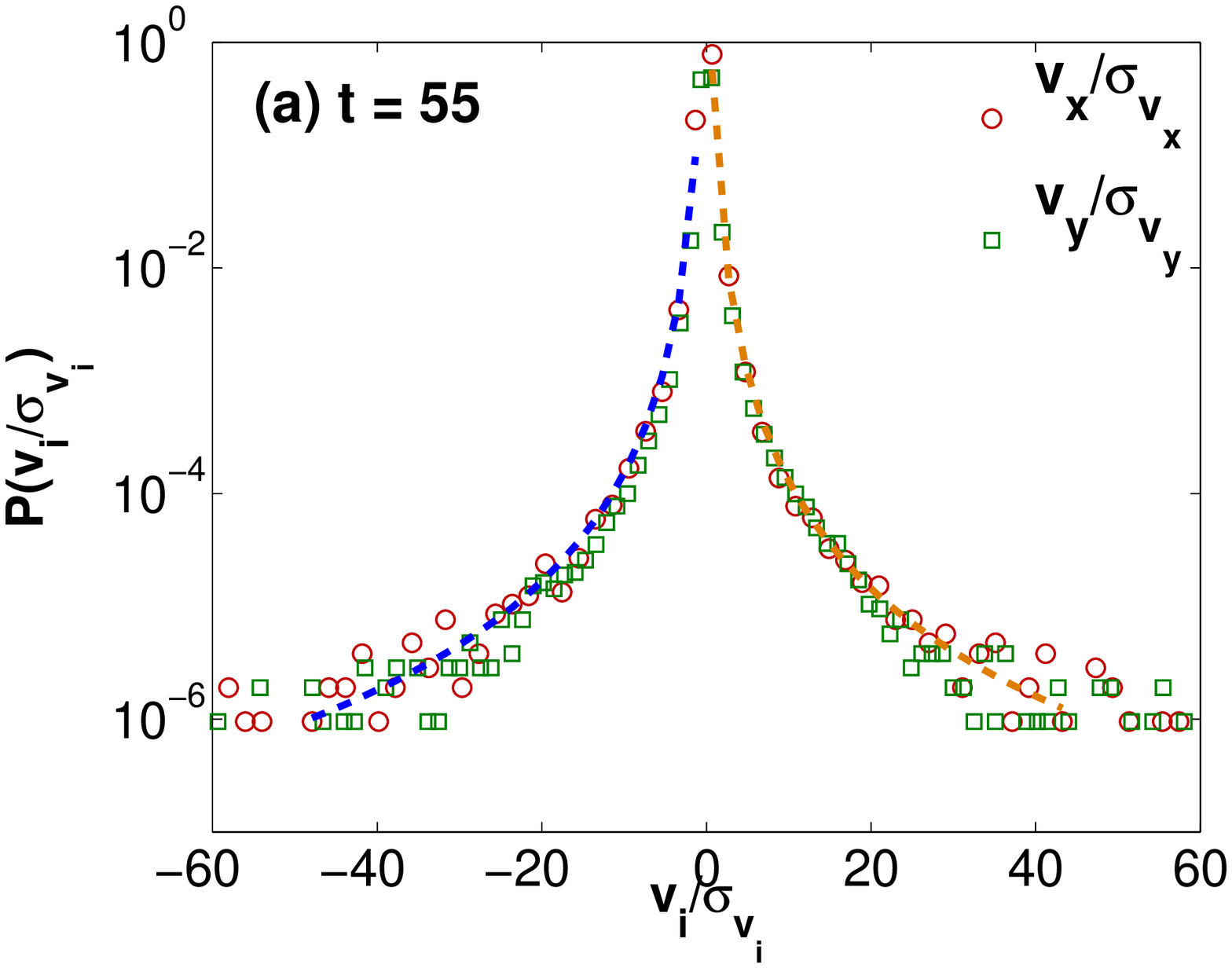}
\includegraphics[height=4.cm]{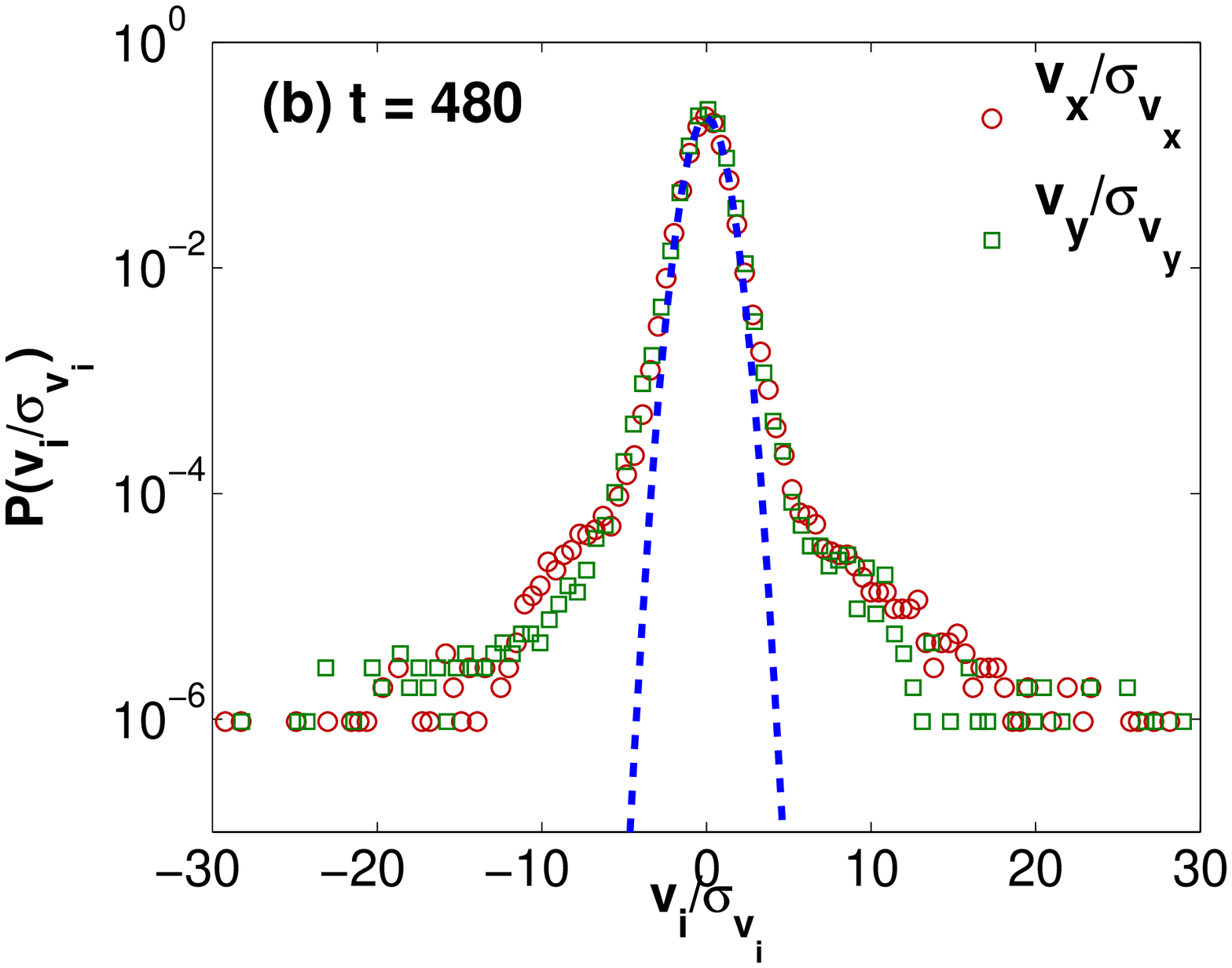}
\includegraphics[height=4.cm]{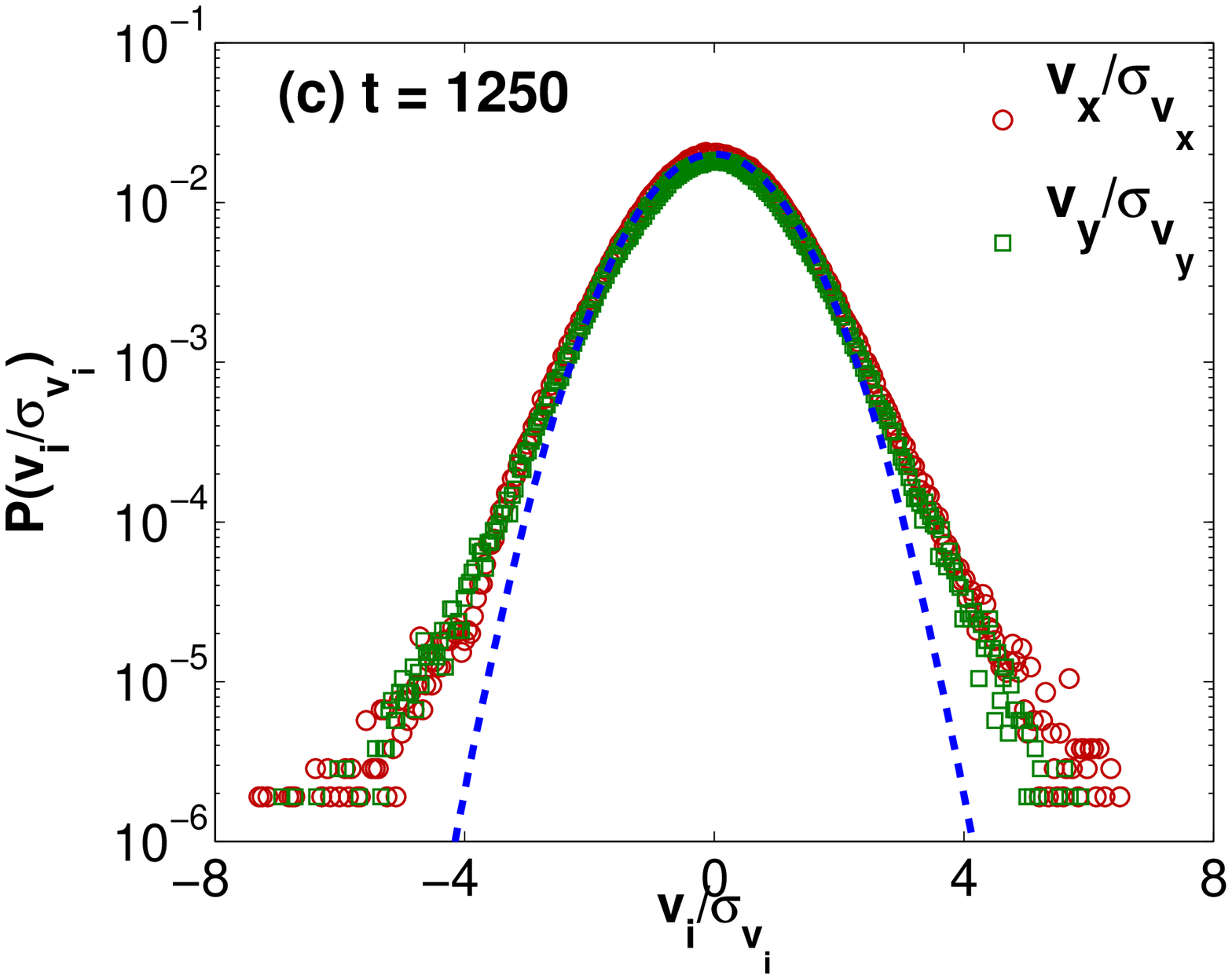}
\\
\includegraphics[height=4.cm]{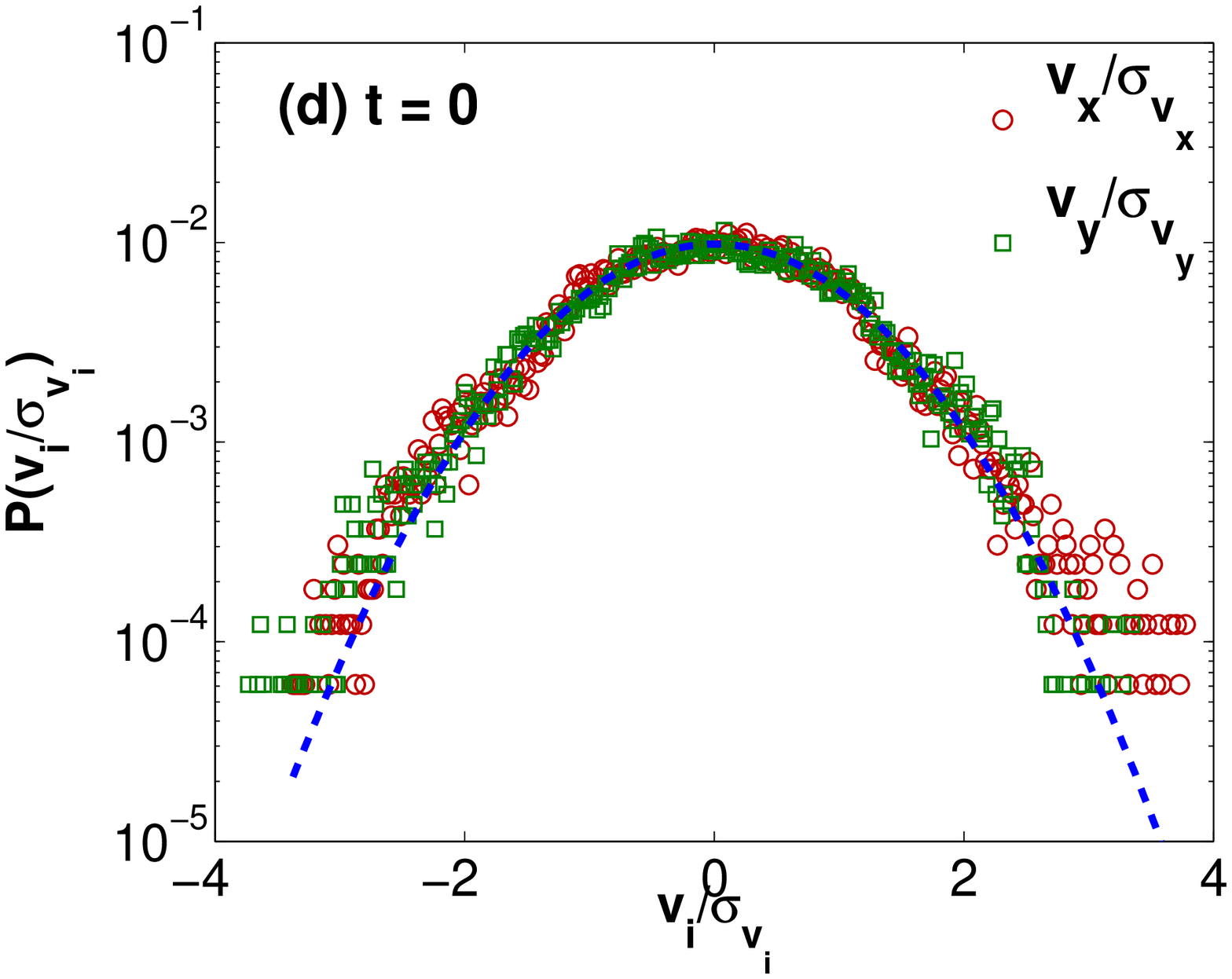}
\includegraphics[height=4.cm]{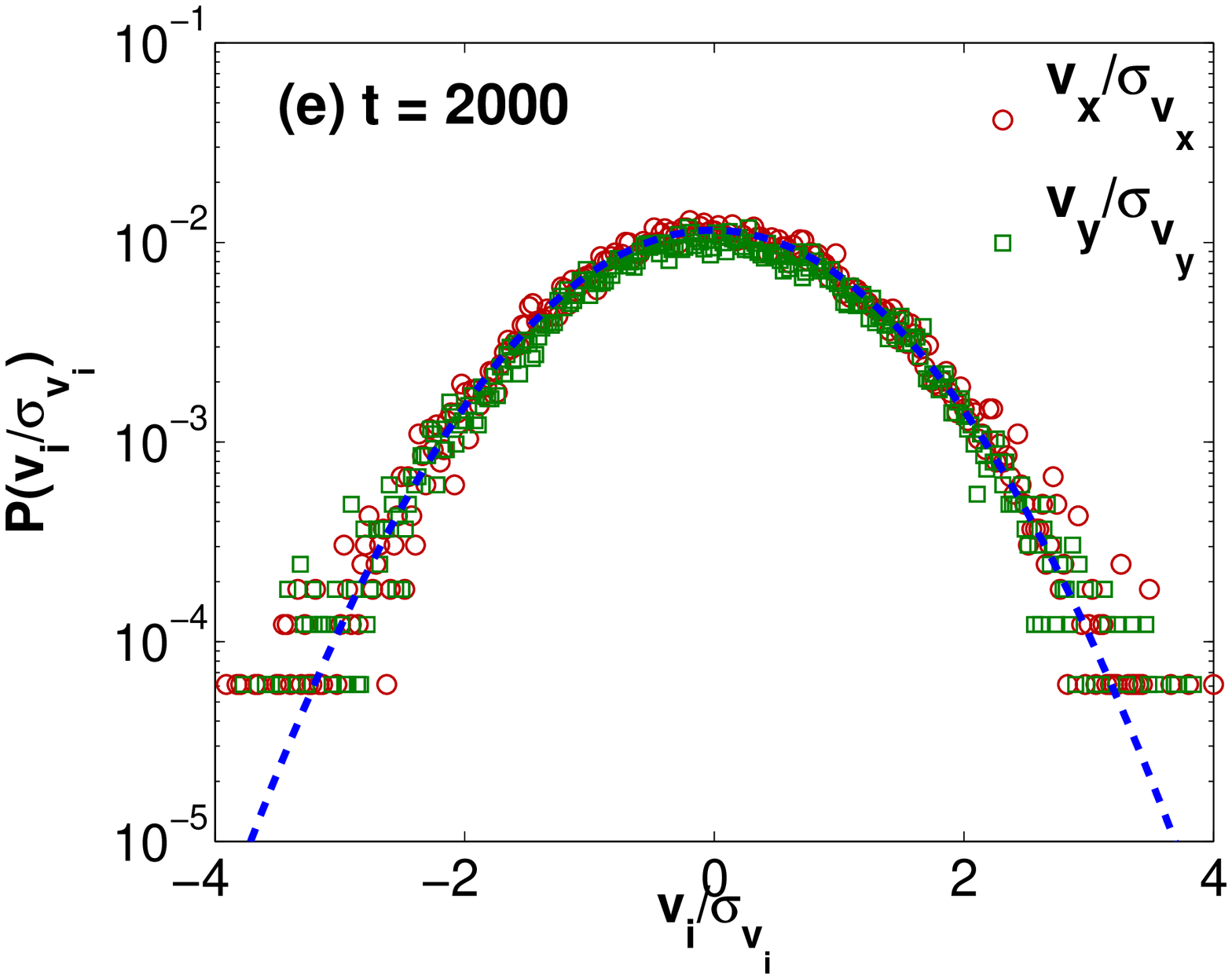}
\includegraphics[height=4.cm]{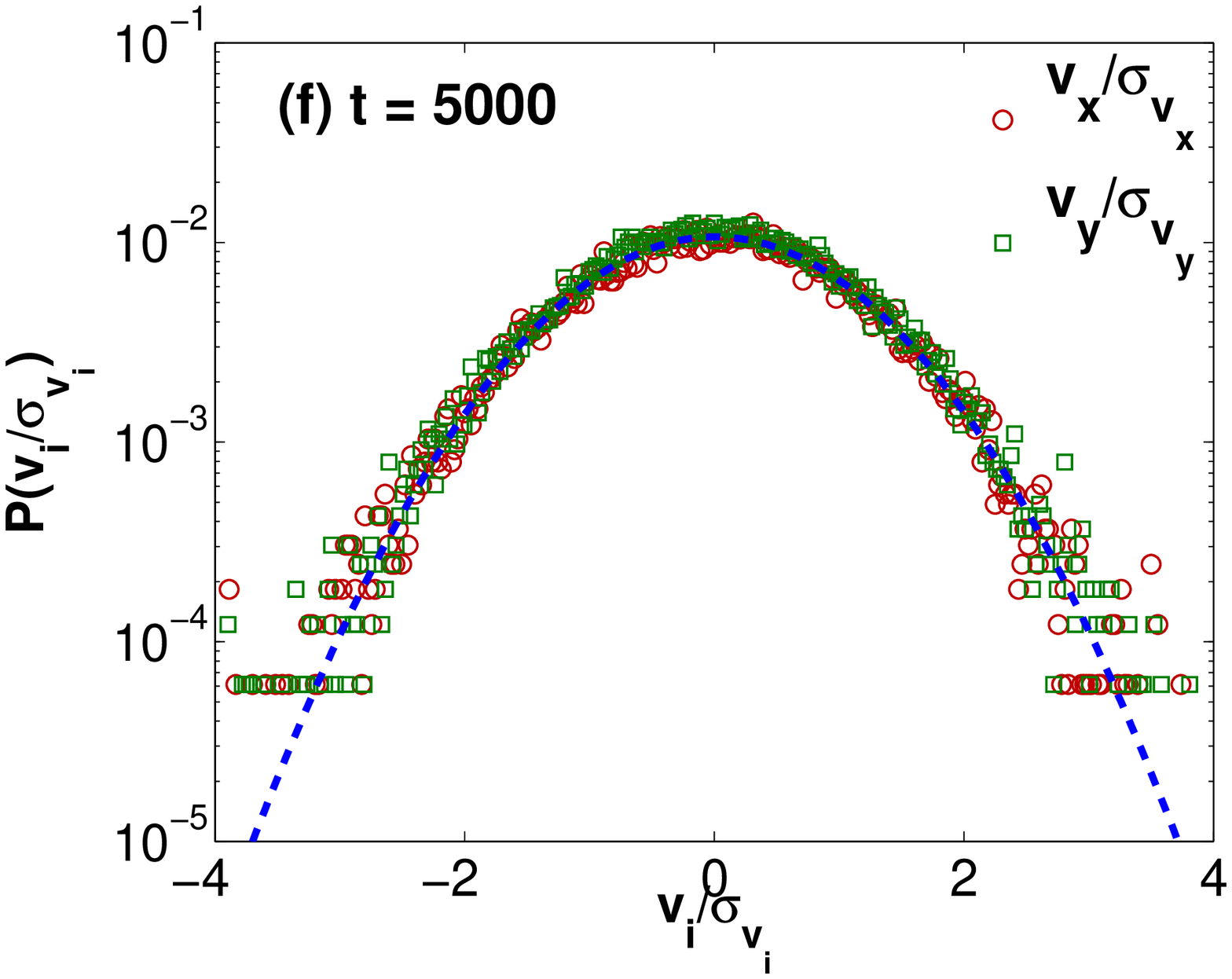}
\\
\includegraphics[height=4.cm]{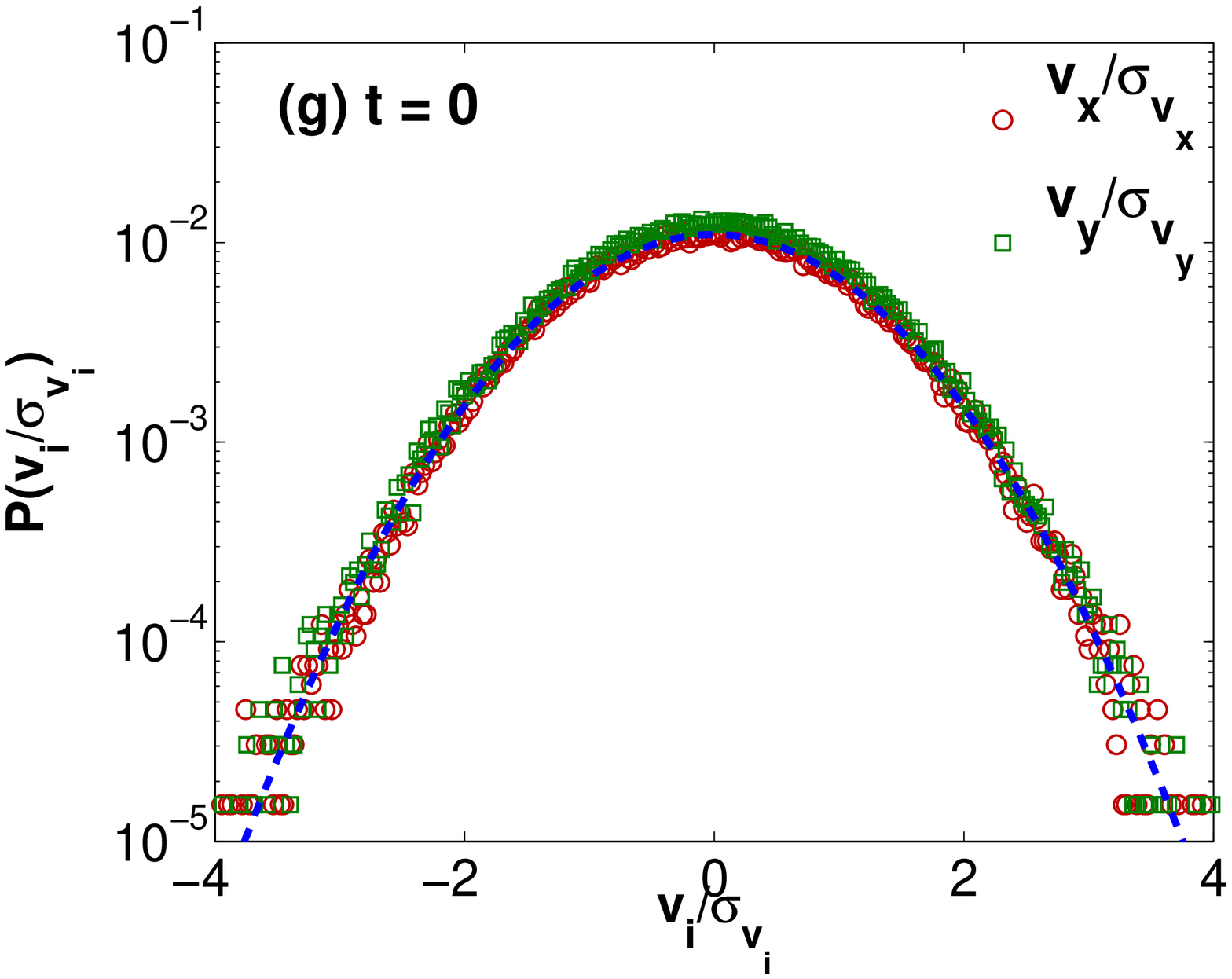}
\includegraphics[height=4.cm]{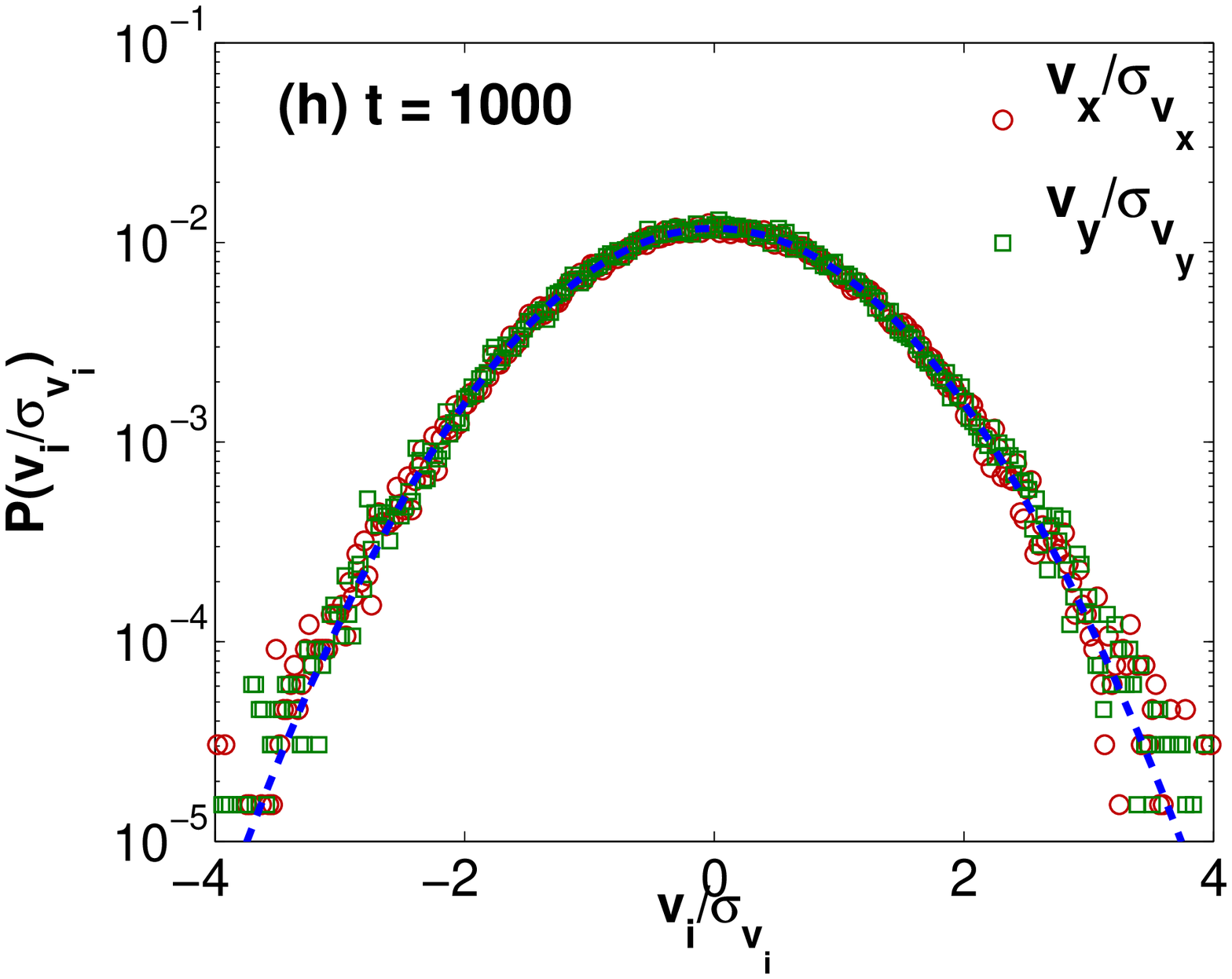}
\includegraphics[height=4.cm]{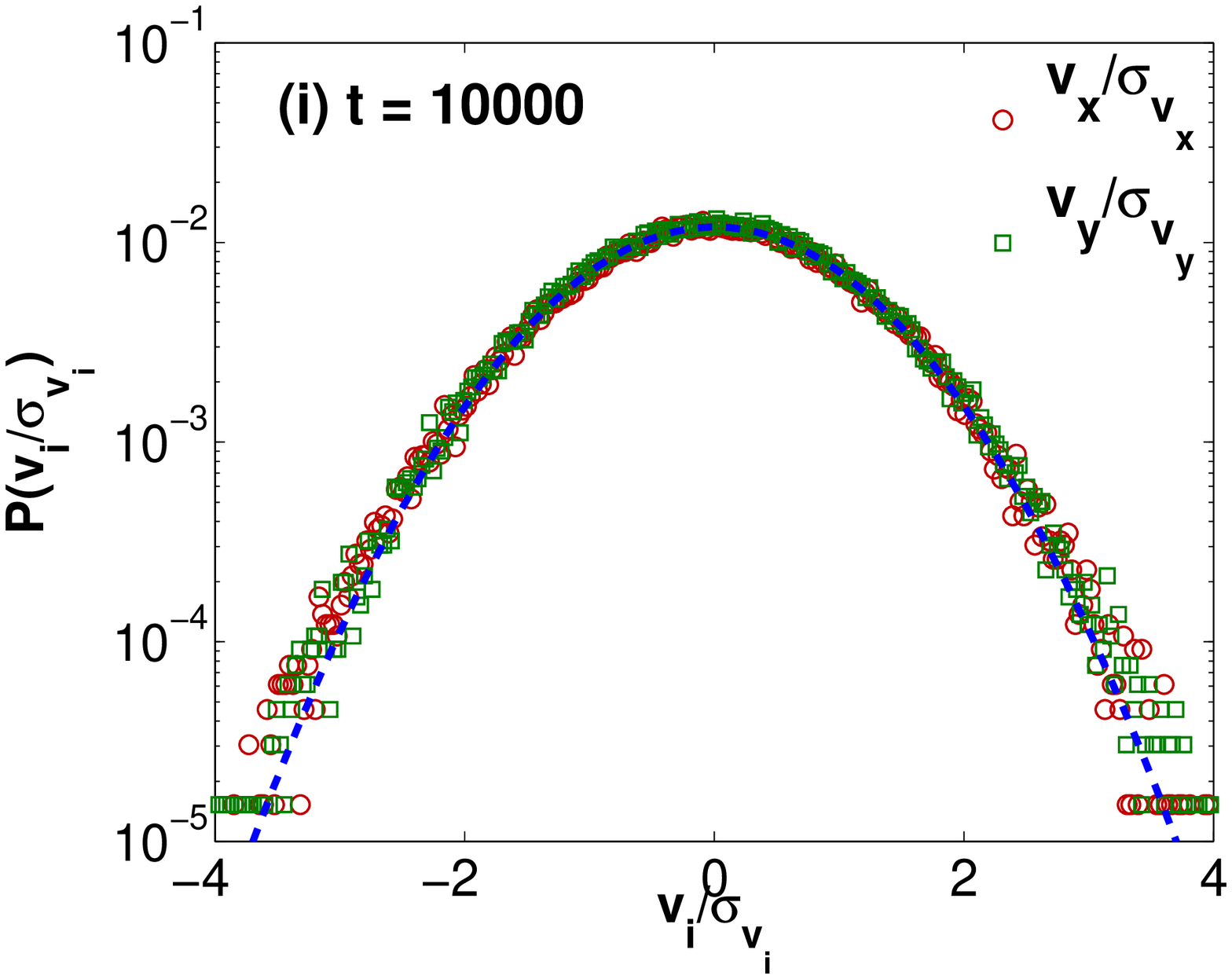}
\end{center}
\caption{\small Semilog (base $10$) plots of the PDFs of the $x$ 
(red circles) and $y$ (green squares) components of the velocity from our
DNS runs: (a)-(c) $\tt A1$, (d)-(f) $\tt B1$, and (g)-(i) $\tt C6$, 
corresponding to each of the three types of initial conditions $\tt IC1$, 
$\tt IC2$, and $\tt IC3$, respectively. The complete
time evolution of the PDFs (a)-(c) for the run $\tt A1$ is 
illustrated in the top-left panel of the video S1 
(supplementary material). The blue-dashed lines in (b)-(i)
indicate fits to Gaussian PDFs; the dashed lines in (a) indicate
power-law fits to the left (blue-dashed line) and right
(orange-dashed line) tails of the PDFs (see text).
}
\label{fig:velpdf}
\end{figure*}

In figure~\ref{fig:velpdf} we plot, at three instants of time,
the PDFs of $v_x$ and $v_y$, the Cartesian components of the
velocity, for our DNS runs $\tt A1$, $\tt B1$, and $\tt C6$,
which correspond, respectively, to  initial conditions of types
$\tt IC1$, $\tt IC2$, and $\tt IC3$.  For the run $\tt A1$, these
PDFs, in figures \ref{fig:velpdf} (a)-(c), show a crossover from
a distribution with power-law tails to one that is Gaussian; the
right and left tails of the PDFs in figure \ref{fig:velpdf} (a) can
be fit to the form $\sim v^{-\gamma}_i$, with $\gamma \simeq
3.2$, and $i=x$ or $y$ (we show fits only for $i=x$).  Such
power-law tails in velocity-component PDFs have been seen in
experiments~\cite{Paoletti2008prl} and some numerical
studies~\cite{White2010prlvpdf,Adachi2011biotsavartvpdf,Proment2012,Nowak2012pra}.
However, it has not been noted hitherto that, for turbulence in
the Fourier-truncated, 2D, GP equation with low-energy initial
conditions, such PDFs evolve, as $t$ increases, from PDFs with
power-law tails (figure~\ref{fig:velpdf} (a) for run $\tt A1$),
to ones with a Gaussian form near the mean, followed by broad
tails (figure~\ref{fig:velpdf} (b) for run $\tt A1$), and then to
more-or-less Gaussian  PDFs (figure~\ref{fig:velpdf} (c) for run
$\tt A1$), but with tails that can be fit to an exponential form.  This evolution towards
Gaussian PDFs is associated with the annihilation of vortices and
anti-vortices. The Video S1 in the Supplementary Material shows
the temporal evolution of this PDF in the left panel and the
spatiotemporal evolution of the pseudocolor plot of the vorticity
in the right panel.  The analogues of
figures~\ref{fig:velpdf}(a)-(c) for runs $\tt B1$ and $\tt C1$,
both of which have a negligibly small value of $E^i_{kin}$ at
$t=0$, are given, respectively, in
figures~\ref{fig:velpdf}(d)-(f) and
figures~\ref{fig:velpdf}(g)-(i).   

\begin{figure*}
\begin{center}
\includegraphics[height=5.5cm]{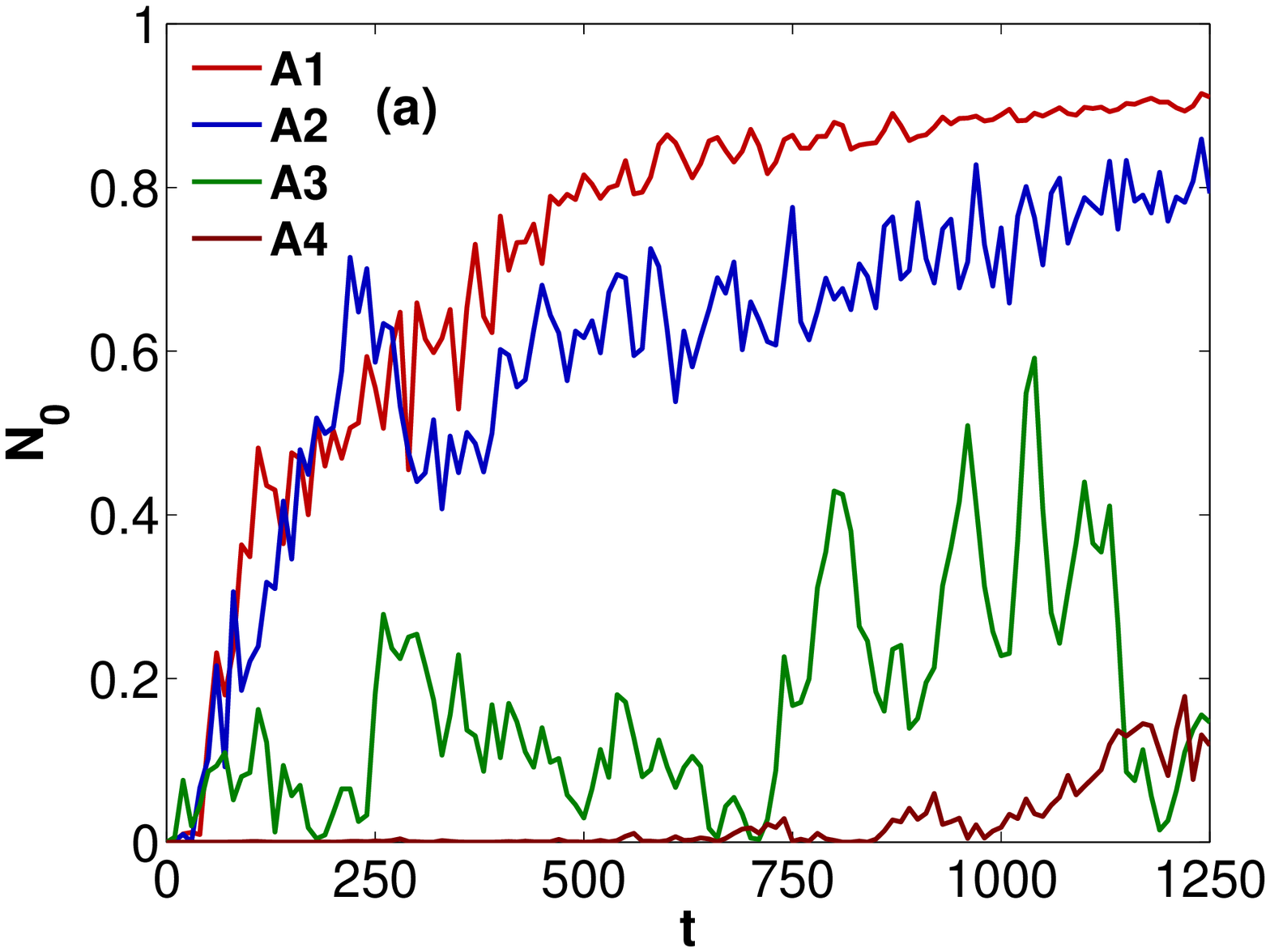}
\includegraphics[height=5.5cm]{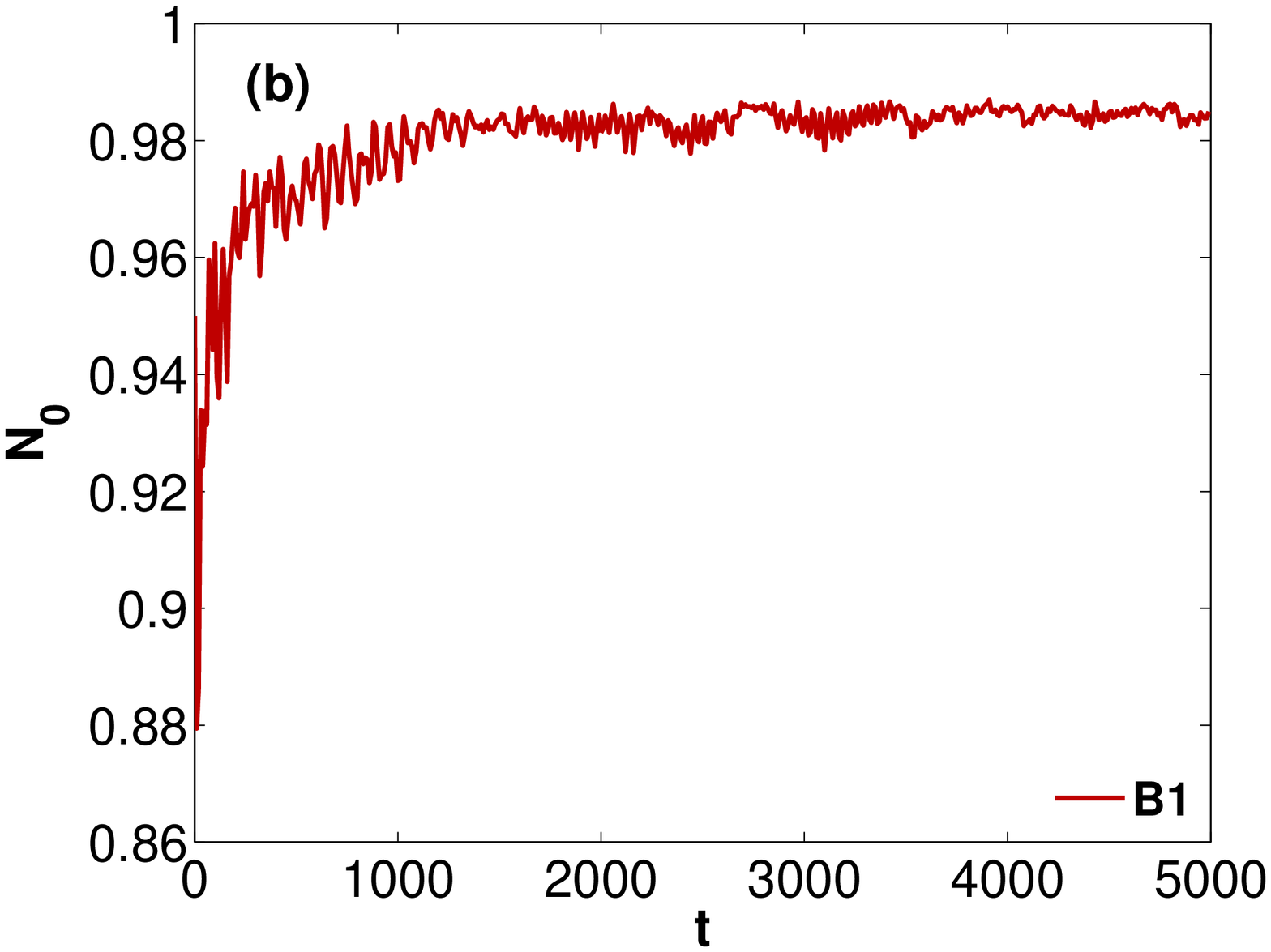}\\
\includegraphics[height=5.5cm]{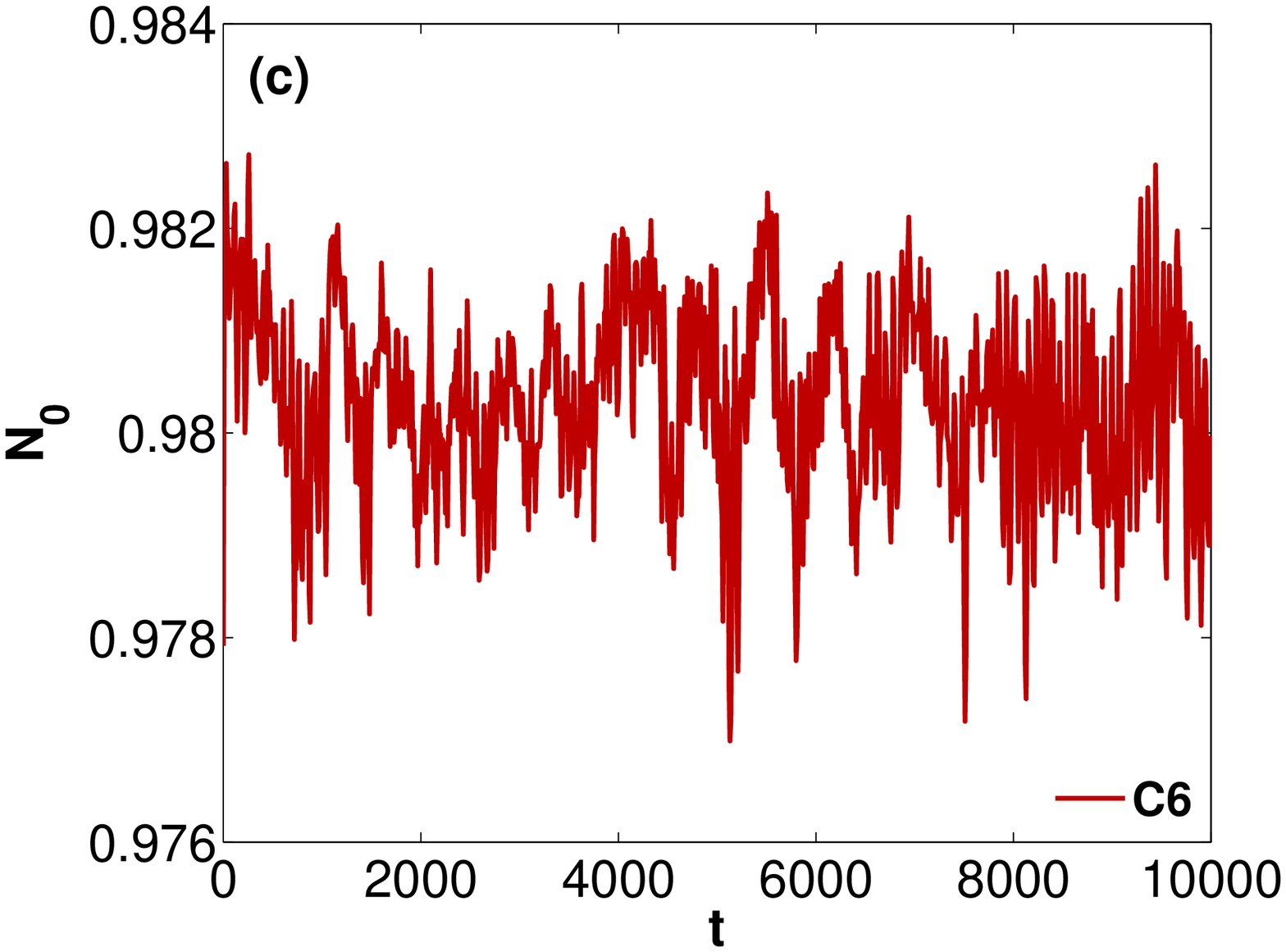}
\includegraphics[height=5.5cm]{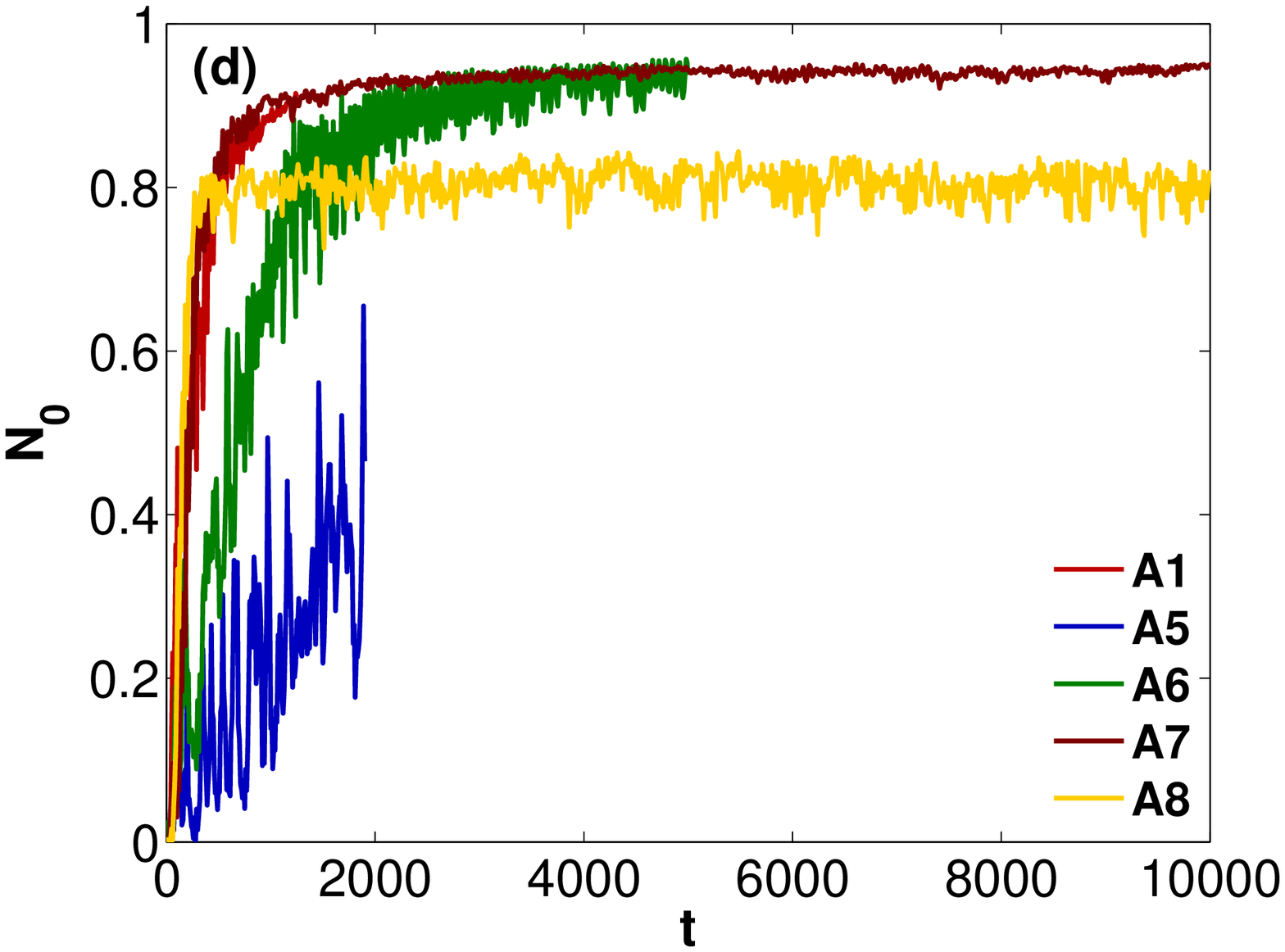}
\end{center}
\caption{\small Plots versus time $t$ of the population $N_0$, in the
zero-wave-number mode, from our DNS runs (a) $\tt A1$-$\tt A4$ (initial 
condition of type $\tt IC1$), (b) $\tt B1$ (initial 
condition of type $\tt IC2$), (c) $\tt C6$ (initial 
condition of type $\tt IC3$), and 
(d) $\tt A1$ and $\tt A5$-$\tt A8$, for five values for the
number of collocation points $N^2_c$, namely, $1024^2,\, 512^2,\,
256^2,\, 128^2$, and $64^2$. 
}
\label{fig:nk0}
\end{figure*}

We turn now to the time evolution of the population $N_0(t)$, in
the $k=0$ mode~\cite{Berloff2002pra,Connaughton2005prl,RicaphysicaD2009},
and its dependence on the initial conditions.  In
figure~\ref{fig:nk0} (a) we plot $N_0$ versus $t$ for the runs
$\tt A1$-$\tt A4$ (red, blue, green, and brown curves, respectively),
which use initial configurations of type $\tt IC1$; these figures
show that $N_0(t)$ increases with $t$, on average, and depends on
$E$, $g$, $k_0$, and $\sigma$. For the runs $\tt A1$ and $\tt A2$
(red and blue curves in figure~\ref{fig:nk0} (a)), $N_0(t)$
approaches a saturation value for the time scales probed by our
simulations; figure~\ref{fig:nk0} (a) also shows that, as we
increase $g$ (red, blue, and green lines in figure~\ref{fig:nk0}
(a)), the fluctuations in $N_0$ are enhanced and its large-$t$ value, which
it seems to approach asymptotically, diminishes.  By comparing
the runs $\tt A1$ and $\tt A4$ (red and brown lines in
figures~\ref{fig:nk0} (a)), we see that the latter has a higher
value of $E$ than the former, because both $k_0$ and $\sigma$ are
smaller for $\tt A1$ than for $\tt A4$; thus, $N_0(t)$ grows more
slowly in $\tt A4$ than in $\tt A1$; and, after an equal amount
of simulation time, its value in $\tt A4$ is nearly an order of
magnitude lower than in $\tt A1$; the former shows large
fluctuations in $N_0(t)$ and no sign of saturation.  The run $\tt
B1$ (figure \ref{fig:nk0} (e)) uses an initial configuration of
type $\tt IC2$, with a large value of $N_0(t=0)=0.95$; in this
case, after a period of initial transients, $N_0(t) \to 0.98$
over our simulation time. The run $\tt C6$ (figure \ref{fig:nk0}
(f)) uses an initial condition of type $\tt IC3$; here $N_0(t)$
fluctuates slightly but remains close to its initial value (cf.
~\cite{Connaughton2005prl,RicaphysicaD2009}).  

To study the dependence of $N_0(t)$ on the number of collocation
points $N_c^2$, we evolve the initial configuration of $\tt
A1$ for $N_c=512$ (run $\tt A5$), $N_c=256$ (run
$\tt A6$), $N_c=128$ (run $\tt A7$), and $N_c=64$ (run $\tt A8$). Figure
\ref{fig:nk0} (g) shows plots of $N_0(t)$ versus $t$ for 
these five runs; clearly, the initial evolution of $N_0(t)$
depends significantly on $N_c$; however, the large-$t$ 
values of $N_0(t)$, on the time scales of our runs, are 
comparable ($\simeq 0.9$) for the runs wth $N_c=128$  
(run $\tt A7$), $N_c=256$  (run
$\tt A6$), and $N_c=1024$ (run $\tt A1$). In contrast, the saturation
value for the run with $N_c=64$ (run $\tt A8$) is $\simeq 0.8$. 
For the run $\tt A5$ ($N_c=512$), $N_0(t)$ shows 
large fluctuations and no sign of saturation over the 
time scale that we have covered; this suggests that
$N_0(t)$ also depends on the realisation of the random phases
$\Theta(k_x,k_y)$ in \eref{eq:ic1inconfig04}. These plots of $N_0(t)$
illustrate that complete thermalization proceeds very slowly
for $N_0$; in the completely thermalized state of the
Fourier-truncated, 2D, GP system, $N_0$ must vanish in the 
thermodynamic limit by virtue of the Hohenberg-Mermin-Wagner
theorem~\cite{Mermin1966,Hohenberg1967}; however, it is not easy to realize this
limit in finite-size systems and with the limited run times that 
are dictated by computational resources. We discuss these issues 
again in section~\ref{subsection:completethermalization} and also refer the reader 
to \cite{Damle1996,RicaphysicaD2009}.

\subsection{Initial transients and the onset of thermalization}
\label{subsection:initialtransientsandonset}

\begin{figure*} \includegraphics[height=5.5cm]{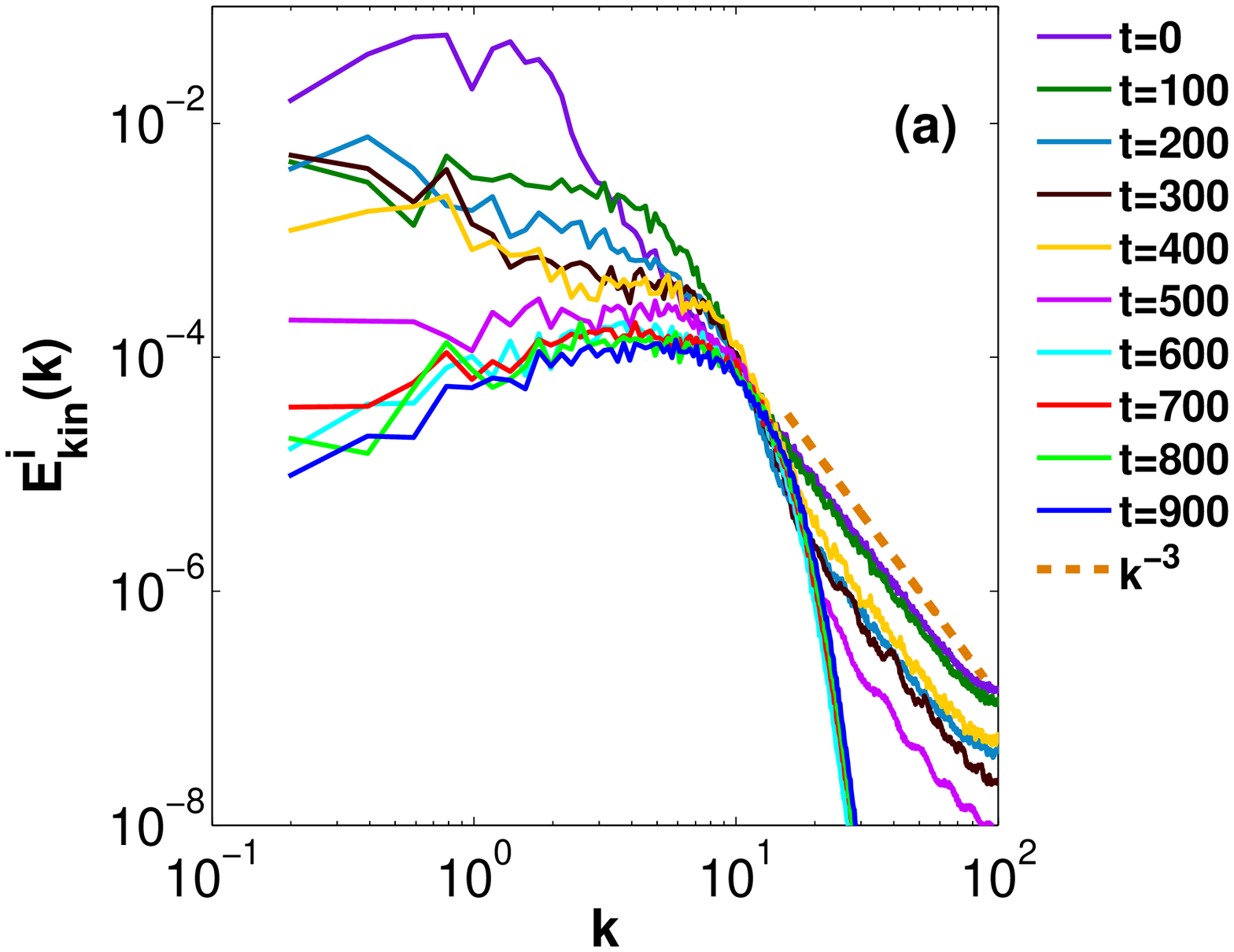}
\includegraphics[height=5.5cm]{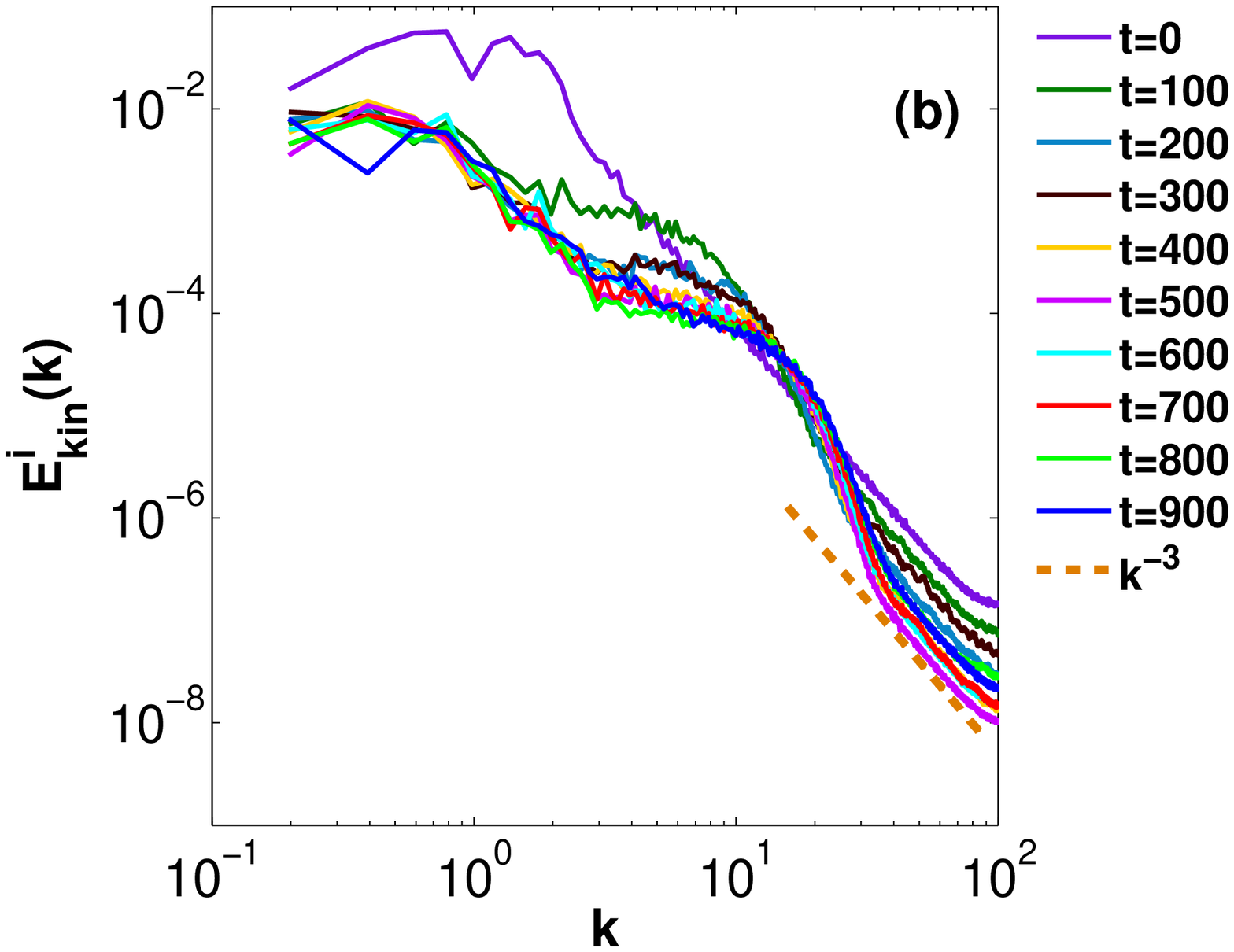}
\\ 
\includegraphics[height=5.5cm]{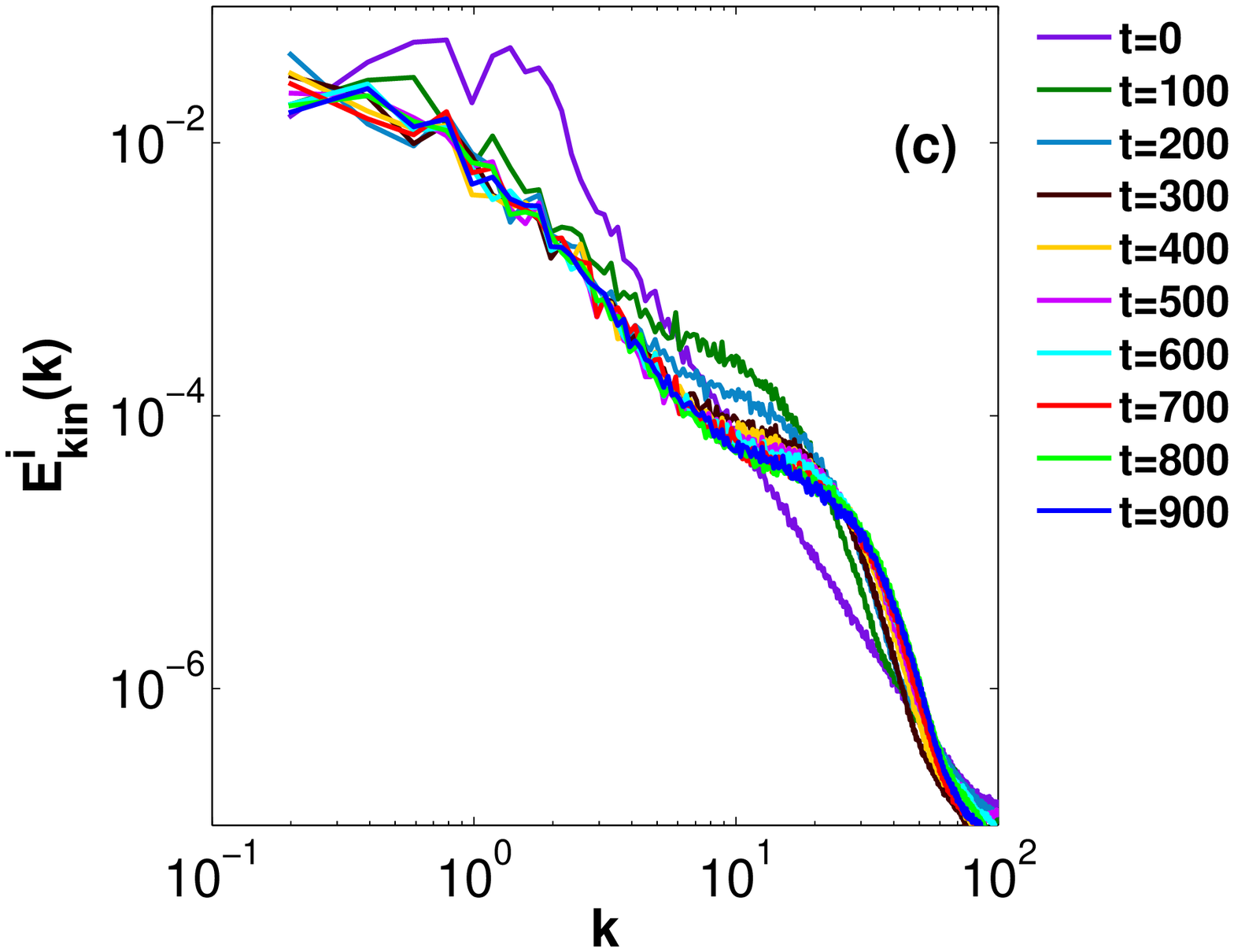}
\includegraphics[height=5.5cm]{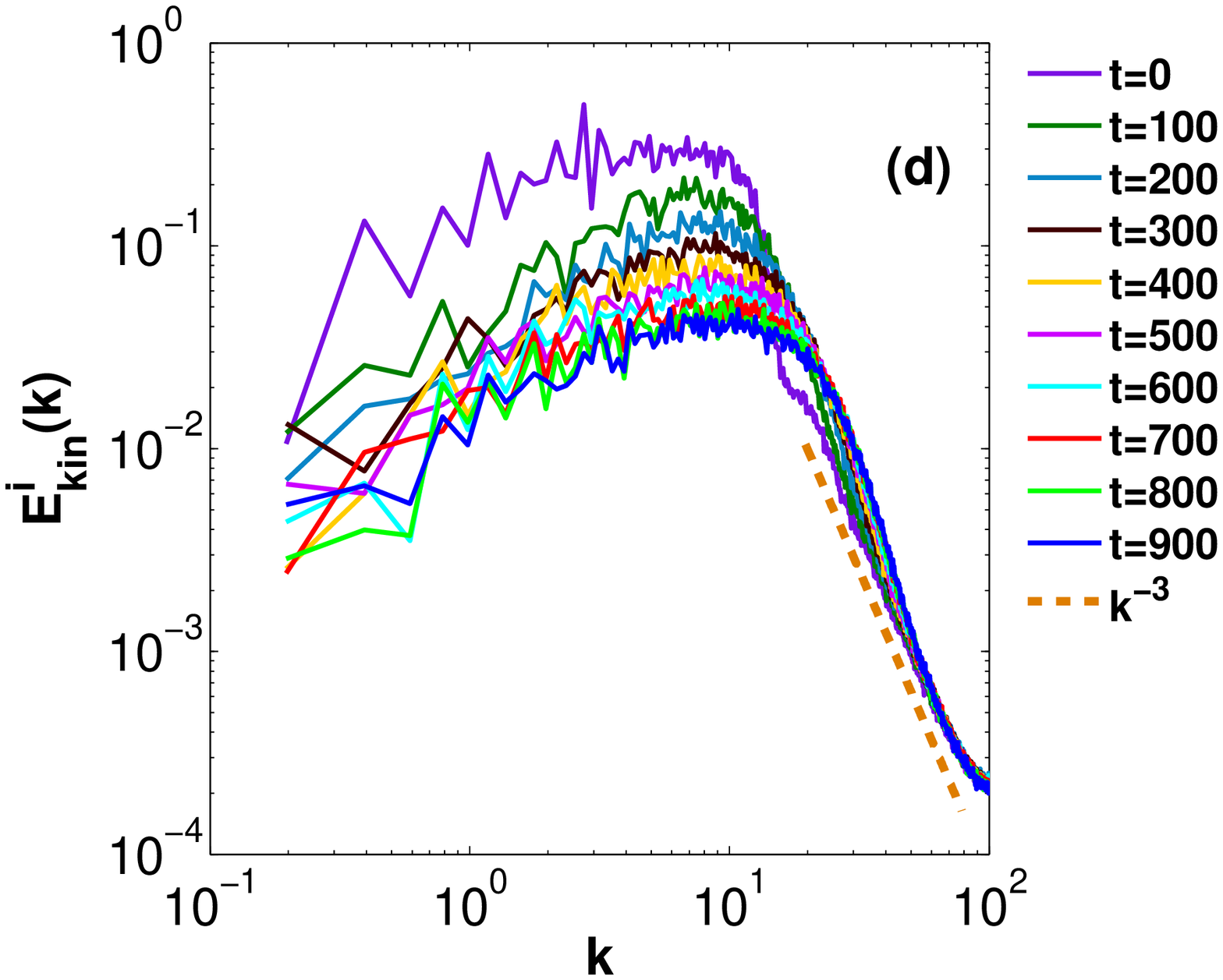}
\caption{\small Log-log (base 10) plots of the spectra $E^i_{kin}(k)$
from our DNS runs (a) $\tt A1$, (b) $\tt A2$, (c) $\tt A3$, and (d)
$\tt A4$ at different times $t$ (indicated by curves of different
colours); a $k^{-3}$ power law is shown by orange-dashed
lines. The complete time evolution of the spectra in (a), (b), (c), and (d) is
illustrated in the video S2 (supplementary material).
}
\label{fig:ikestransients} \end{figure*}

The initial stages of the evolution of energy spectra for the 
Fourier-truncated, 2D, GP equations are qualitatively different for
initial conditions of types $\tt IC1$, $\tt IC2$, and $\tt IC3$.
The first type begins with a sizeable incompressible kinetic
energy spectrum $E^i_{kin}(k)$; and the initial transients are
associated with the annihilation and creation of
vortex-antivortex pairs, the associated depletion of
$E^i_{kin}(k)$, and the growth of the other energy components
~\cite{Nazarenko2007freedecay2d}. In contrast, runs with initial
conditions of types $\tt IC2$ and $\tt IC3$ start with a very
small incompressible-energy component, therefore, even the early
stages of their dynamical evolution are akin to the late stages
of the dynamical evolution with initial conditions of type $\tt
IC1$.  In figures \ref{fig:ikestransients} (a)-(d) we show the
time evolution of the spectra $E^i_{kin}(k)$, for the runs $\tt
A1$, $\tt A2$, $\tt A3$, and $\tt A4$, to ascertain the presence
of scaling behaviour, if any.  We find that, in the low-$k$
region, $E^i_{kin}(k)$ lacks a well-defined scaling region
(unlike in ~\cite{Numasato2Dgp2010}); indeed, this region depends
on the initial configuration, changes continuously with time,
and, in particular, a $k^{-5/3}$ scaling region is tenable (a)
over a range of wave numbers that is very tiny and (b) over a
fleetingly short interval of time (around $t=50$ for the run $\tt
A1$). At large wave numbers, $E^i_{kin}(k) \sim k^{-3}$, during
the initial stages of evolution, because of the presence of the
vortices ~\cite{BradleyPhysRevX2012}; this power-law form holds over the
same time scales for which the PDF $P(v_x/\sigma_{v_x}) \sim
v_x^{-\gamma}$ (figures~\ref{fig:velpdf} (a)-(b)).

\begin{figure*}
\begin{center}
\includegraphics[height=4.cm]{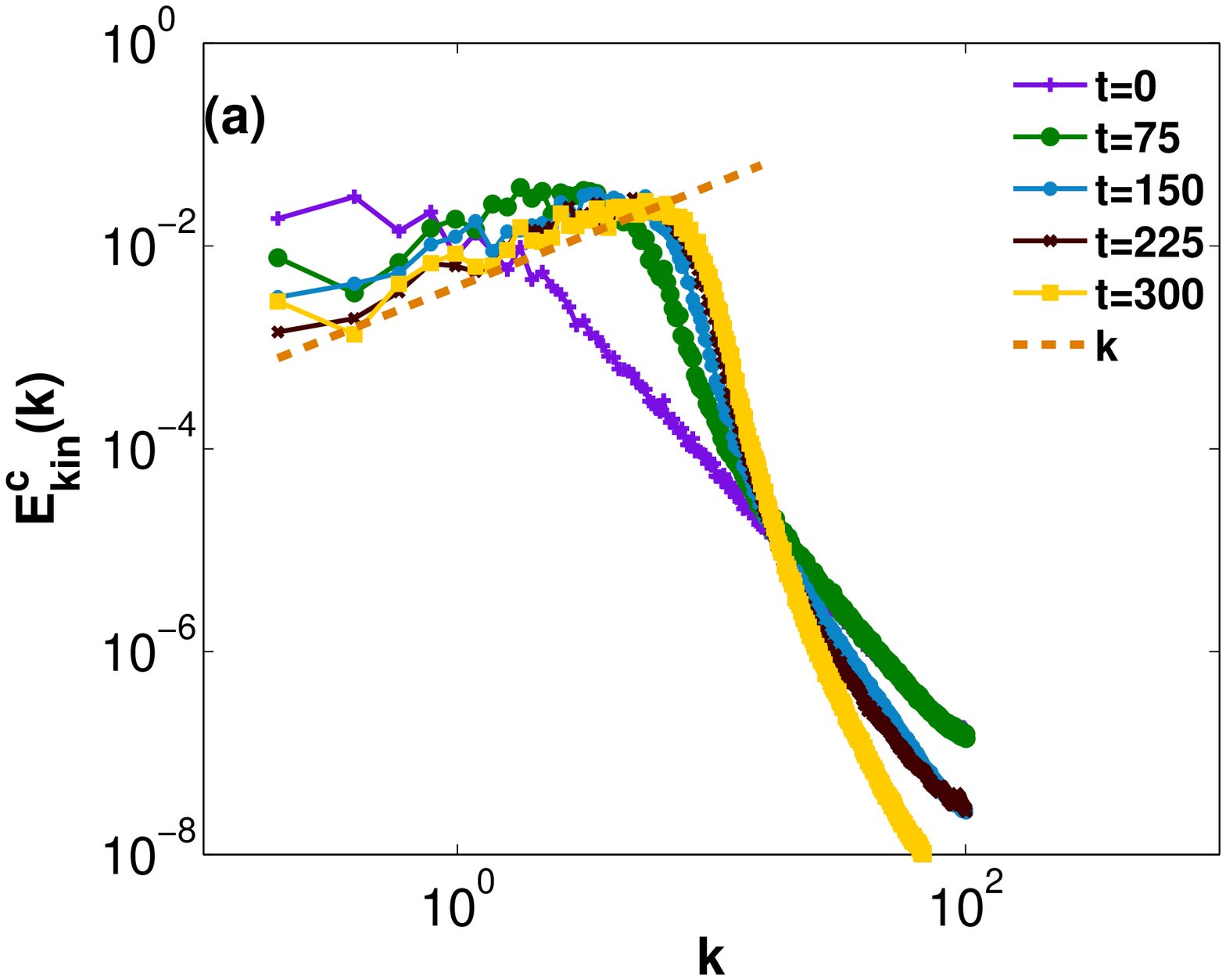}
\includegraphics[height=4.cm]{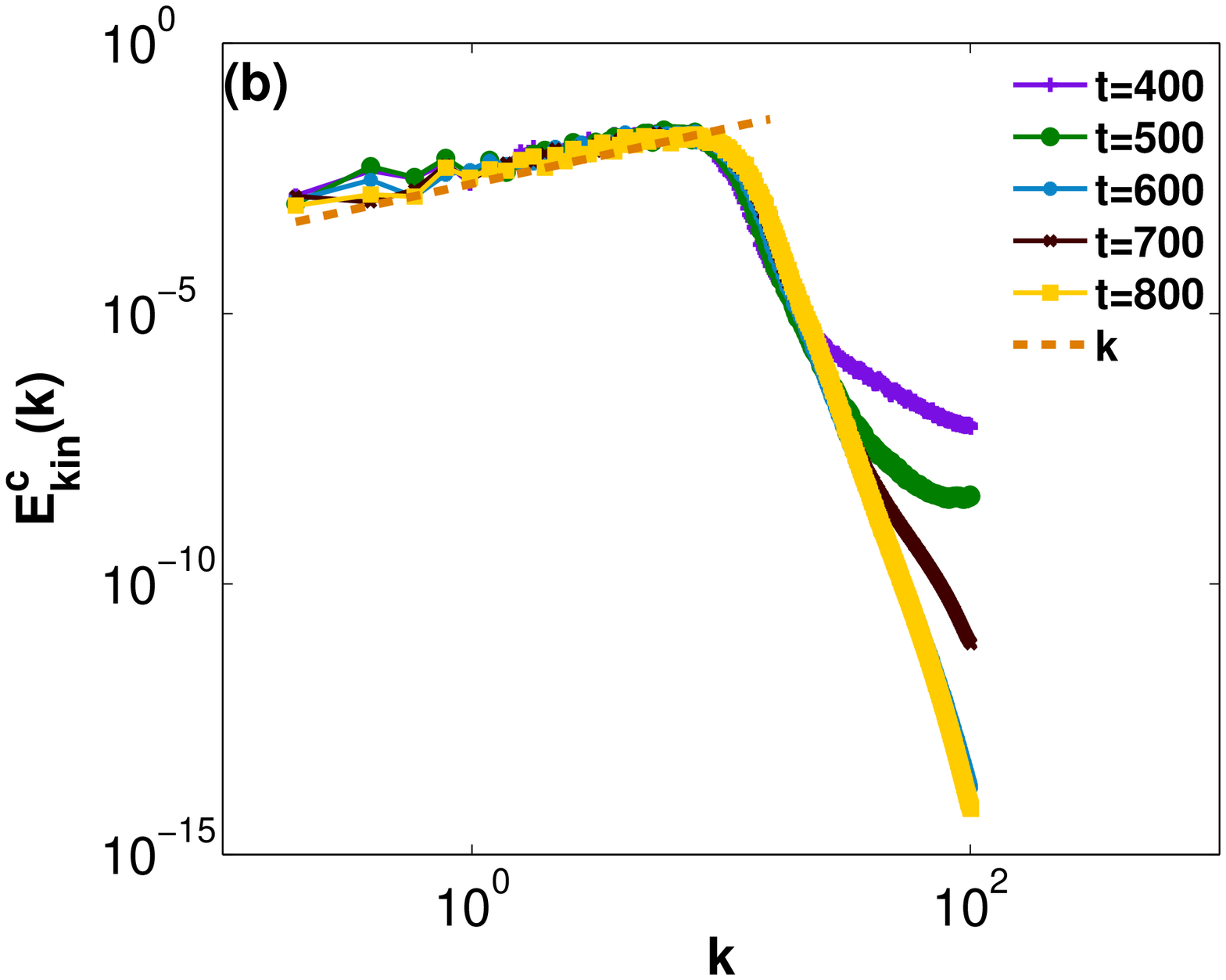}
\includegraphics[height=4.cm]{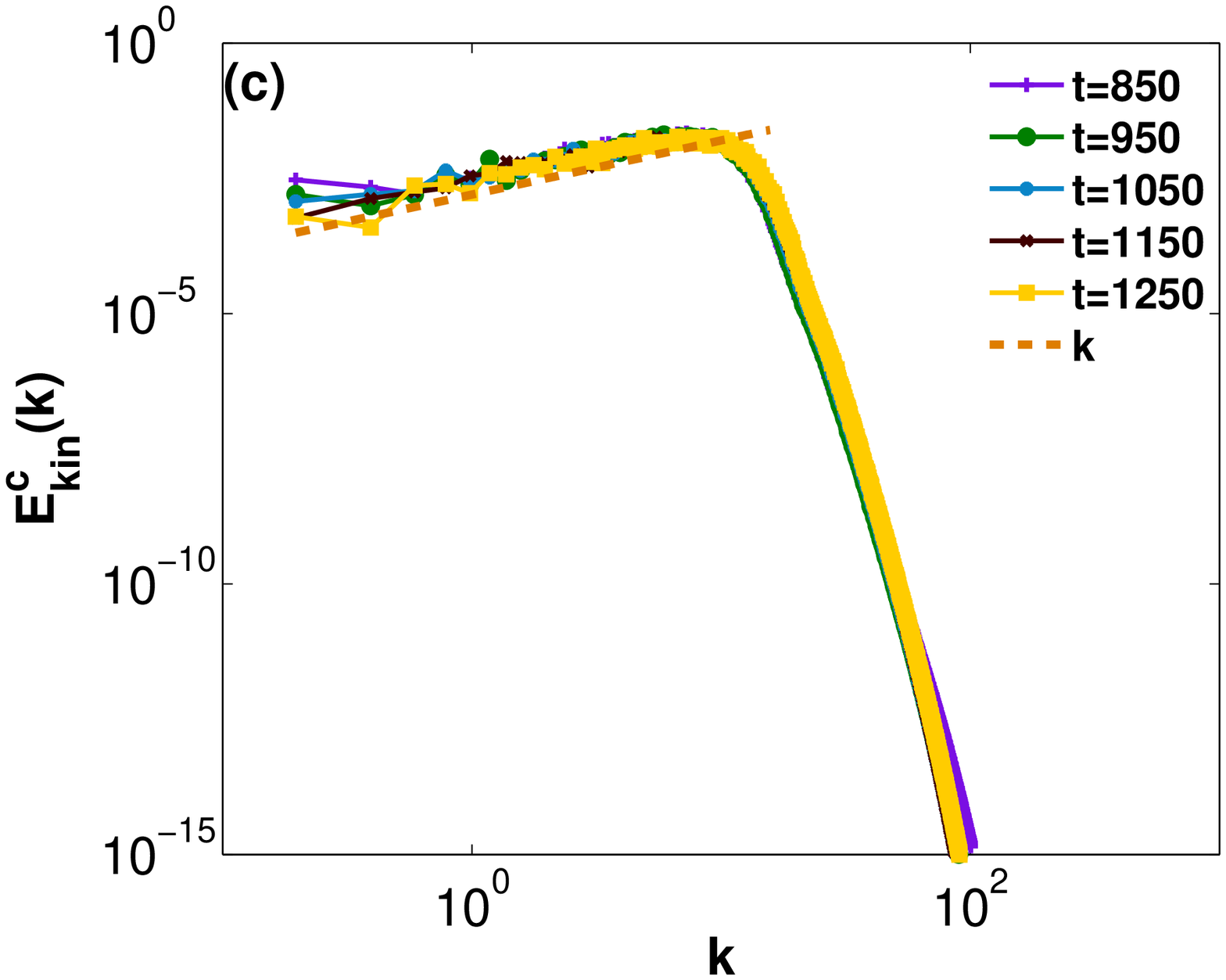}
\includegraphics[height=4.cm]{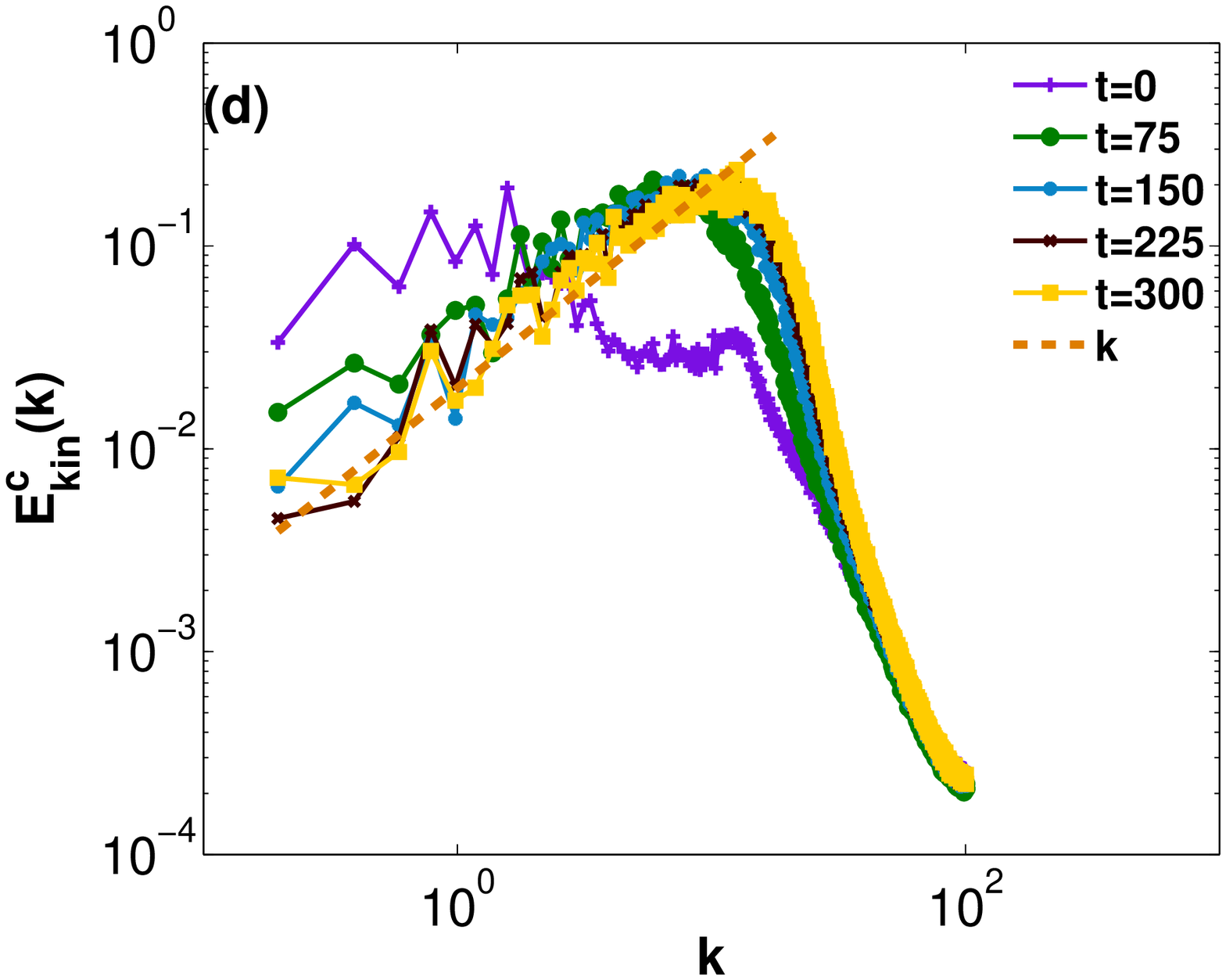}
\includegraphics[height=4.cm]{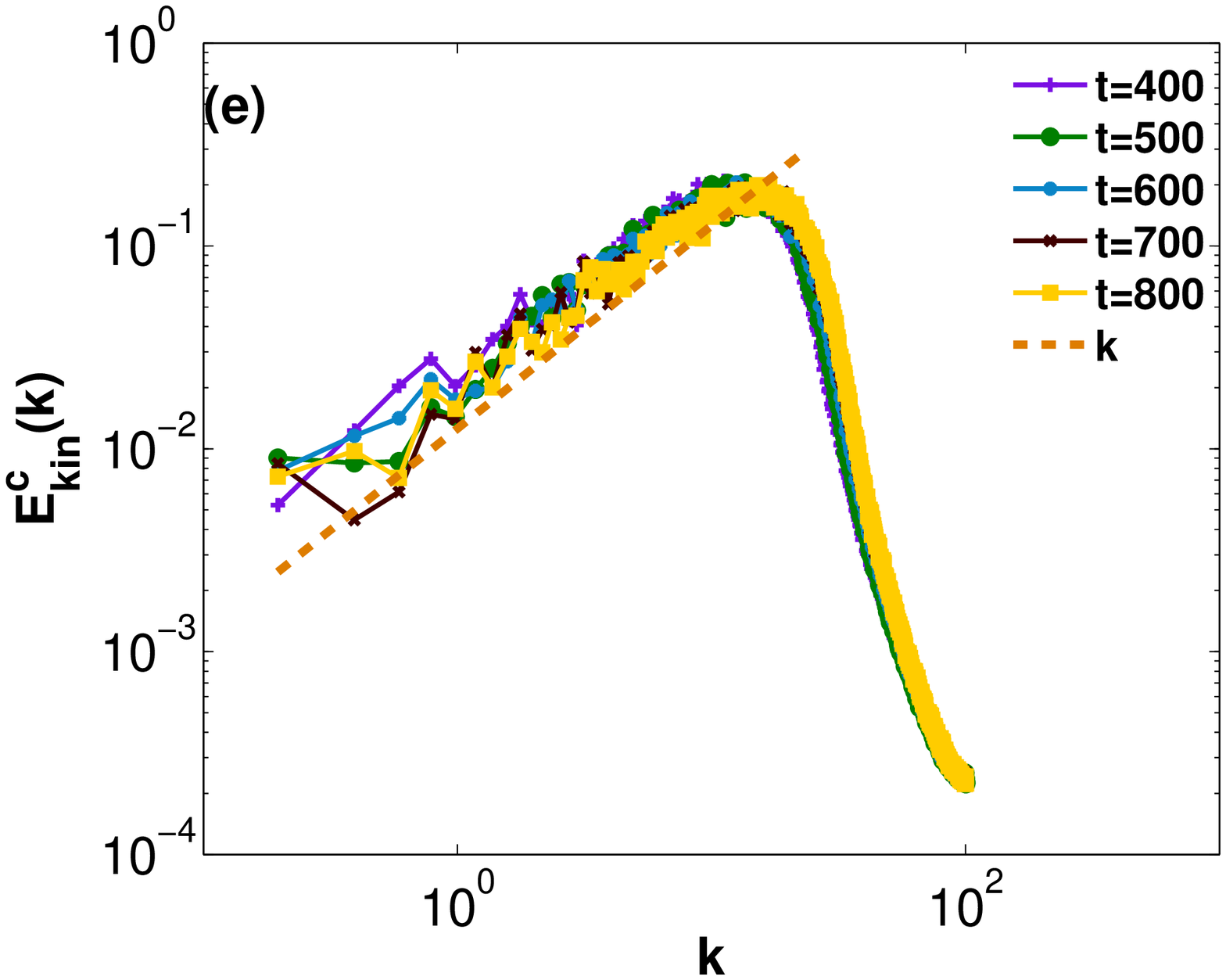}
\includegraphics[height=4.cm]{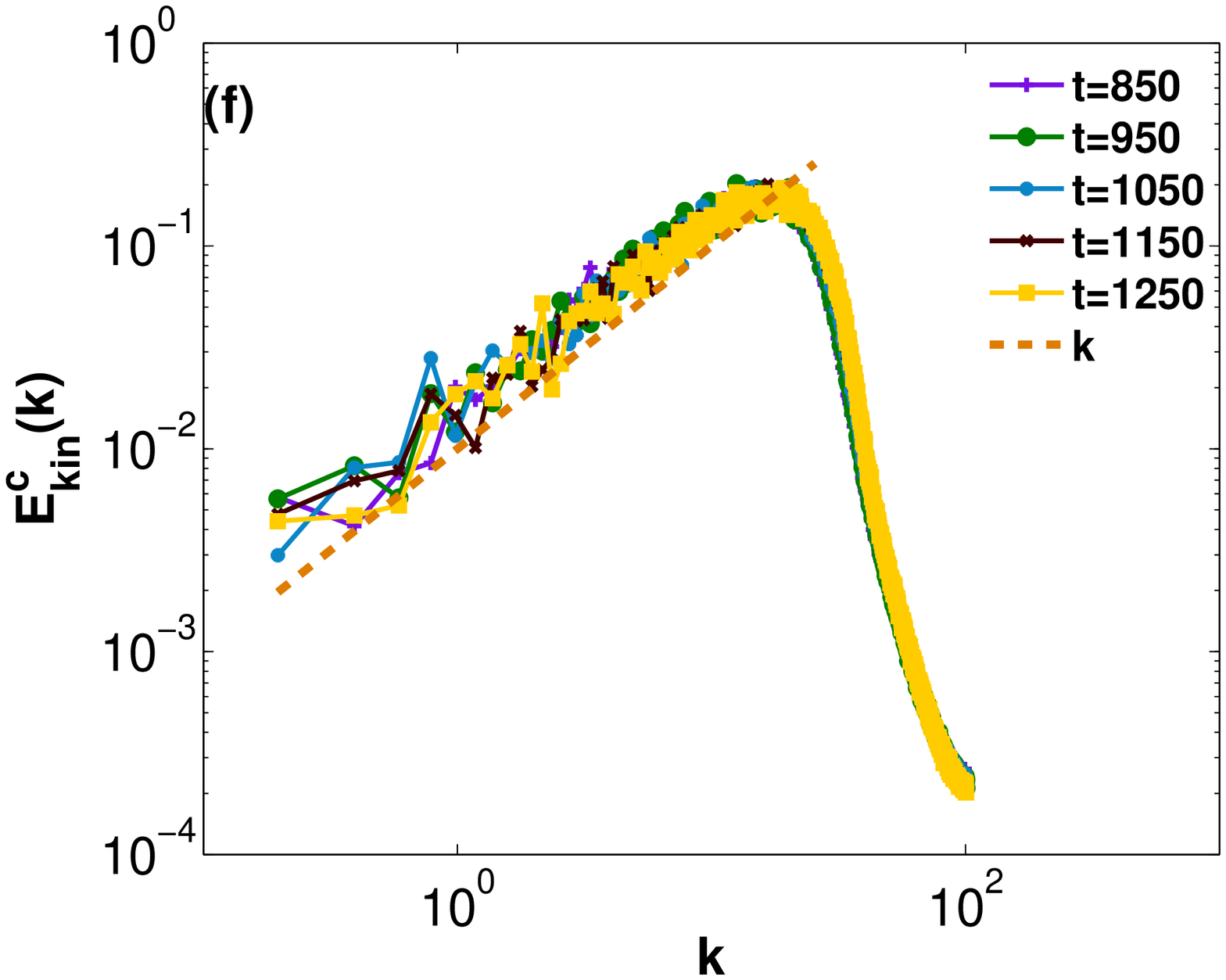}
\includegraphics[height=4.cm]{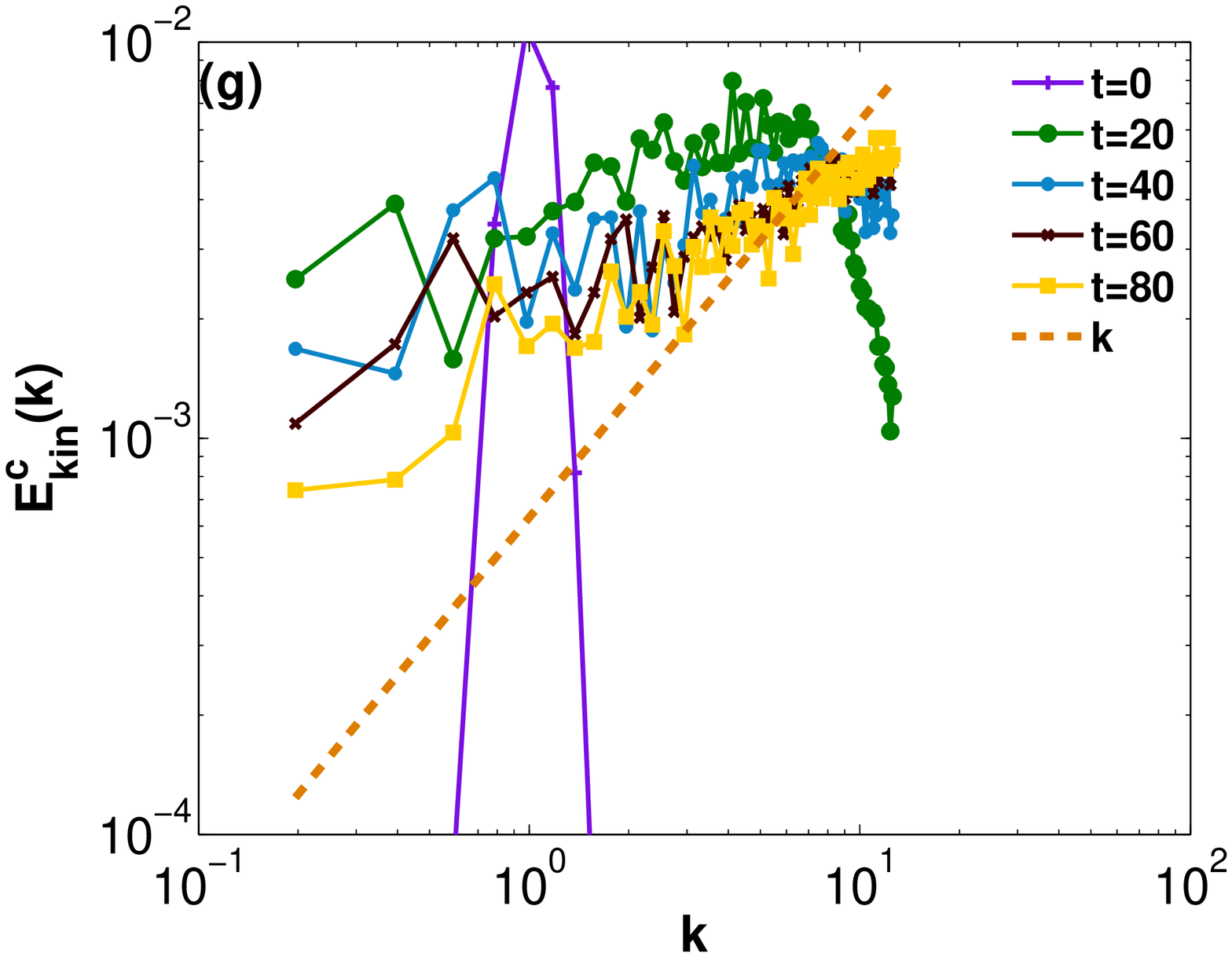}
\includegraphics[height=4.cm]{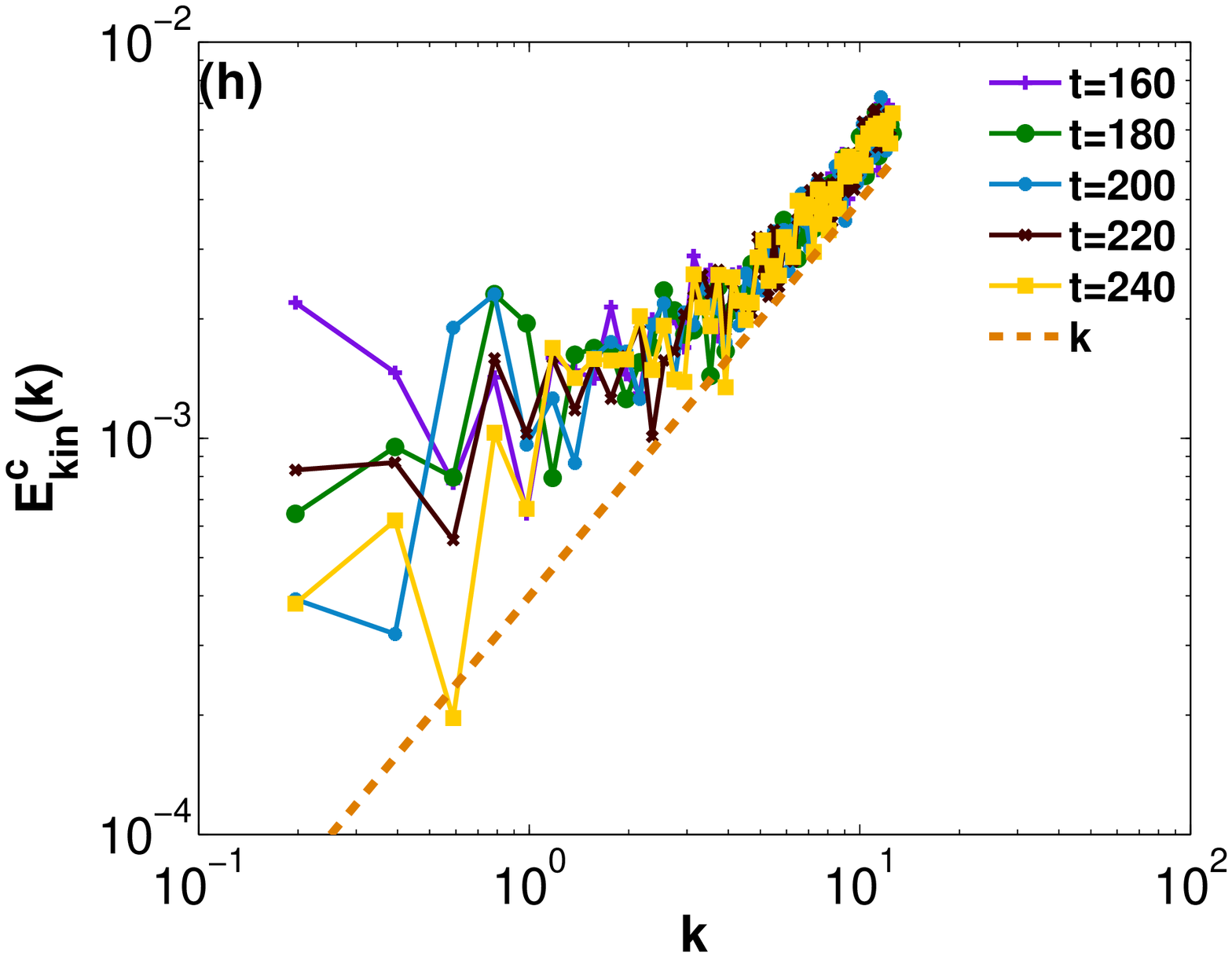}
\includegraphics[height=4.cm]{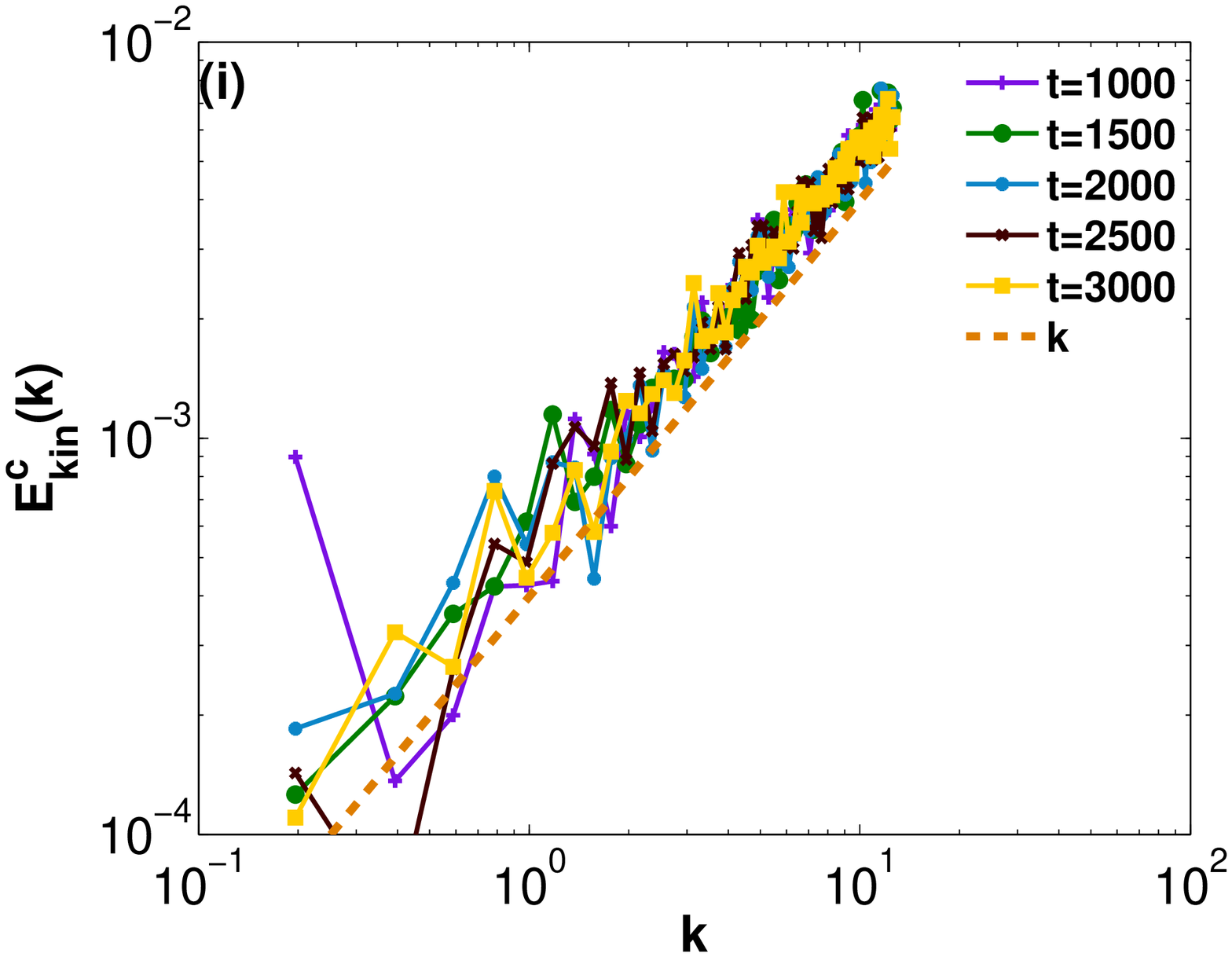}
\end{center}
\caption{\small Log-log (base 10) plots of the spectra $E^c_{kin}(k)$ 
from our DNS runs (a)-(c) $\tt A1$, (d)-(f) $\tt A4$, and 
(g)-(i) $\tt B1$ at different times $t$ (indicated by curves of different
colours); a $k$ power law is shown by orange-dashed
lines.}
\label{fig:A1A4B1ckes}
\end{figure*}

\begin{figure*}
\begin{center}
\includegraphics[height=4.cm]{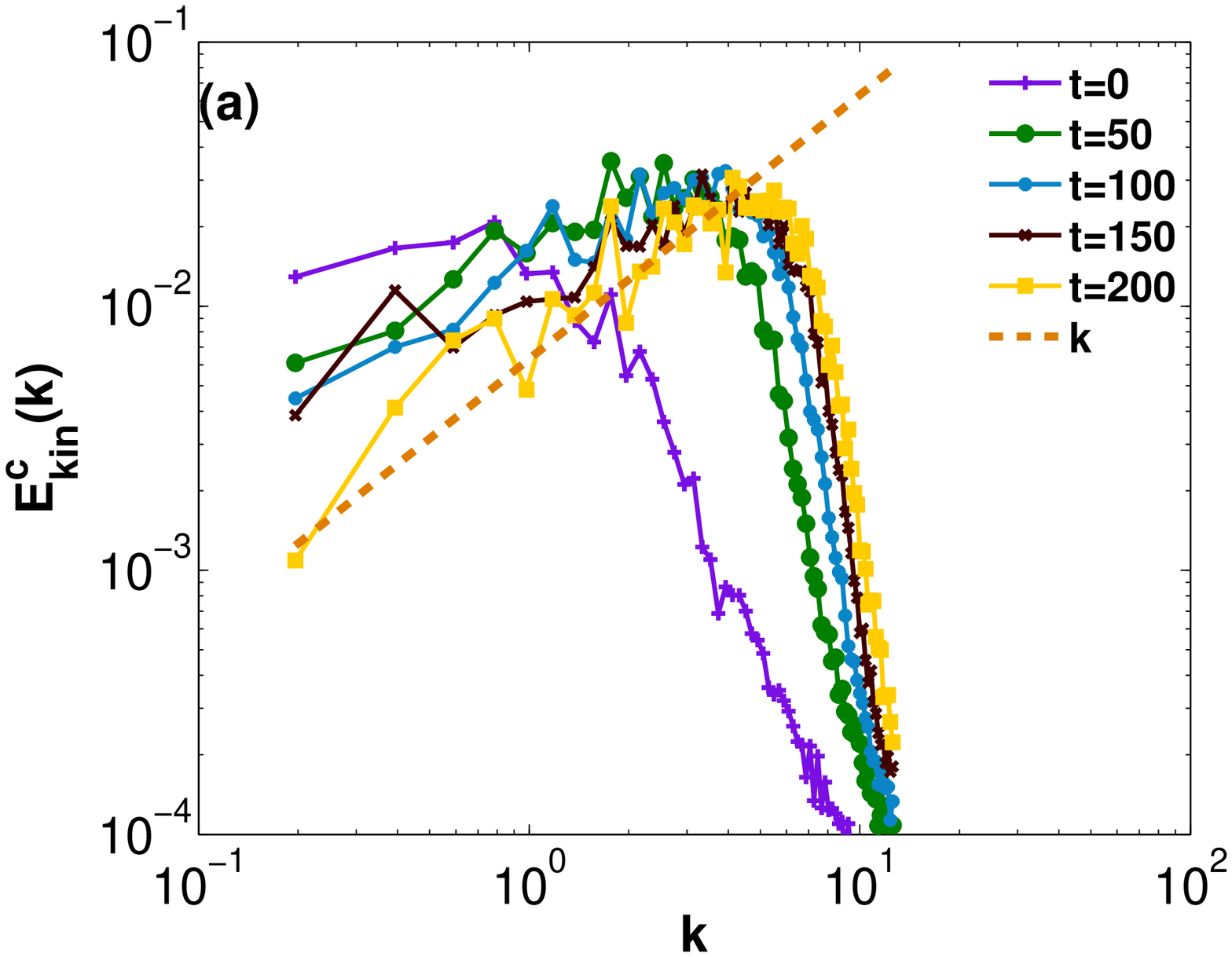}
\includegraphics[height=4.cm]{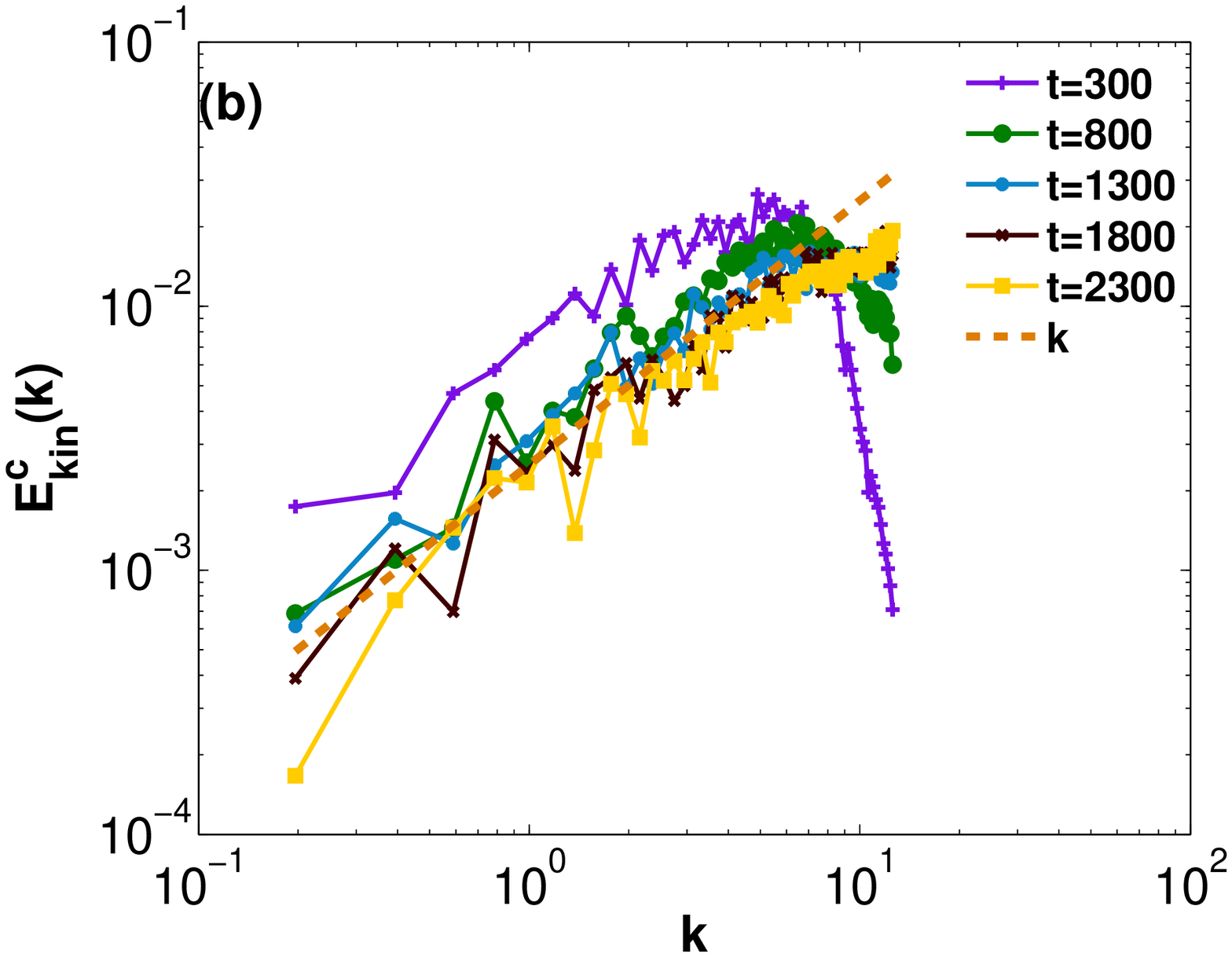} 
\includegraphics[height=4.cm]{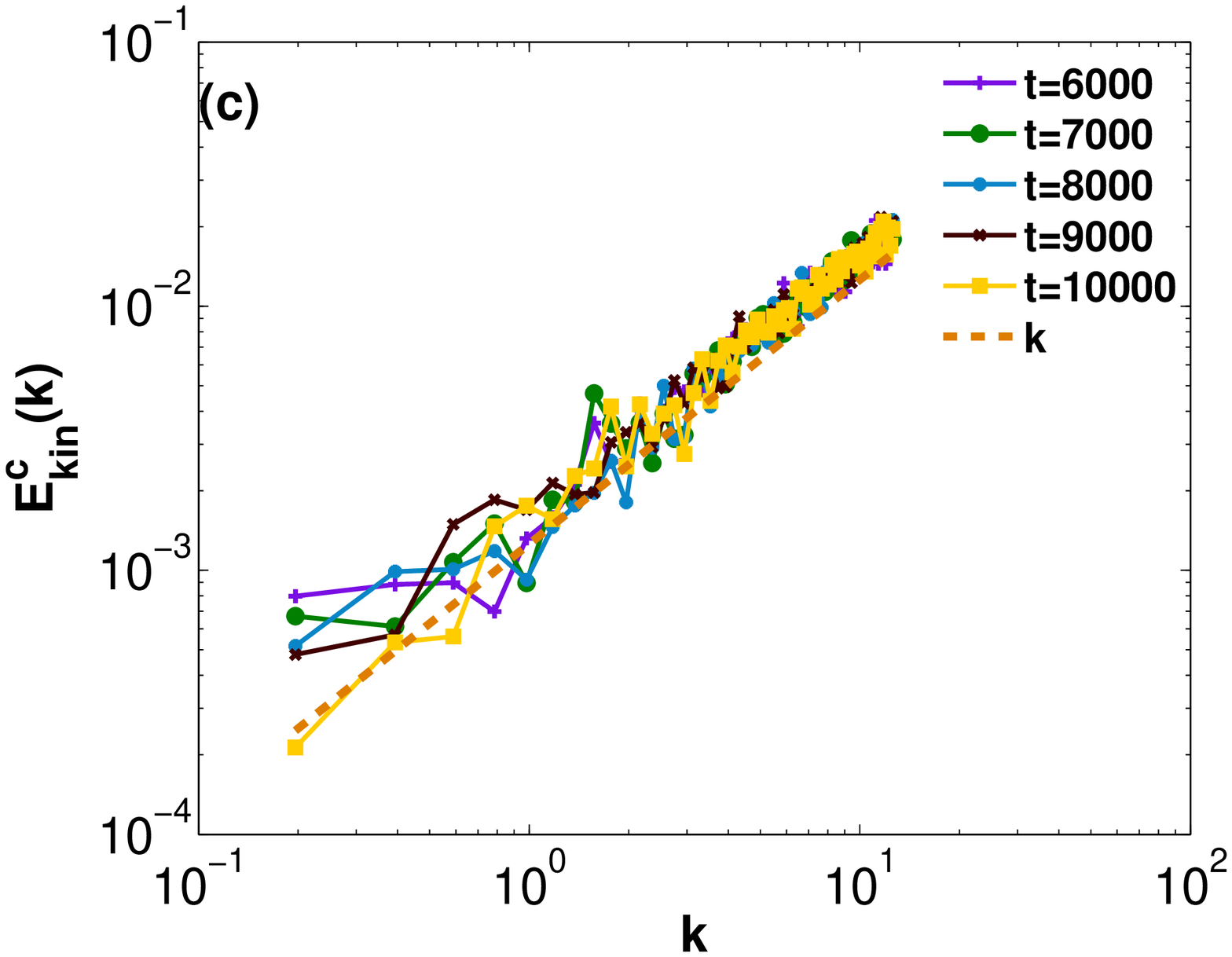} \\
\includegraphics[height=4.cm]{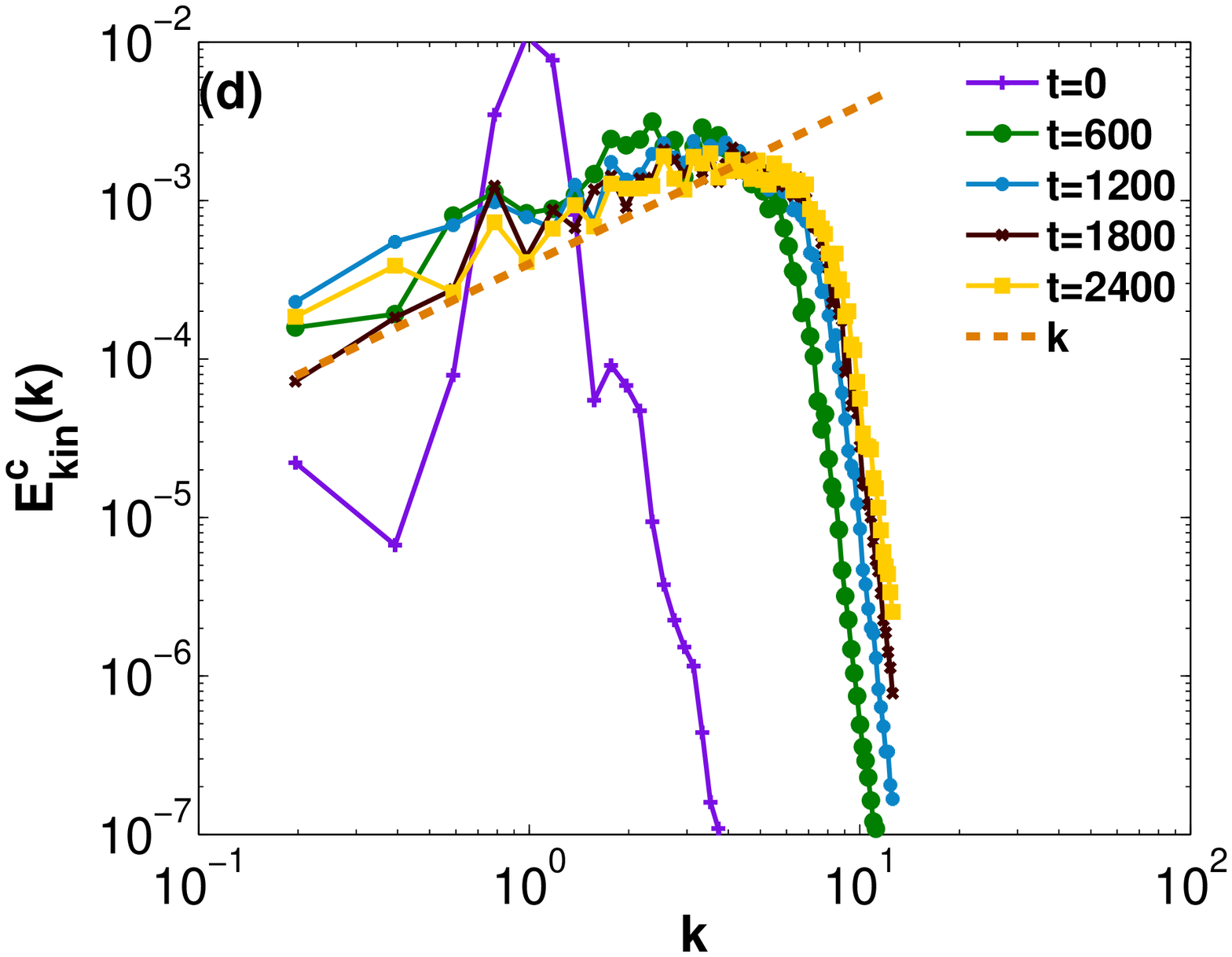}
\includegraphics[height=4.cm]{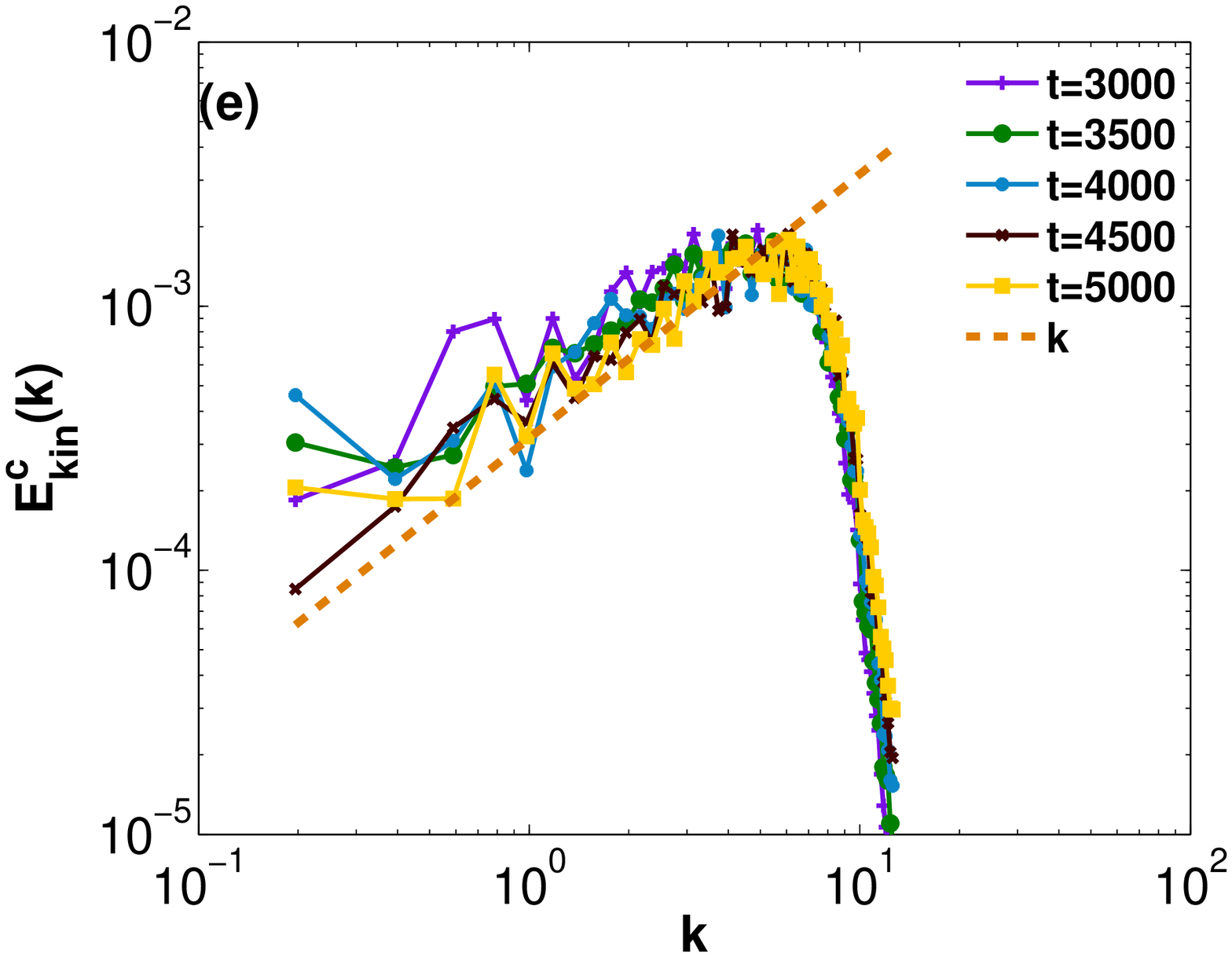}
\includegraphics[height=4.cm]{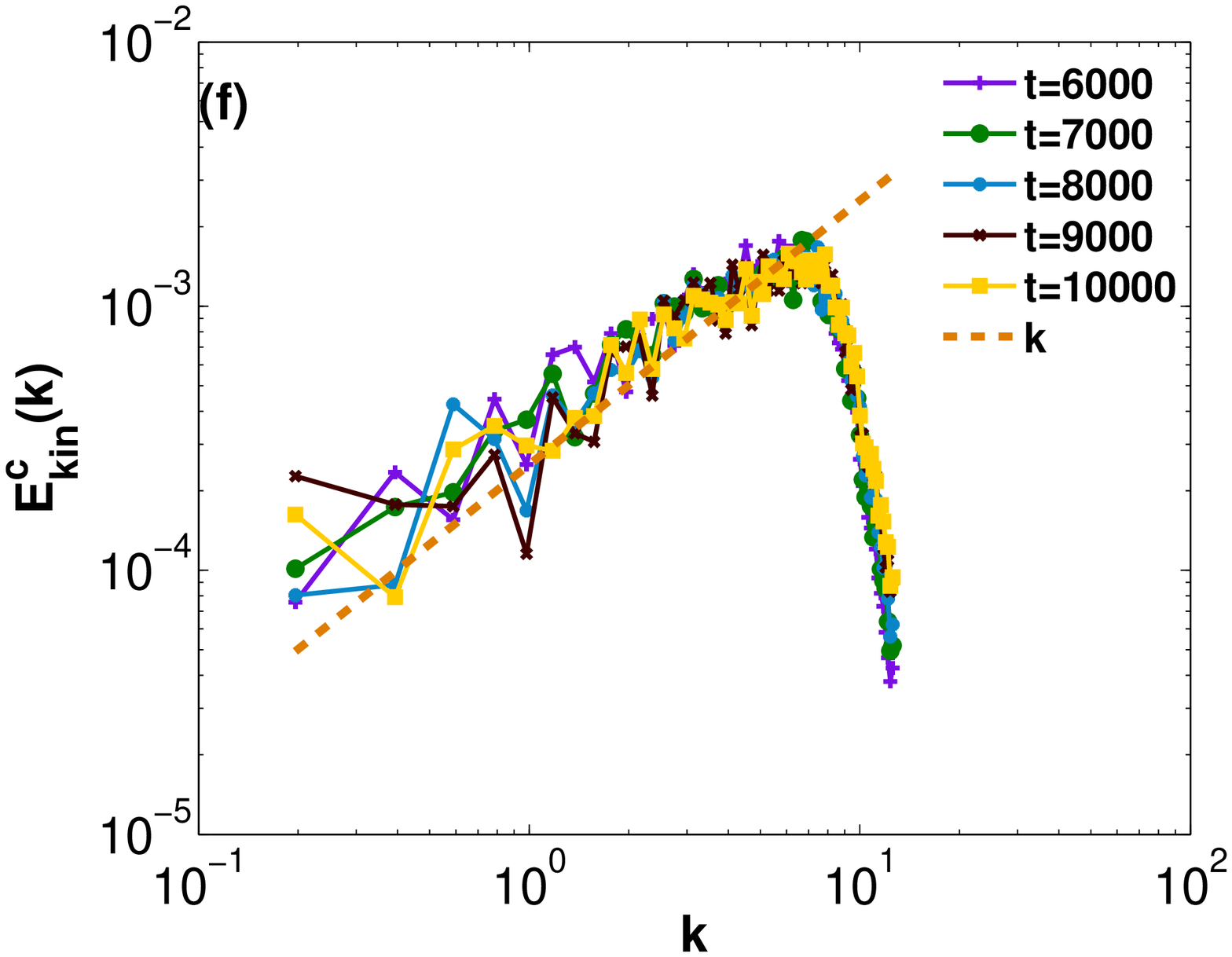}
\end{center}
\caption{\small Log-log (base 10) plots of the spectra $E^c_{kin}(k)$ 
from our DNS runs (a)-(c) $\tt A7$ and (d)-(f) $\tt B2$
at different times $t$ (indicated by curves of different
colours); a $k$ power law is shown by orange-dashed lines. 
}
\label{fig:A7B2ckes}
\end{figure*}

\begin{figure*}
\begin{center}
\includegraphics[height=4.cm]{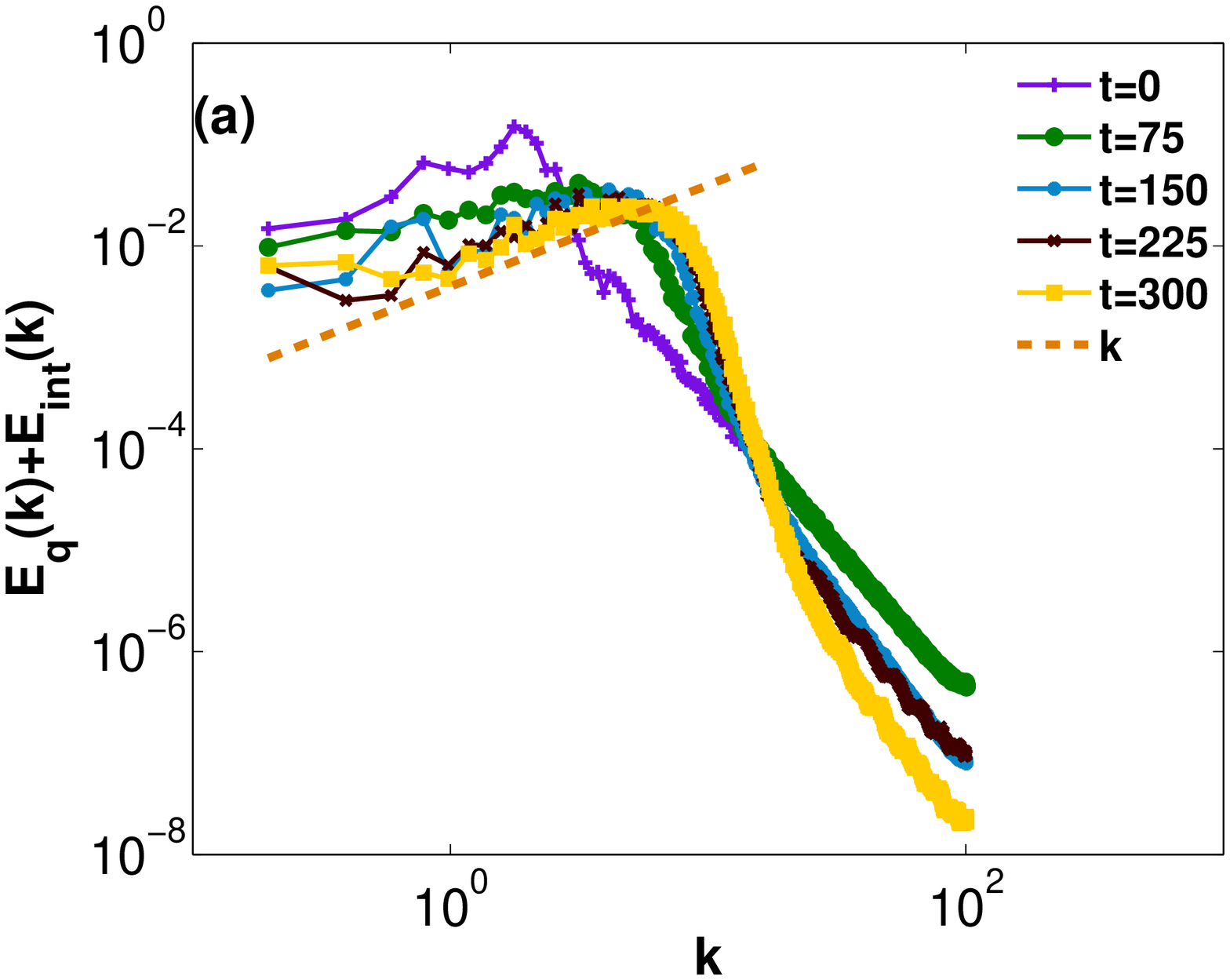}
\includegraphics[height=4.cm]{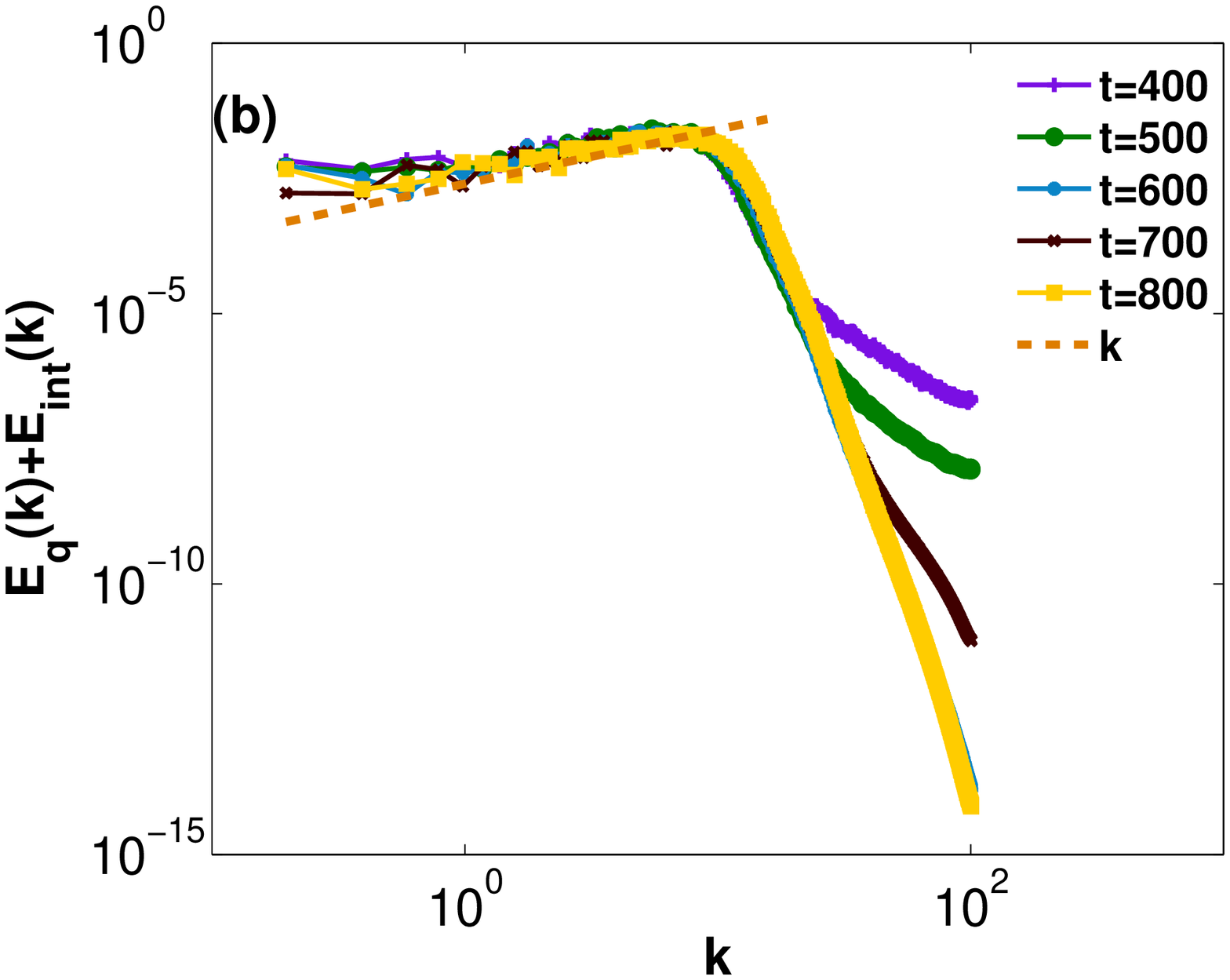}
\includegraphics[height=4.cm]{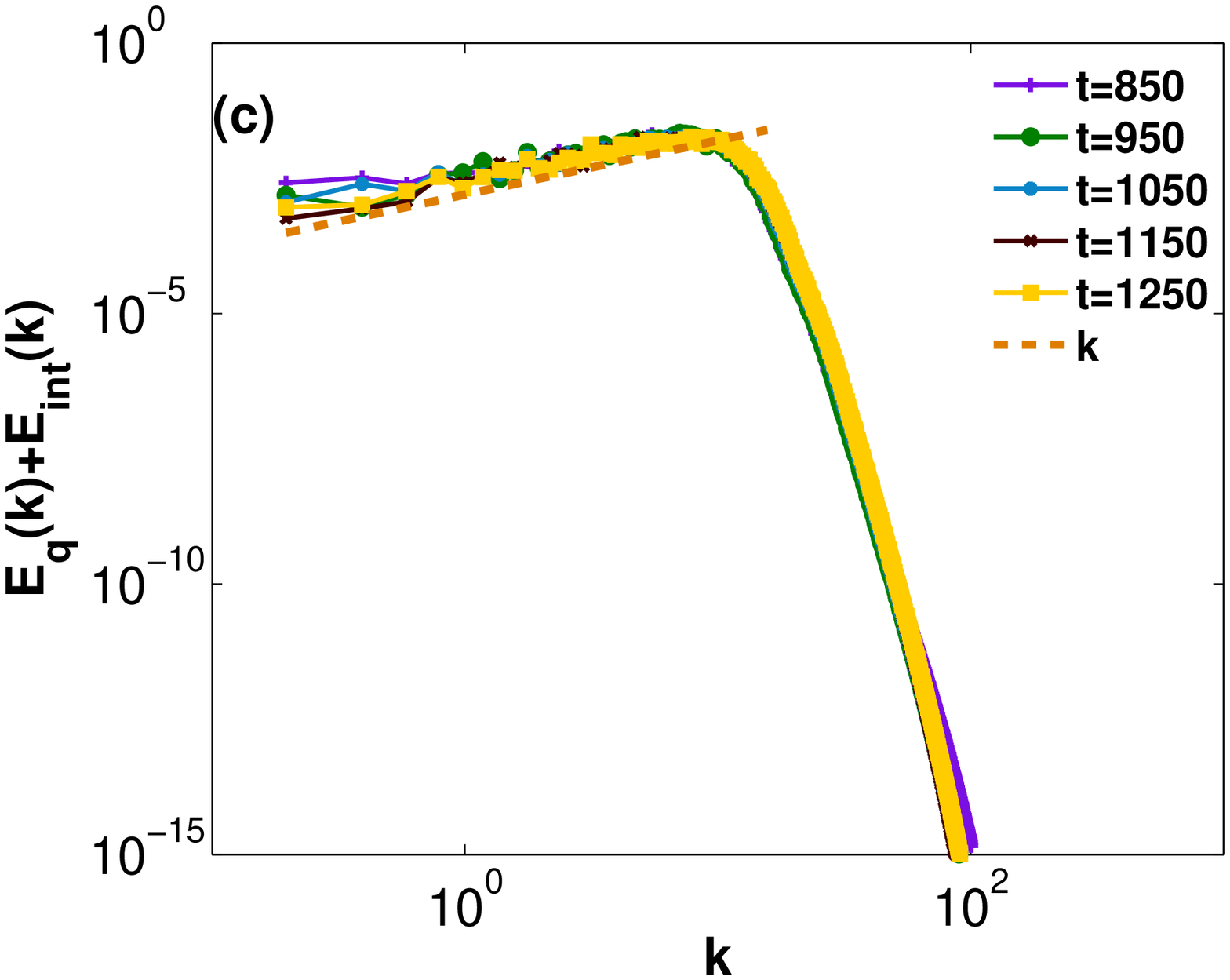}
\includegraphics[height=4.cm]{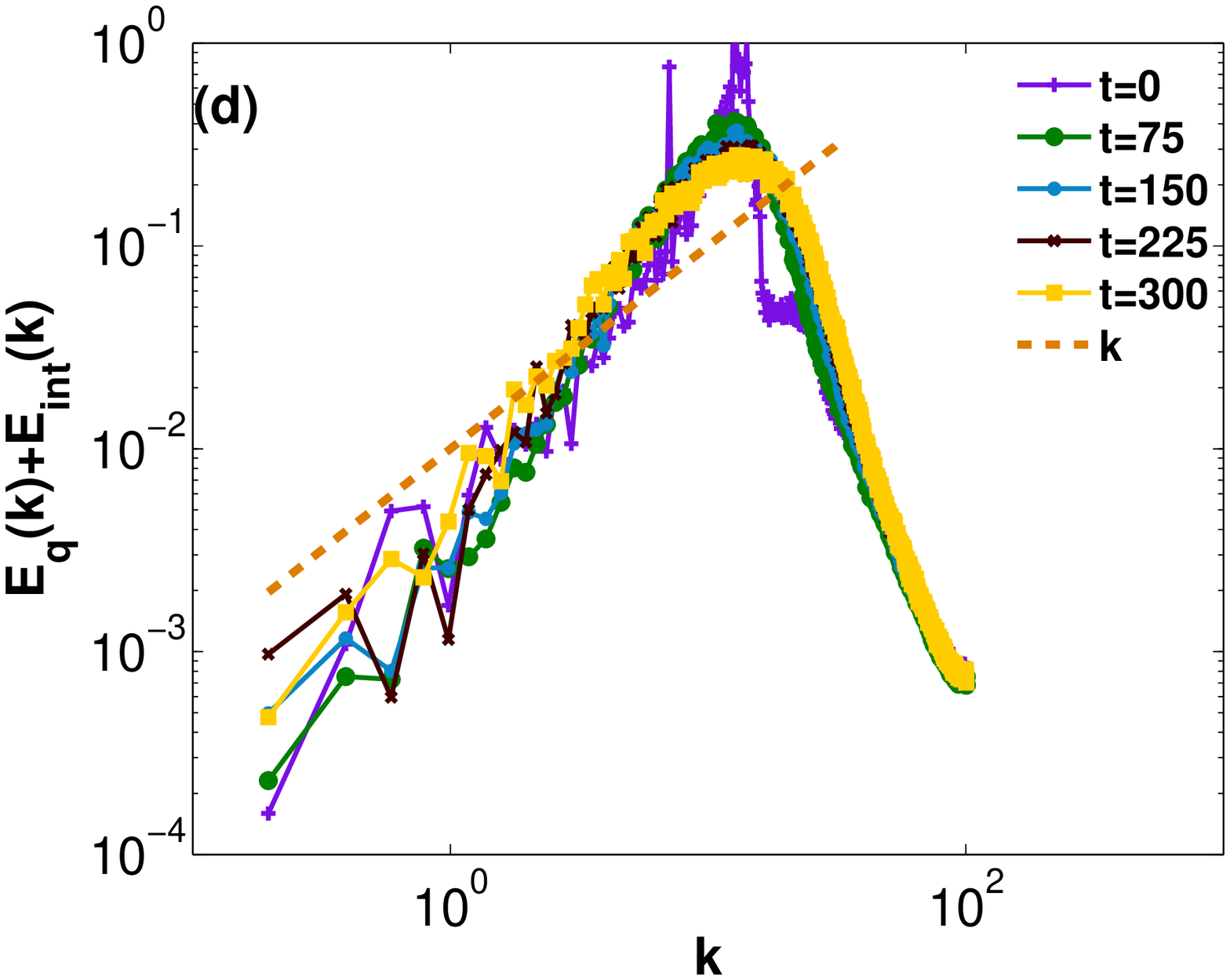}
\includegraphics[height=4.cm]{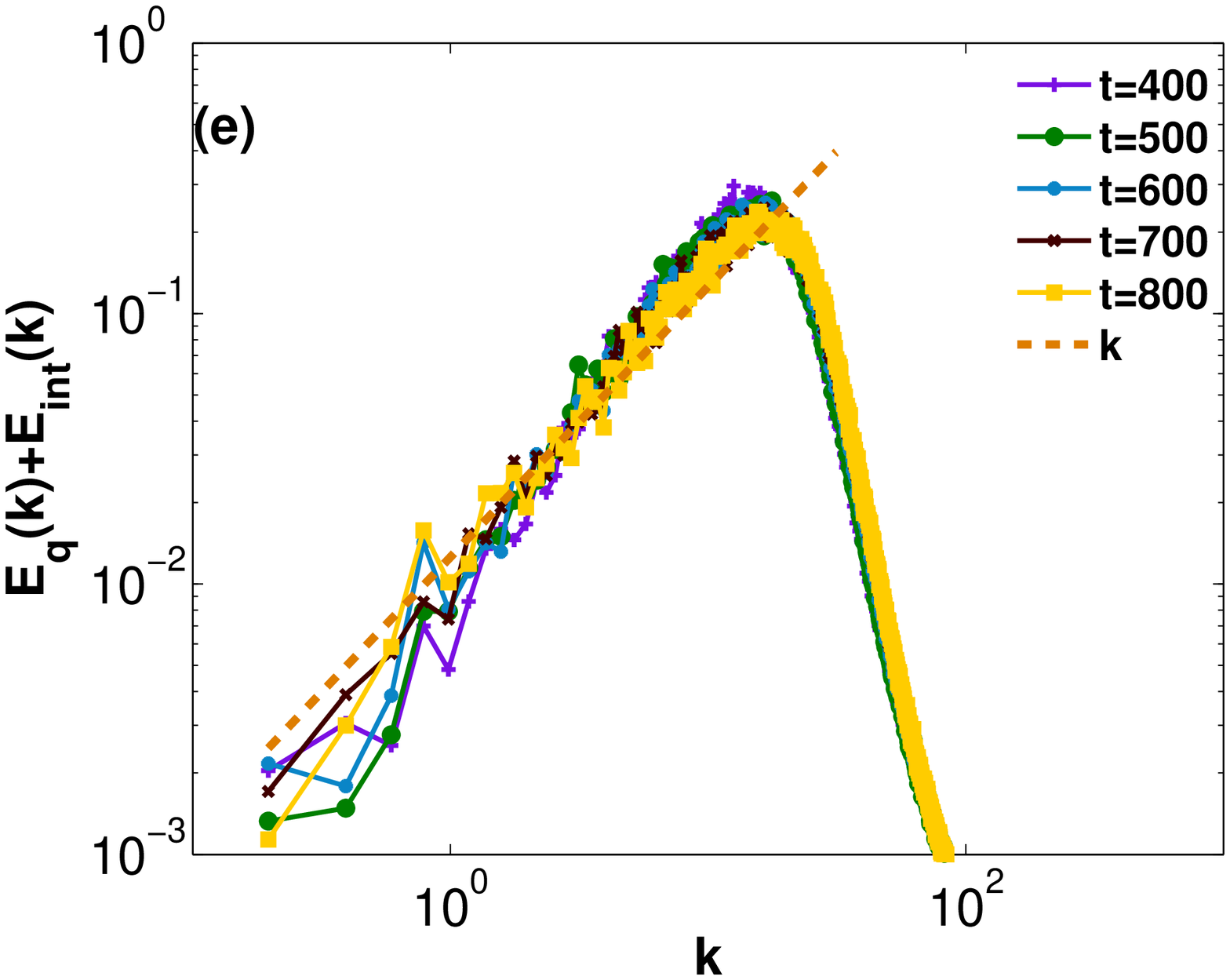}
\includegraphics[height=4.cm]{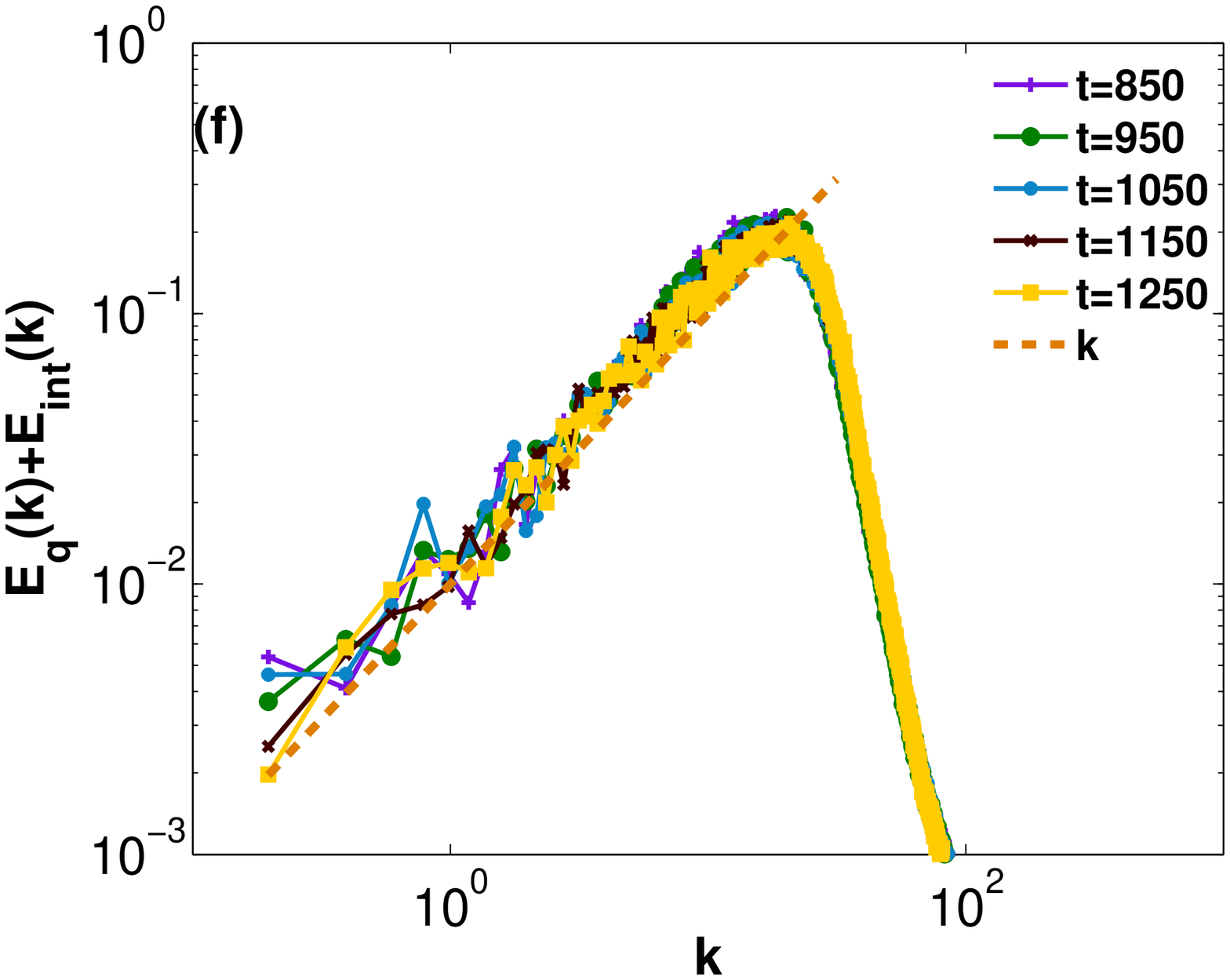}
\includegraphics[height=4.cm]{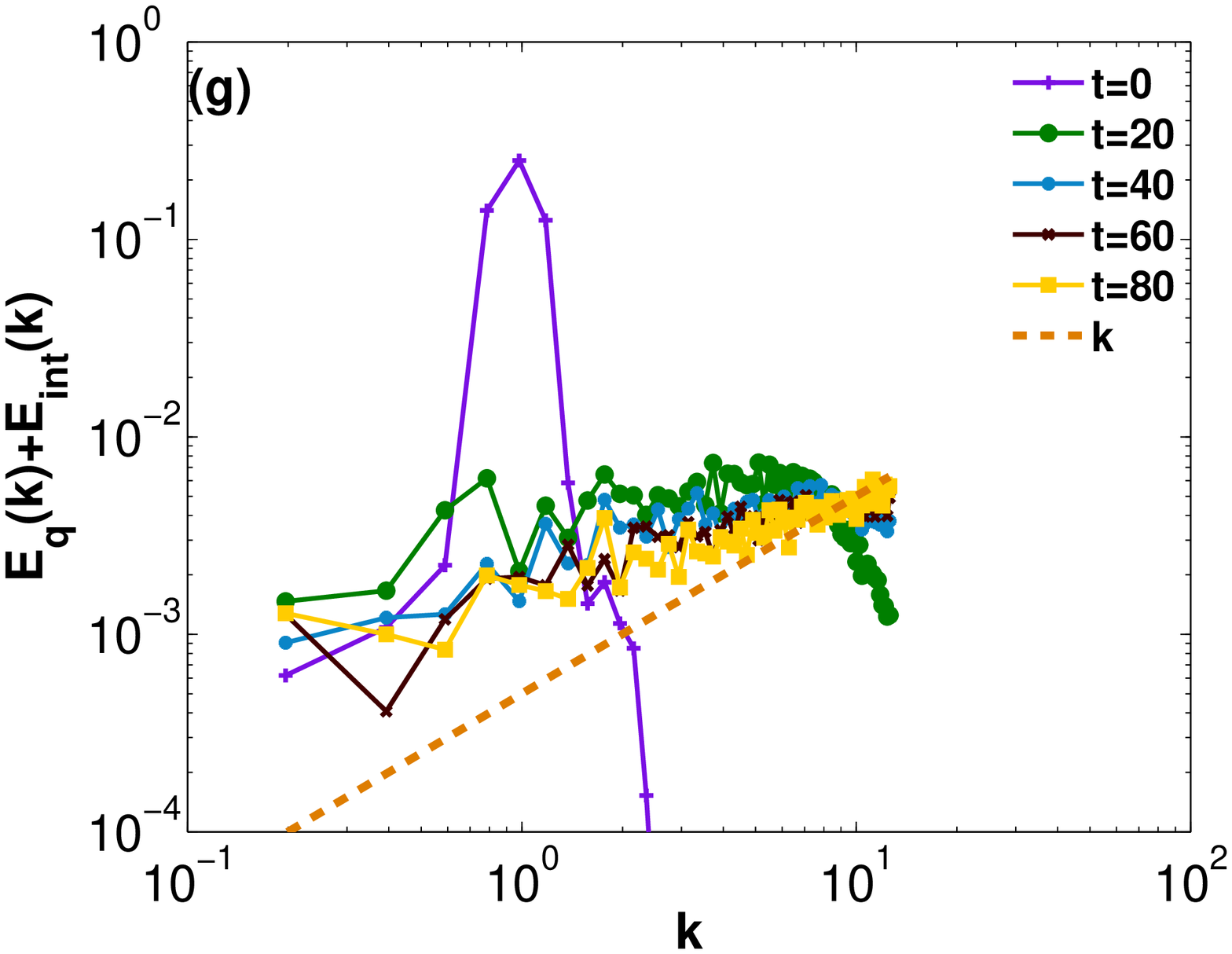}
\includegraphics[height=4.cm]{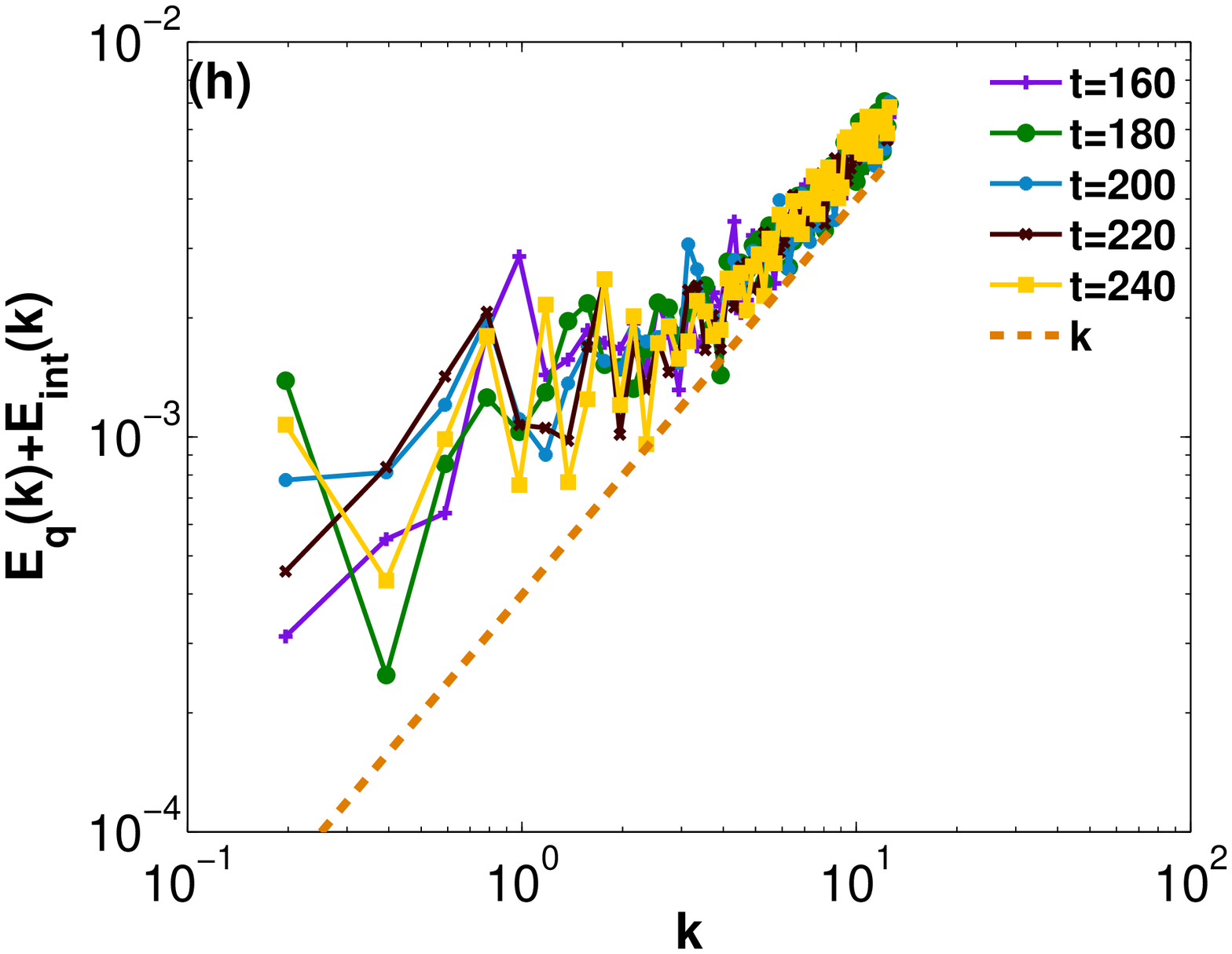}
\includegraphics[height=4.cm]{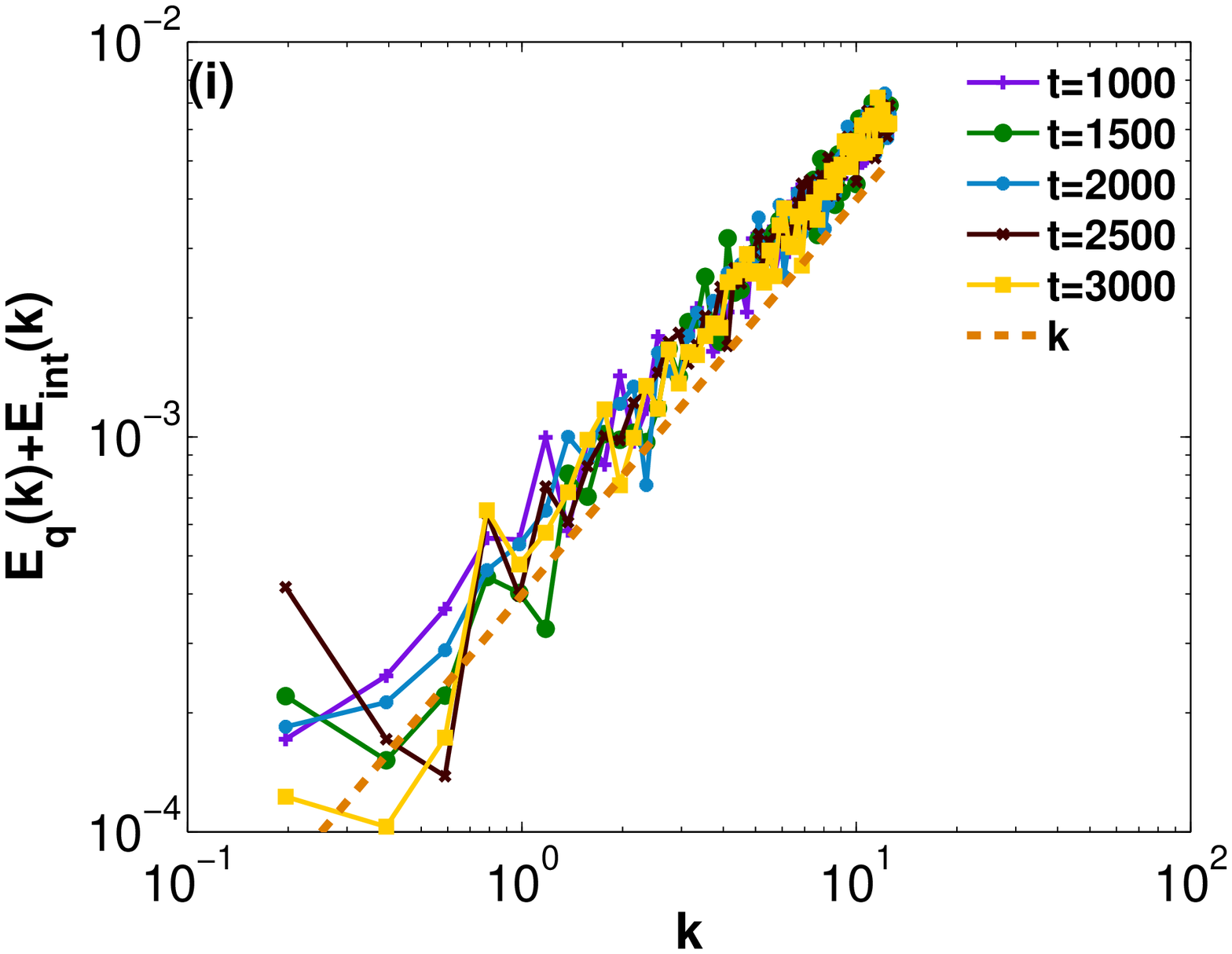}
\end{center}
\caption{\small Log-log (base 10) plots of the spectra $E_{int}(k)+E_q(k)$ 
from our DNS runs (a)-(c) $\tt A1$, (d)-(f) $\tt A4$, and (g)-(i) $\tt B1$
at different times $t$ (indicated by curves of different
colours); a $k$ power law is shown by orange-dashed lines.
}
\label{fig:A1A4B1iqes}
\end{figure*}

\begin{figure*}
\begin{center}
\includegraphics[height=3.9cm]{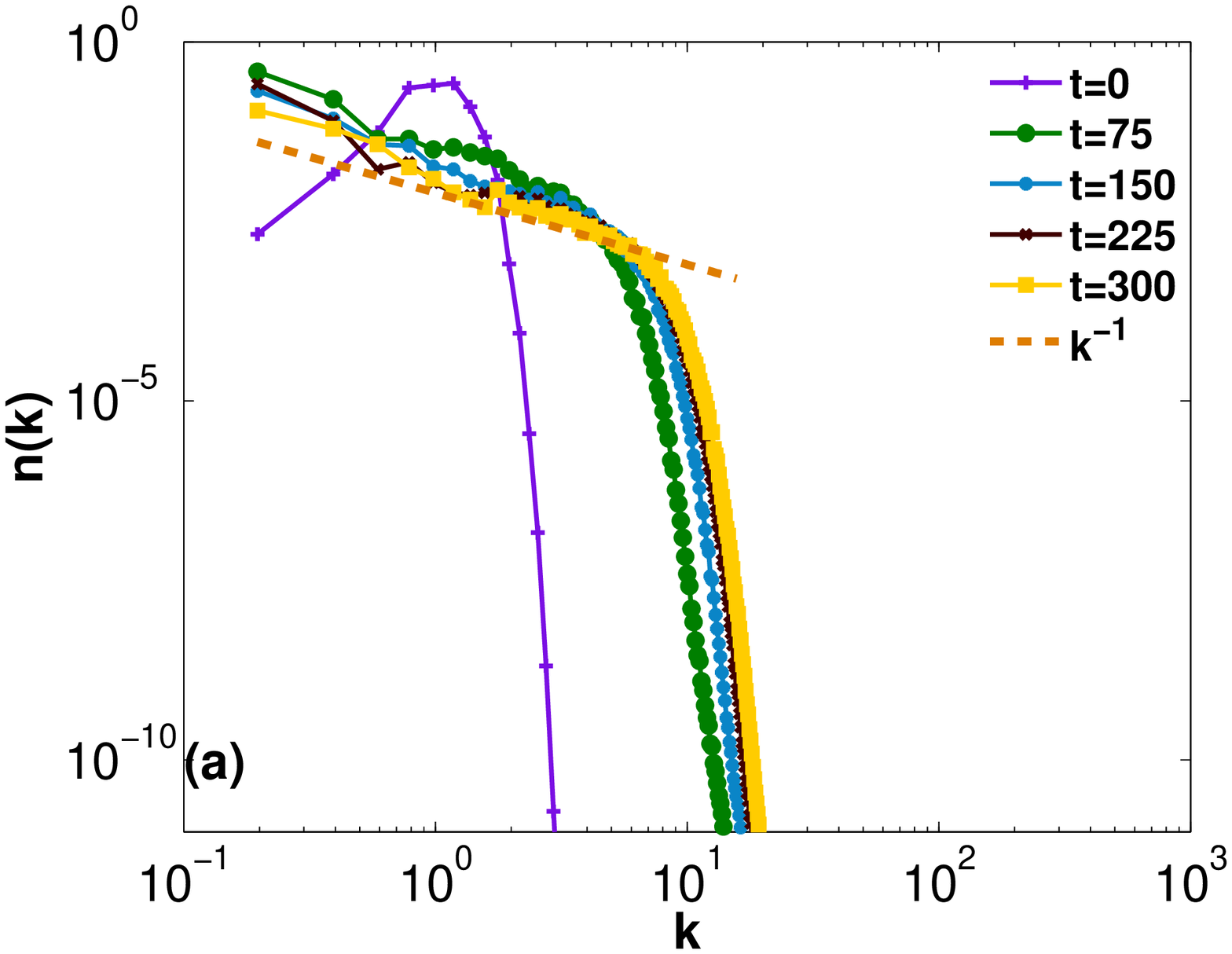}
\includegraphics[height=3.9cm]{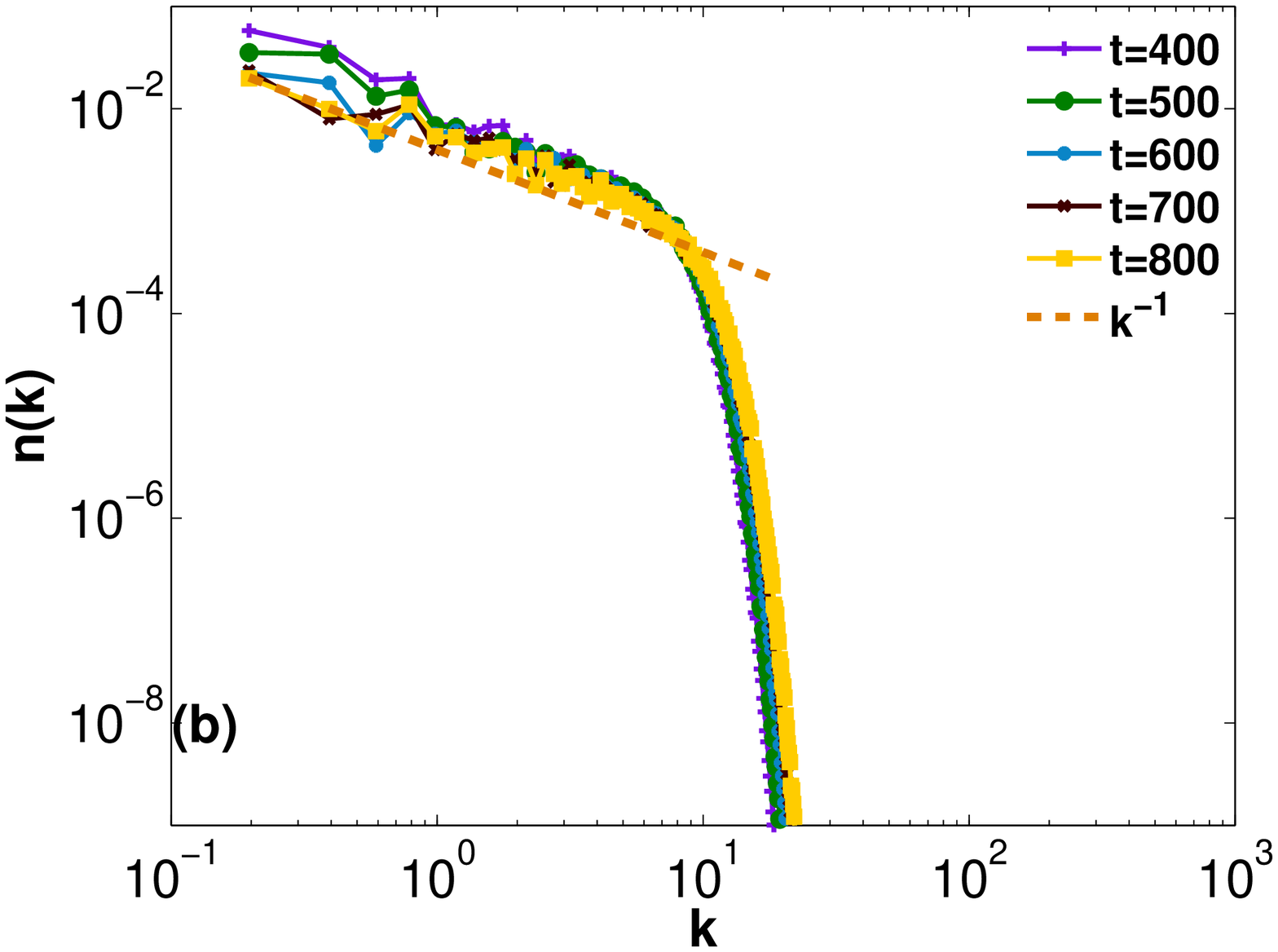}
\includegraphics[height=3.9cm]{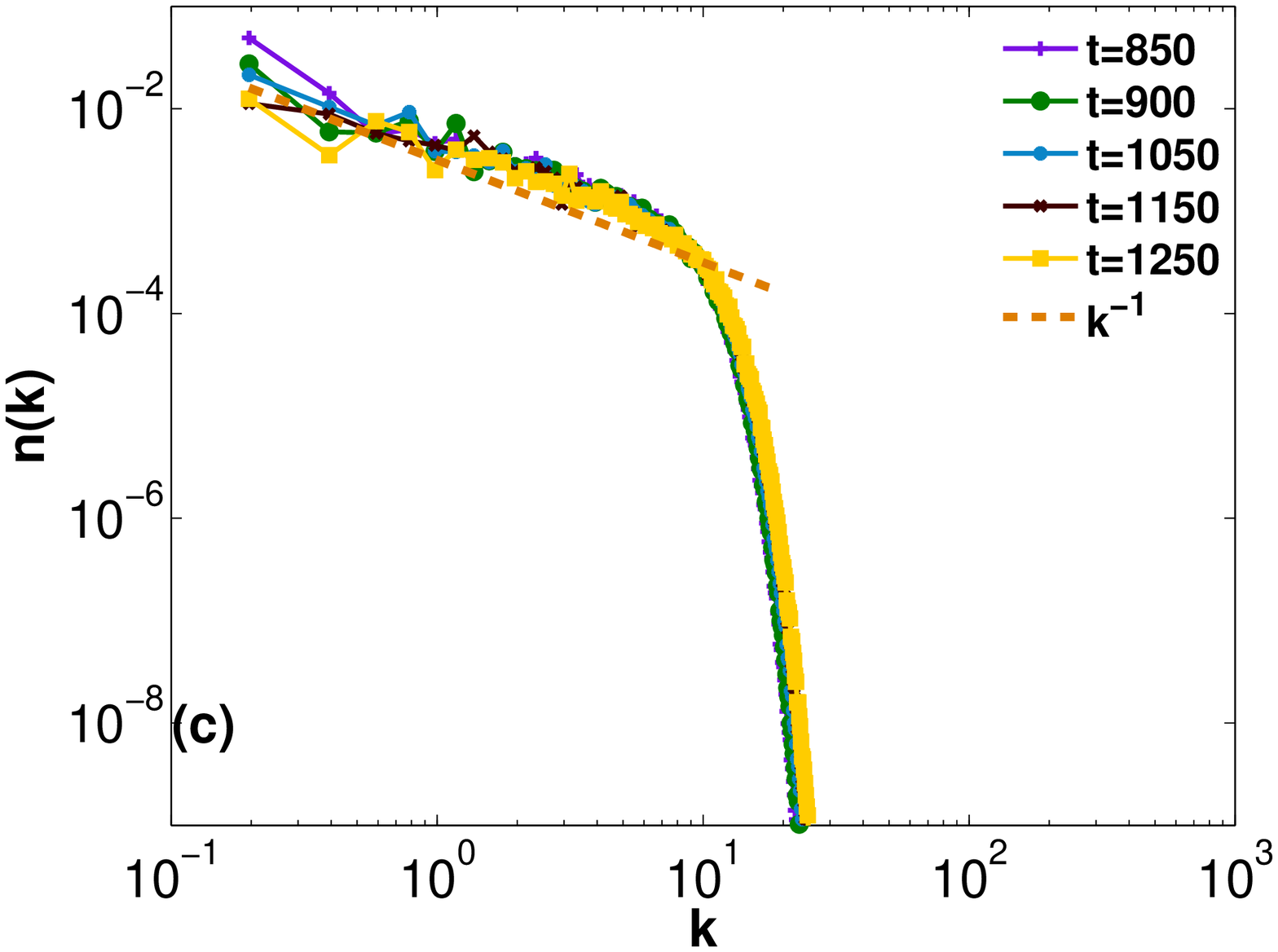}
\includegraphics[height=3.9cm]{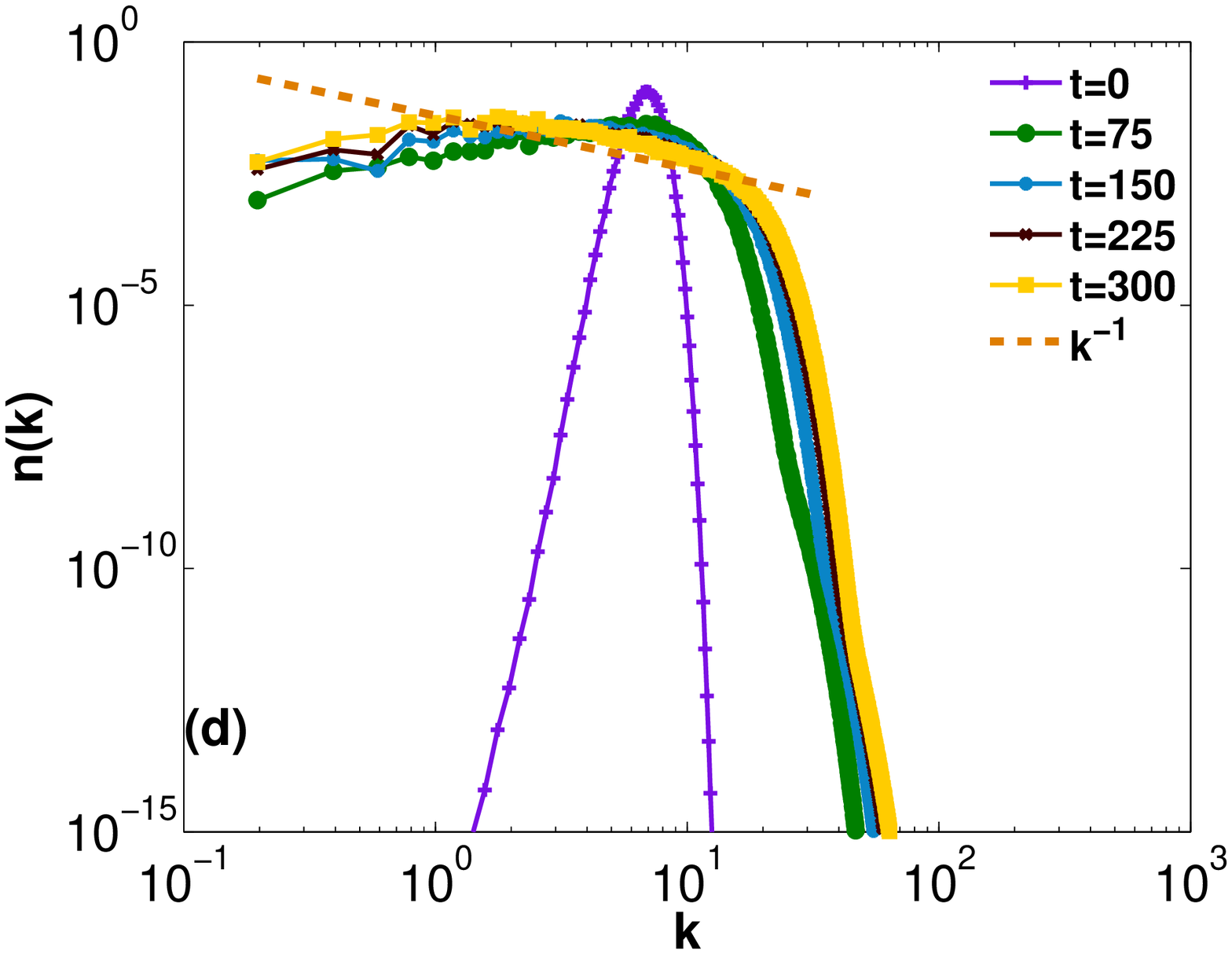}
\includegraphics[height=3.9cm]{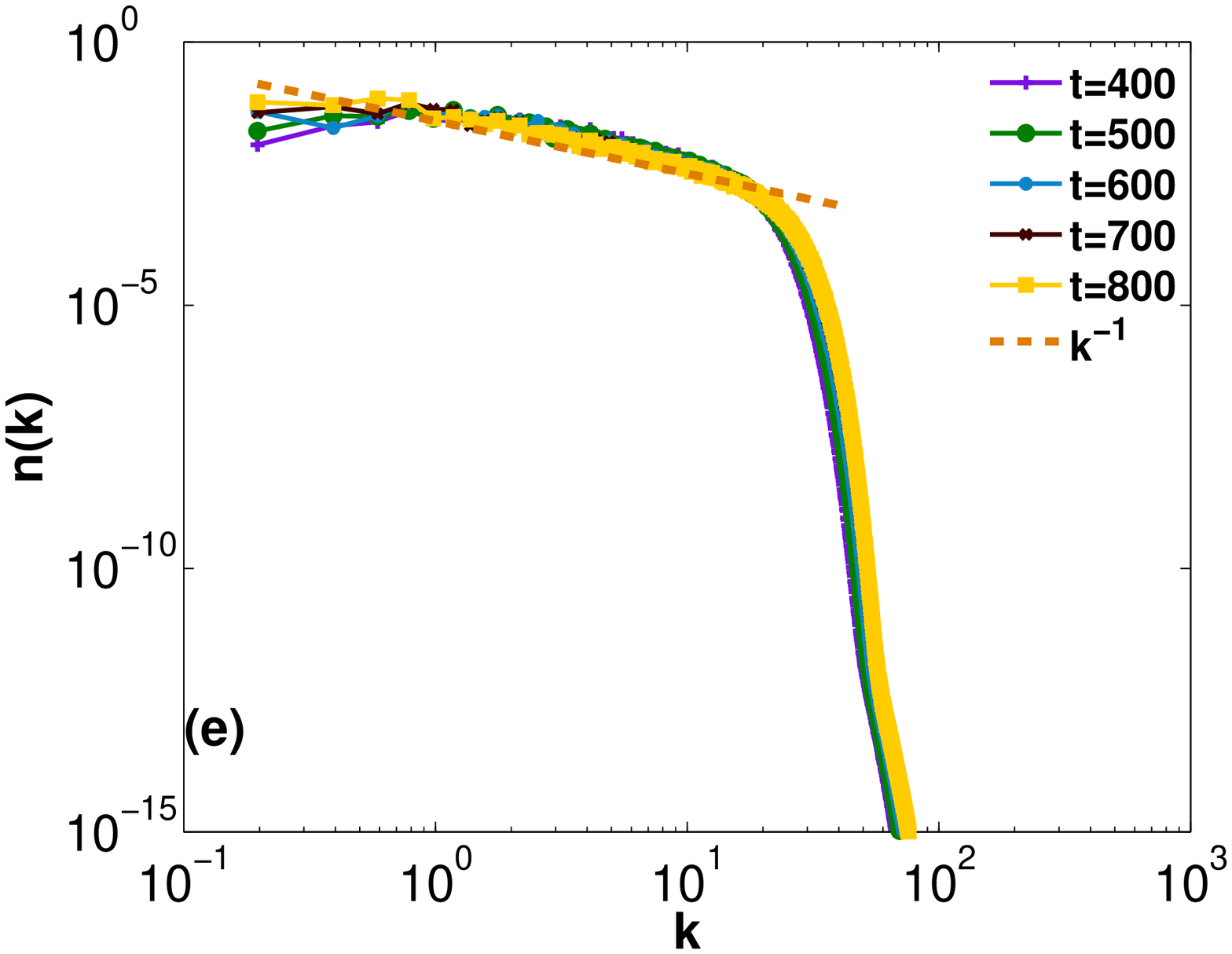}
\includegraphics[height=3.9cm]{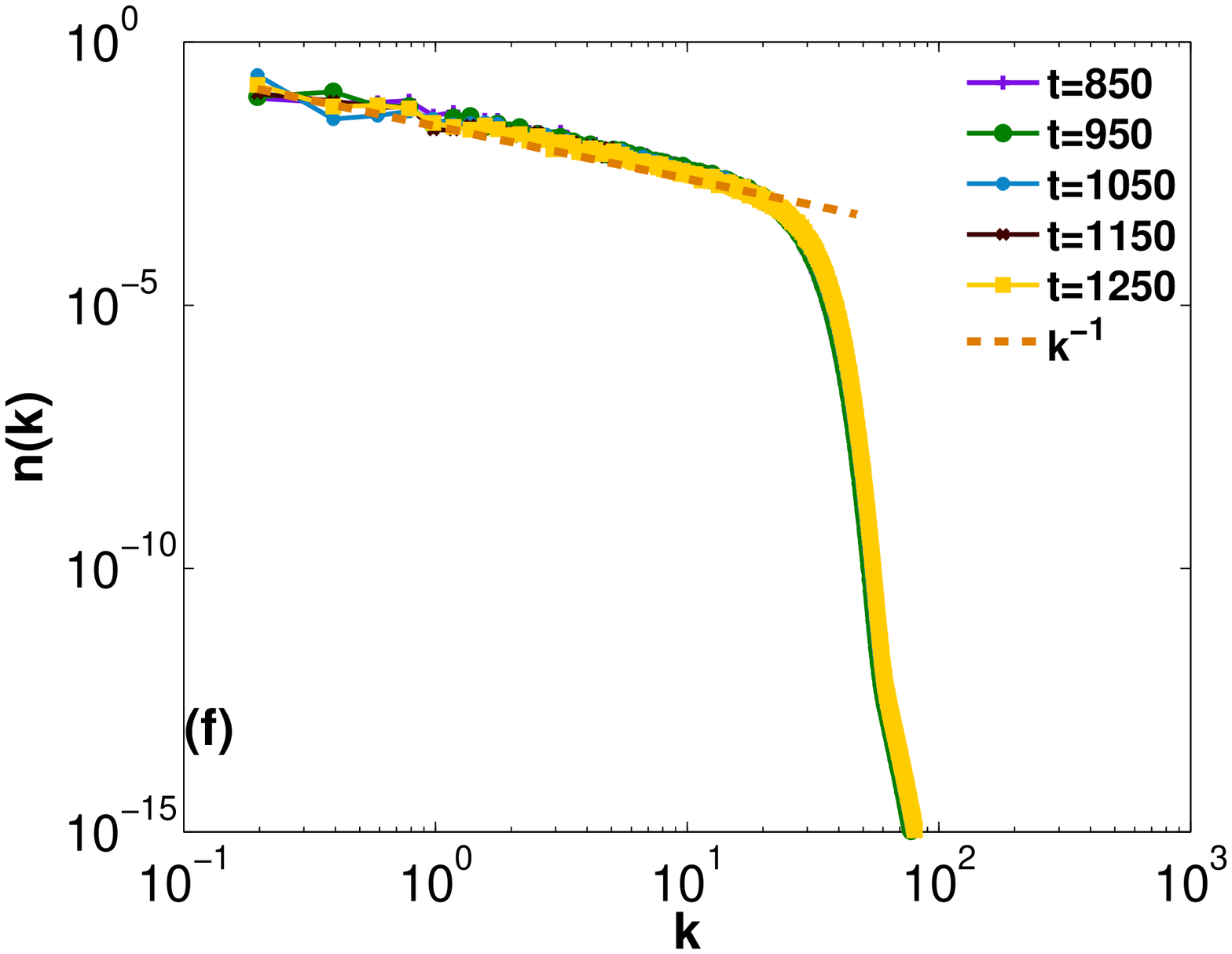}
\includegraphics[height=3.9cm]{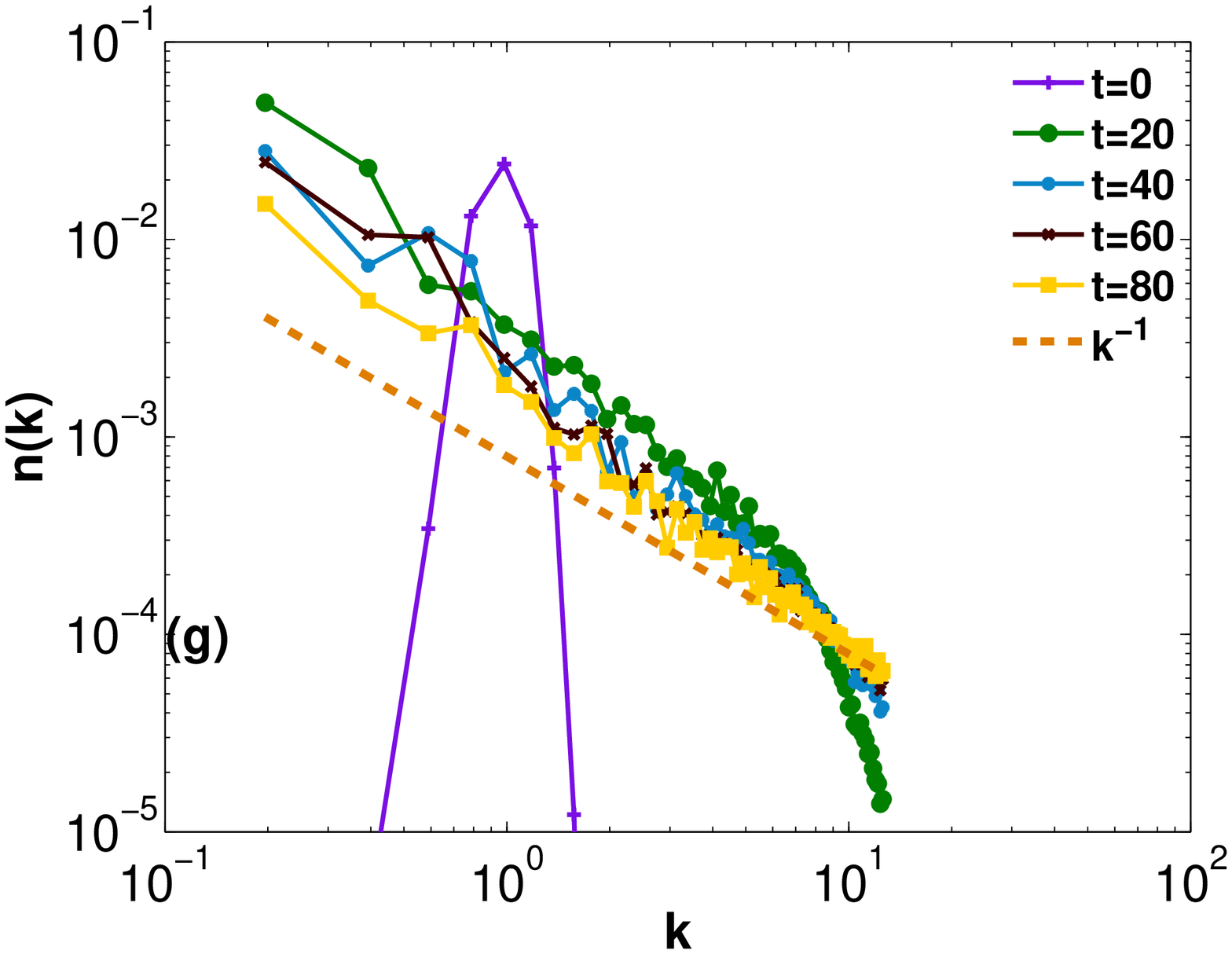}
\includegraphics[height=3.9cm]{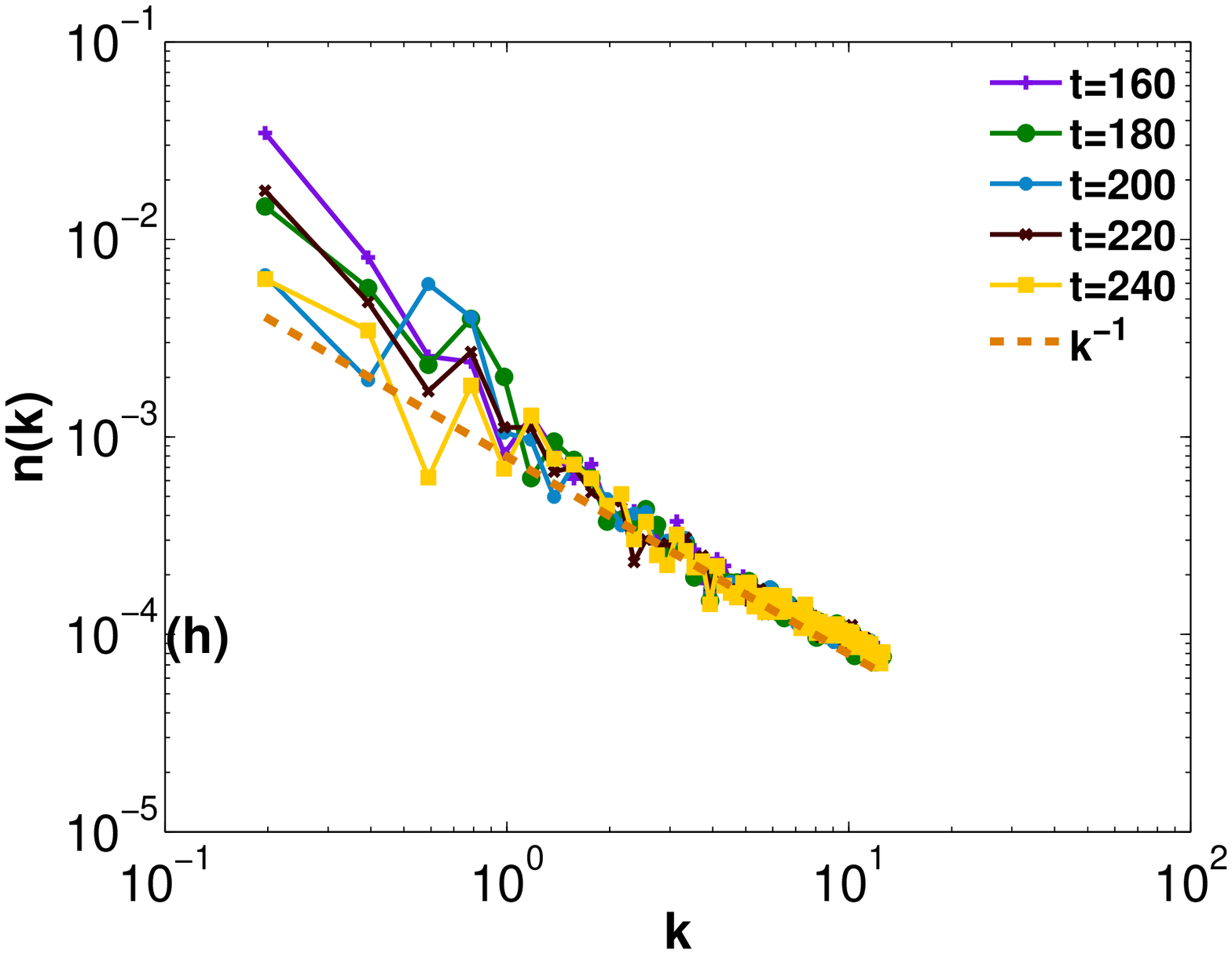}
\includegraphics[height=3.9cm]{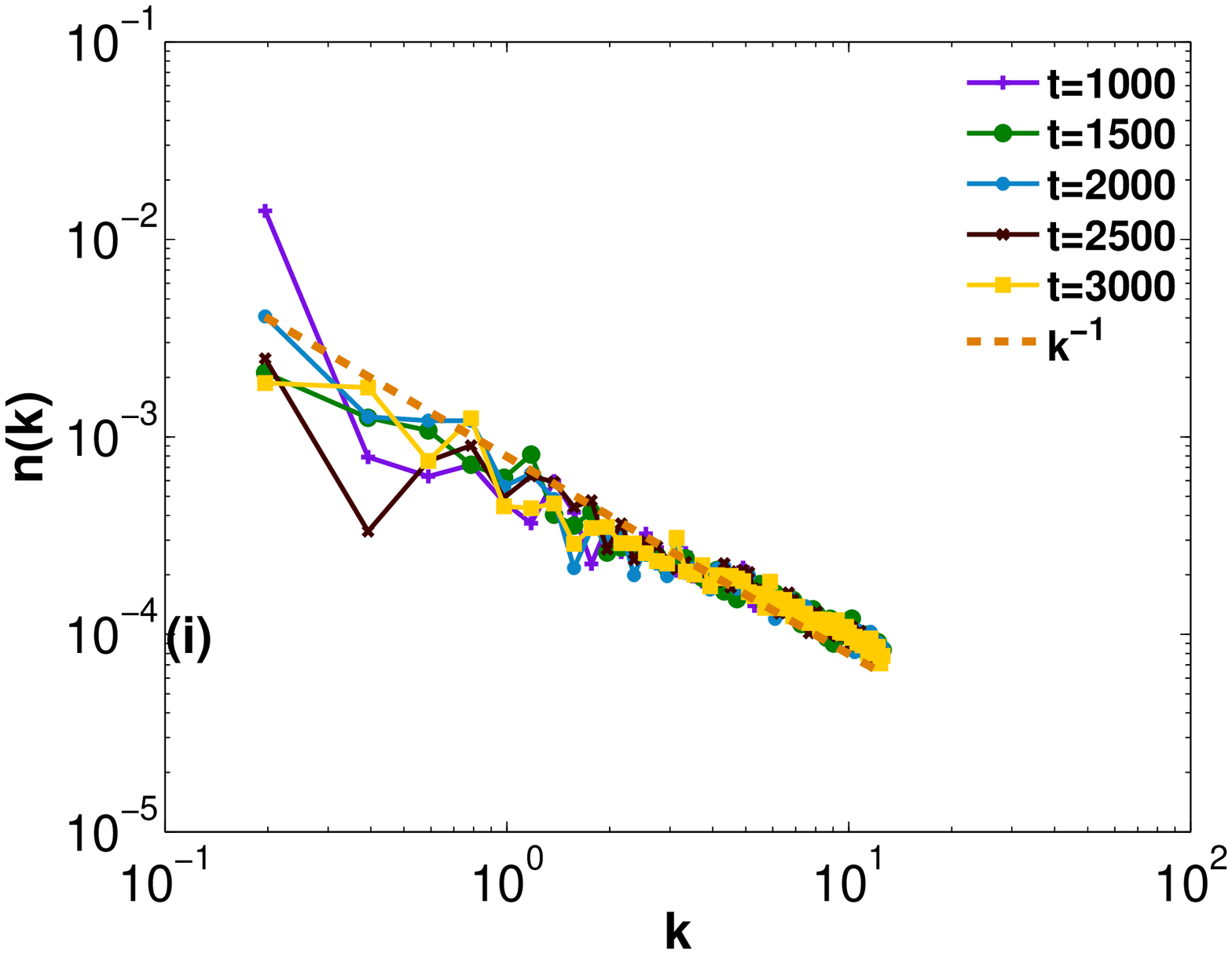}
\end{center}
\caption{\small Log-log (base 10) plots of the spectra $n(k)$ from our DNS runs (a)-(c)
$\tt A1$, (d)-(f) $\tt A4$, and (g)-(i) $\tt B1$
at different times $t$ (indicated by curves of different
colours); a $k^{-1}$ power law is shown by orange-dashed lines. 
}
\label{fig:A1A4B1occps}
\end{figure*}

The initial transients described above are followed by a regime
in which the energy and occupation-number spectra begin to show
power-law-scaling behaviours, but the power-law exponent and the
extent of the scaling region change with time and depend on the
initial conditions; we regard this as the onset of
thermalization, which is shown in figures \ref{fig:A1A4B1ckes}
and \ref{fig:A7B2ckes}, where we illustrate the time evolution of
$E^c_{kin}$. Figure \ref{fig:A1A4B1ckes} (a) shows $E^c_{kin}(k)$
for the run $\tt A1$; we begin to see a power-law region here with
$E^c_{kin}(k) \sim k$, on the low-$k$ side of the peak after
which the spectrum falls steeply.  A similar $E^c_{kin}(k) \sim
k$ behaviour starts to emerge in the region $k\lesssim k_{max}$
for the run $\tt B1$ (figure \ref{fig:A1A4B1ckes} (g)).  In this
onset-of-thermalization regime, we also see the development of
the following power laws: $E_{int}(k)+E_q(k) \sim k$ (figure
\ref{fig:A1A4B1iqes}) and $n(k) \sim 1/k$ (figure
\ref{fig:A1A4B1occps}).

\subsection{Partial thermalization and self-truncation}
\label{subsection:partialandselftruncation}

\subsubsection{Partial thermalization}

In the third stage of the dynamical evolution of the 2D,
Fourier-truncated, GP equation, which we refer to as the
partial-thermalization stage, well-defined, power-law-scaling
behaviours appear in energy and occupation-number spectra, with
exponents that are independent of the initial conditions as
illustrated by the compressible-kinetic-energy spectra in
figures~\ref{fig:A1A4B1ckes} (b), (c) (e), (f), and
\ref{fig:A7B2ckes} (b) for initial conditions of type $\tt IC1$,
and figures~\ref{fig:A1A4B1ckes} (h) and \ref{fig:A7B2ckes} (e),
and (f), for initial conditions of type $\tt IC2$. It is
important to distinguish between ({\tt I}) spectra that fall steeply at
large values of $k$, e.g., the spectra in
figures~\ref{fig:A1A4B1ckes} (b), (c) (e), (f), and
\ref{fig:A7B2ckes} (e) and (f), and ({\tt II}) spectra that increase
all the way to $k_{max}$, e.g., the spectra in
figures~\ref{fig:A1A4B1ckes} (h) and \ref{fig:A7B2ckes} (b) and
(c). In case ({\tt I}), we have spectral convergence to the 2D GP
partial differential equation (PDE); in case ({\tt II}), the effects of
Fourier truncation are so pronounced that our truncated 2D, GP
system does not provide a good representation of the 2D, GP PDE.
As we show below, case ({\tt I}) can be further subdivided into ({\tt A}) a
subclass in which the maximum, at $k=k_c$ in  $E_{kin}(k)=
(E^c_{kin}(k) + E^i_{kin}(k))$, referred to as the
self-truncation wave number~\cite{Krstulovic2011pre}, moves out
to $k_{max}$ as a power of $t$ and ({\tt B}) a subclass in which $k_c$
moves out to $k_{max}$ at a rate that is slower than a power of
$t$.

Figures~\ref{fig:A1A4B1ckes} (g)-(i), from the run $\tt B1$,
show how $E^c_{kin}(k)$ evolves as the spectral convergence to 
the GP PDE is lost in case ({\tt II}); note that the scaling region with 
$E^c_{kin} \sim k$ sets in at high wave numbers close to
$k_{max}$ and then extends to the low-wave-number regime. 
For case ({\tt IA}) analogous plots of  $E^c_{kin}(k)$ are given in,
e.g., figures~\ref{fig:A7B2ckes} (a)-(c). We give plots for 
case ({\tt IB}) in the next subsection, where we study in detail the 
time dependence of $k_c$. Illustrative plots of the spectra
$(E_{i}(k) + E_{q}(k))$ and $n(k)$ in this regime of partial 
thermalization are given in figures~\ref{fig:A1A4B1iqes} 
and~\ref{fig:A1A4B1occps}, respectively.

\subsubsection{Self-truncation}

We now present a detailed characterization of the
partial-thermalization regime, when energy
spectra display self-truncation at wave-numbers beyond
$k_c(t)$, which can be defined as follows: 
\begin{equation} \label{eq:kc}
k_c = \sqrt{\frac{2\int^{k_{max}}_0 k^2 E_{kin}(k) dk}
{\int^{k_{max}}_0 E_{kin}(k) dk}};
\end{equation}
as the system approaches complete thermalization, $k_c(t) \to
k_{max}$. In particular, we explore how the scaling ranges in
energy spectra grow with $t$ for different values of $g$, with
the initial  configuration and number of collocation points $N_c$
held fixed. For an initial condition of type $\tt IC1$, with
$k_0=5\Delta k$, $\sigma=2\Delta k$, and $N_c=256$, we obtain the
time evolution of energy spectra for $g=1000$ (run $\tt A6$),
$g=2000$ (run $\tt A9$), and $g=5000$ (run $\tt A10$) in figures
\ref{fig:kediffgNc} (a), (b), and (c), respectively, and their
video analogues (Videos S3 (panel V2) in the Supplementary Material). The
larger the value of $g$, the more rapid is the thermalization,
and the consequent loss of spectral convergence, as we can see by
comparing the sky-blue (run $\tt A10$), green (run $\tt A9$), and
purple  (run $\tt A6$) spectra in figures~\ref{fig:kediffgNc}
(a)-(c); run $\tt A6$ loses spectral convergence around $t=2500$.
We obtain the same qualitative $g$ dependence, with $k_0=15\Delta
k$, $\sigma=2\Delta k$, and $N_c=256$, for $g=1000$, $2000$, and
$5000$, i.e., runs $\tt A11$, $\tt A12$, and $\tt A13$,
respectively, for which energy spectra are portrayed in figures
\ref{fig:kediffgNc} (d)-(f) and Video S3 (panel V3) in the Supplementary
Material.

In figures \ref{fig:kediffgNc} (g)-(i) we explore
the $N_c$ dependence of the self-truncation of energy 
spectra, for initial conditions, with $k_0=5\Delta k$, 
$\sigma=2\Delta k$, and $g=1000$, and five different values
of $N_c$, namely, $N_c=1024$ (run $\tt A1$),
$512$(run $\tt A5$), $256$ (run $\tt A6$), $128$ (run $\tt A7$), 
and $64$ (run $\tt A8$). We find, not surprisingly, that the
lower the value of $N_c$ the more rapidly does the system lose
spectral convergence.
 
Initial conditions of type $\tt IC2$ lead to energy spectra whose
time evolution, and their dependence on $g$ and $N_c$, is similar
to those that are obtained from intial conditions of type $\tt
IC1$.

\begin{figure*}
\begin{center}
\includegraphics[height=4.cm]{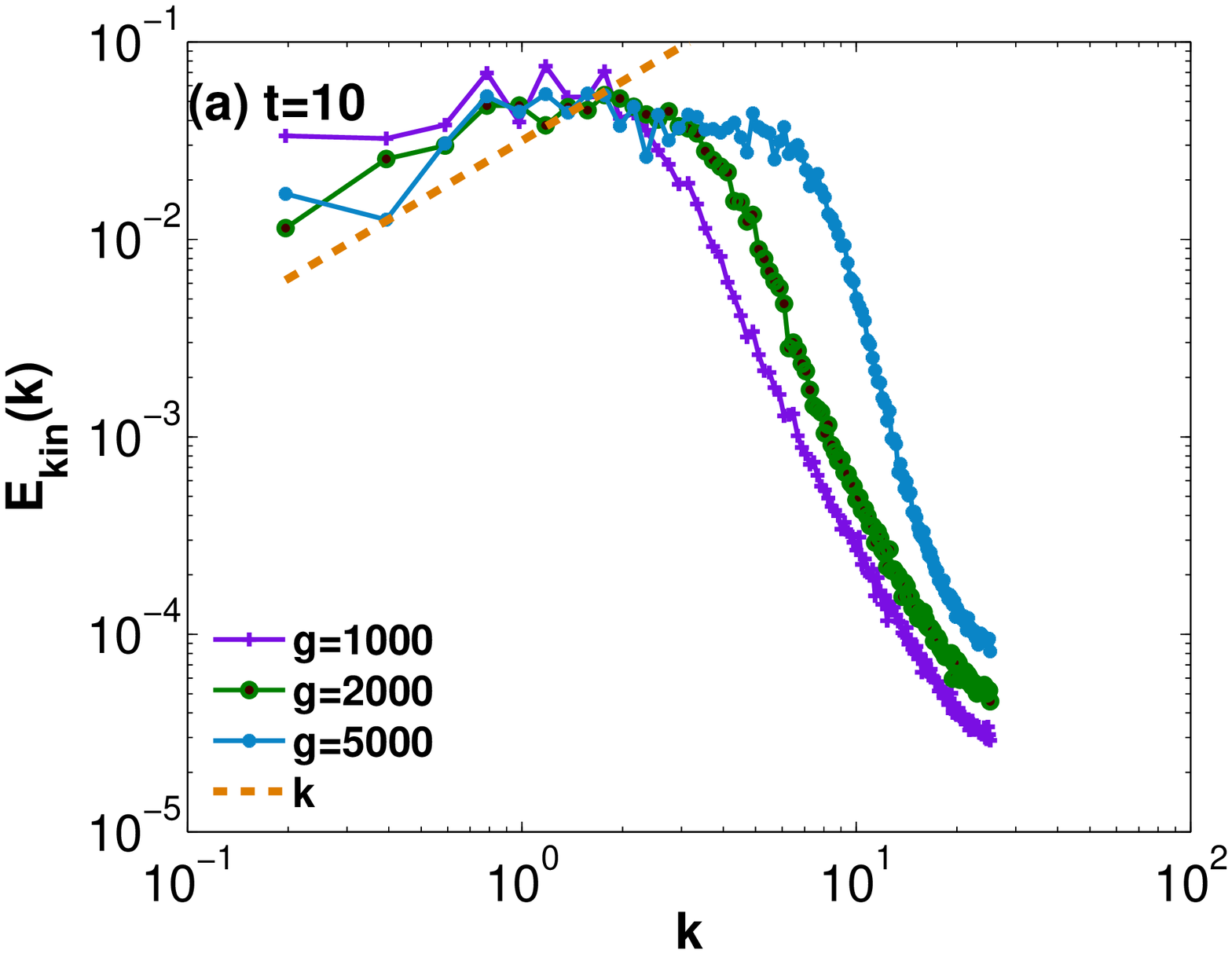}
\includegraphics[height=4.cm]{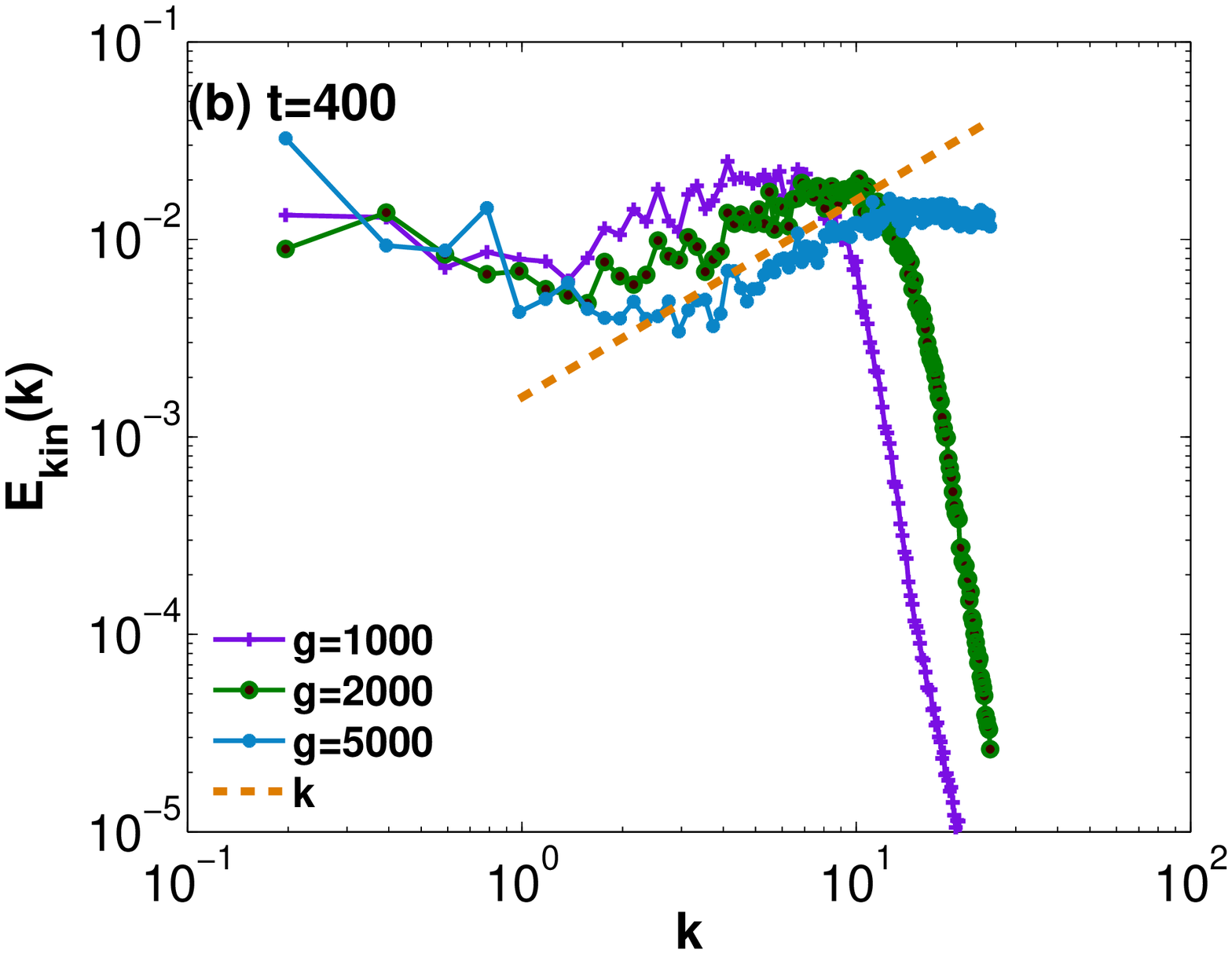}
\includegraphics[height=4.cm]{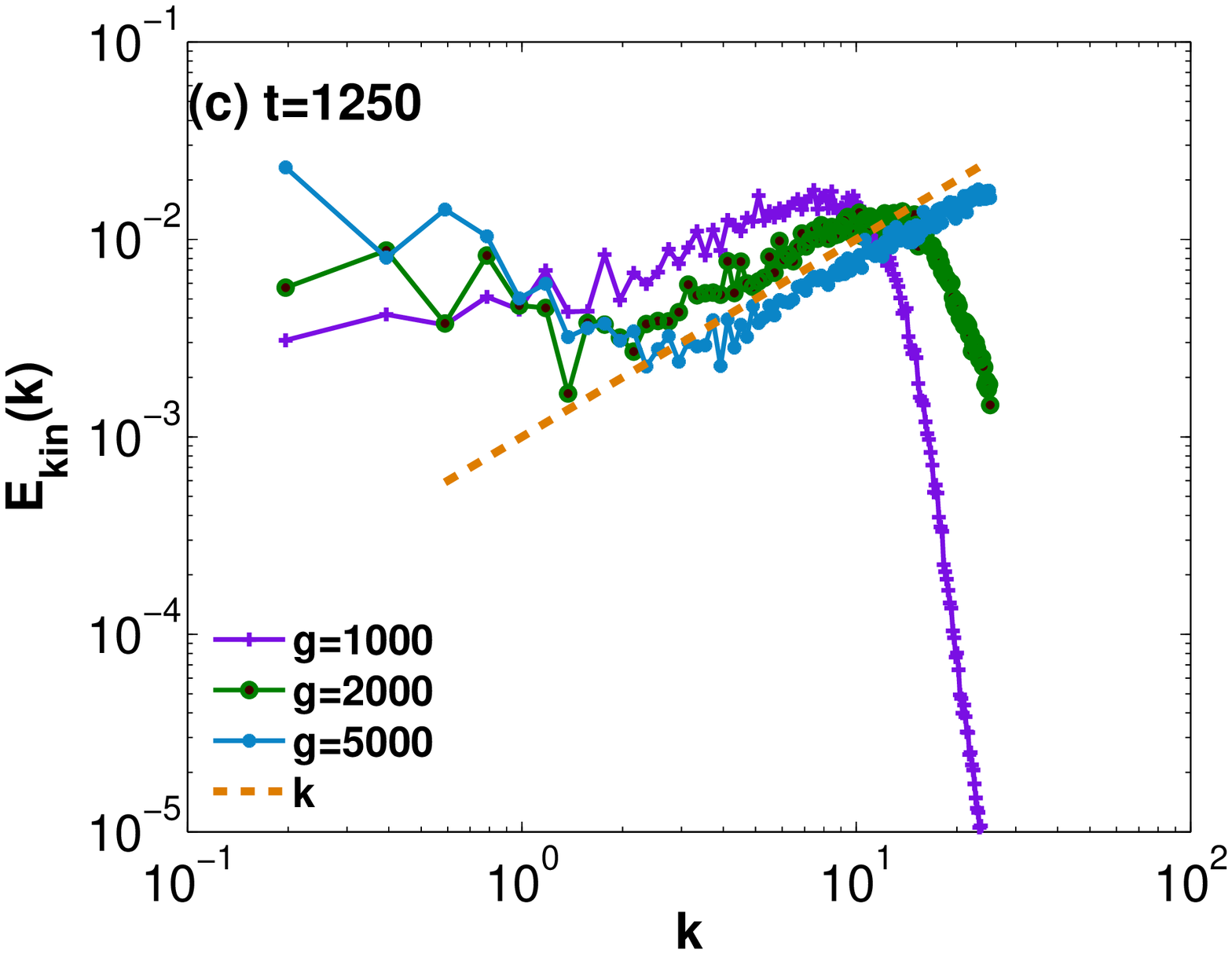}
\includegraphics[height=4.cm]{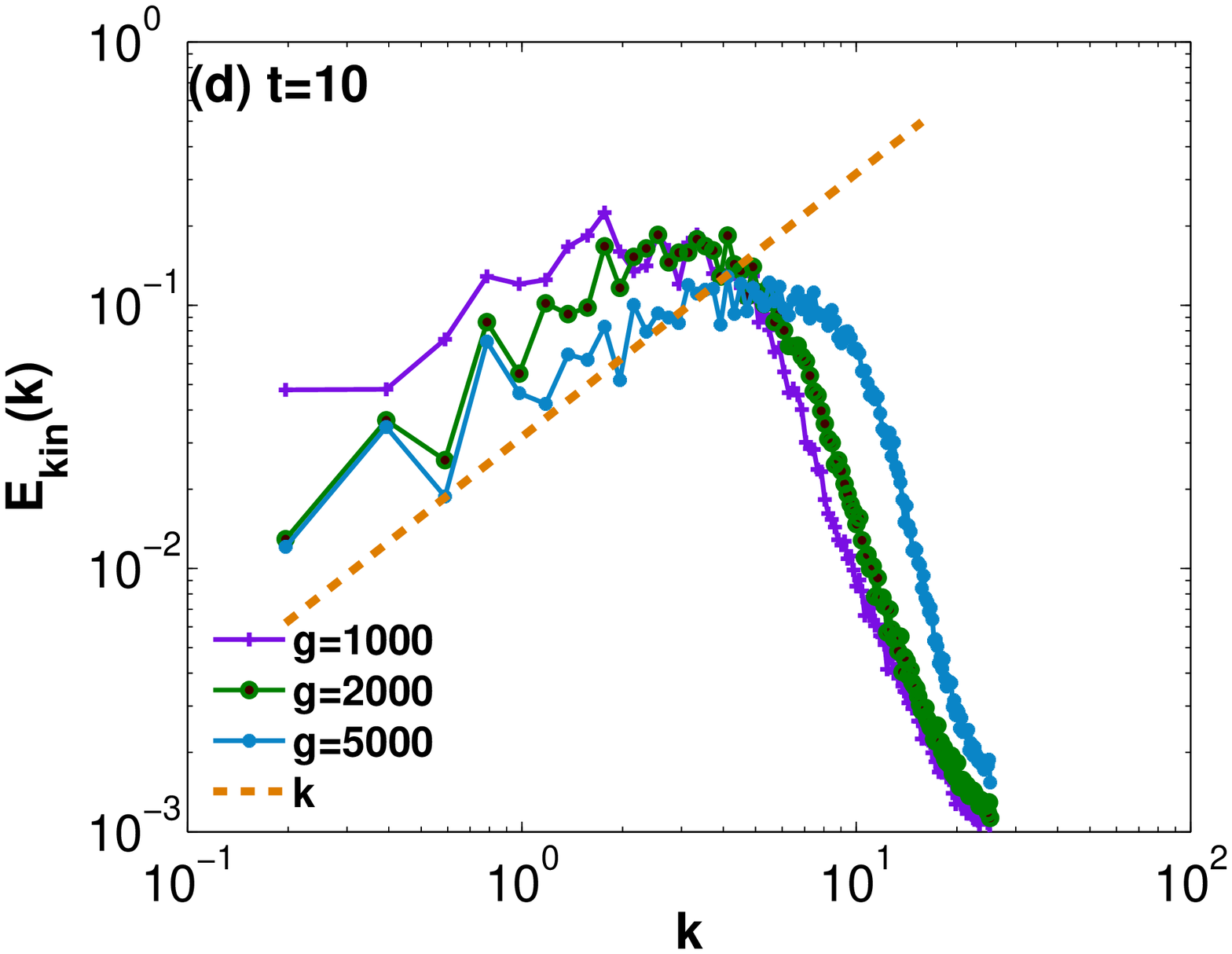}
\includegraphics[height=4.cm]{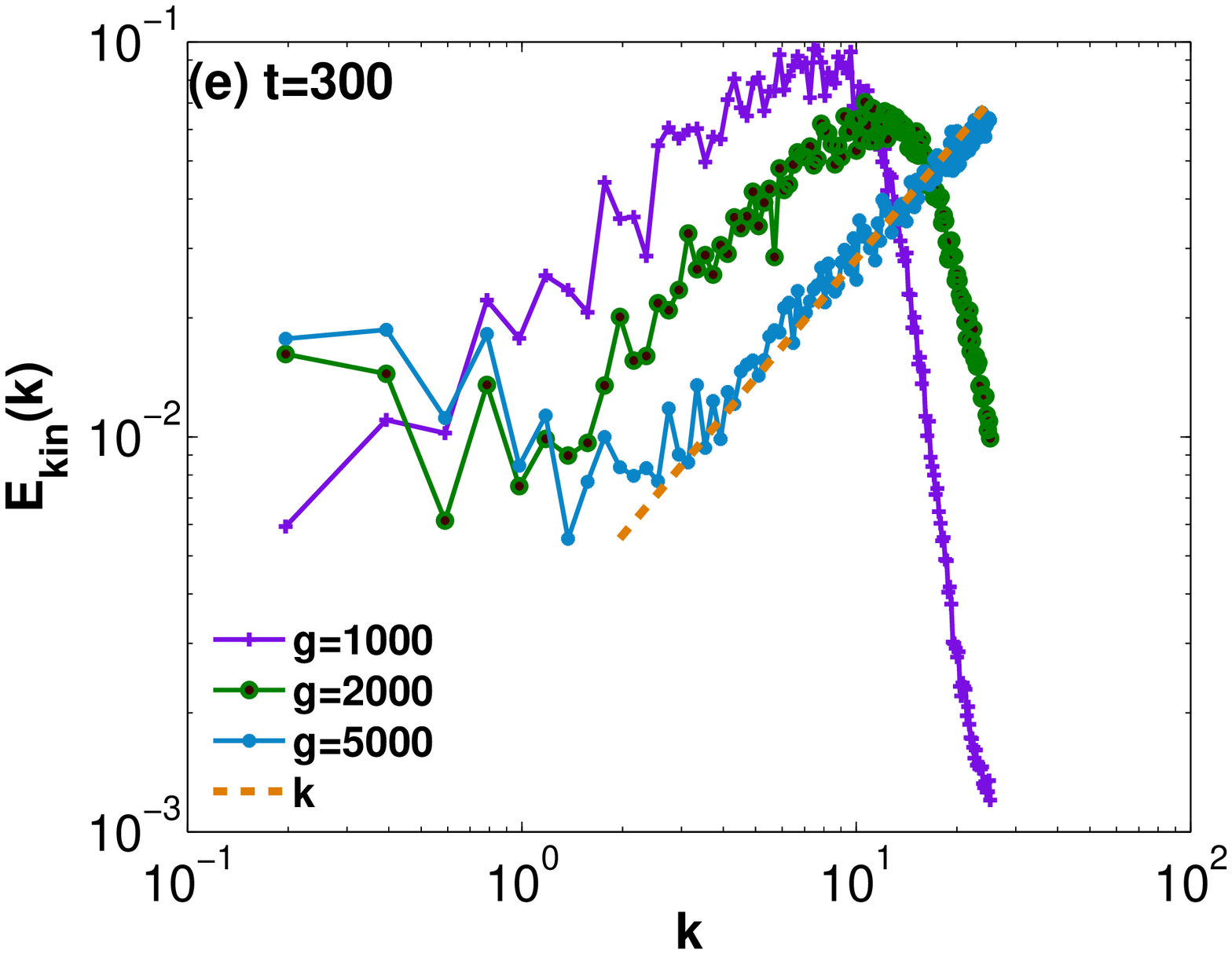}
\includegraphics[height=4.cm]{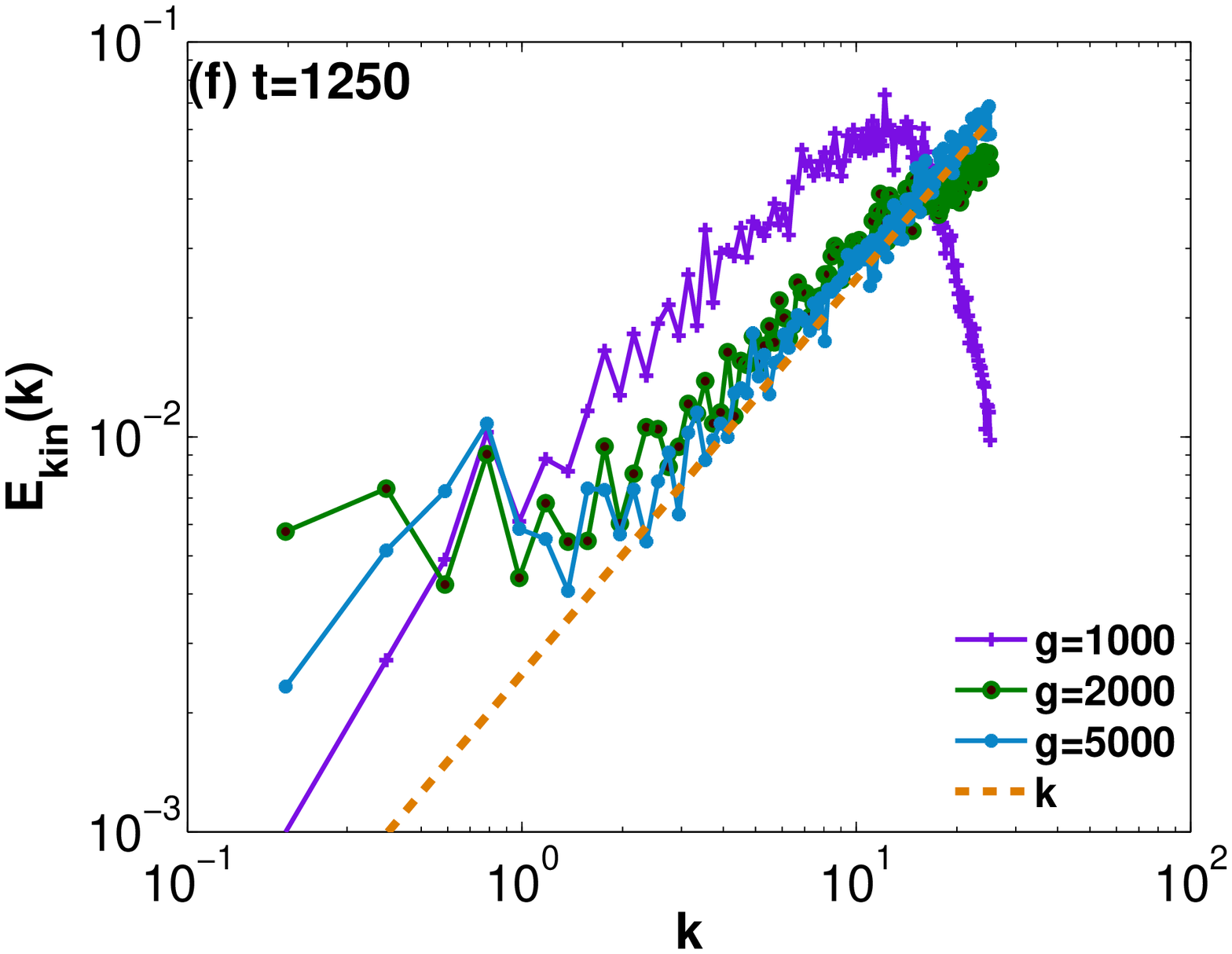}
\includegraphics[height=3.9cm]{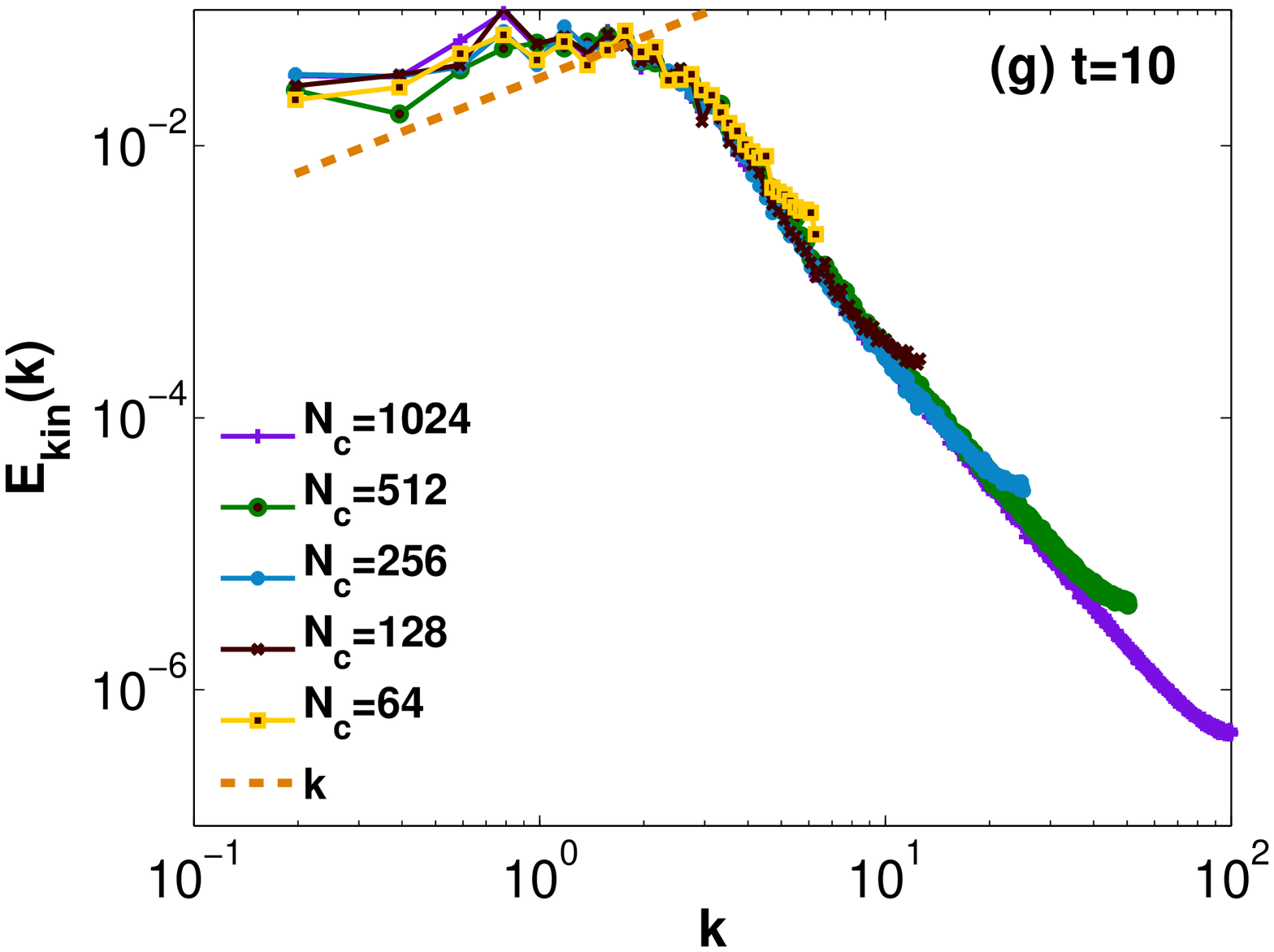}
\includegraphics[height=3.9cm]{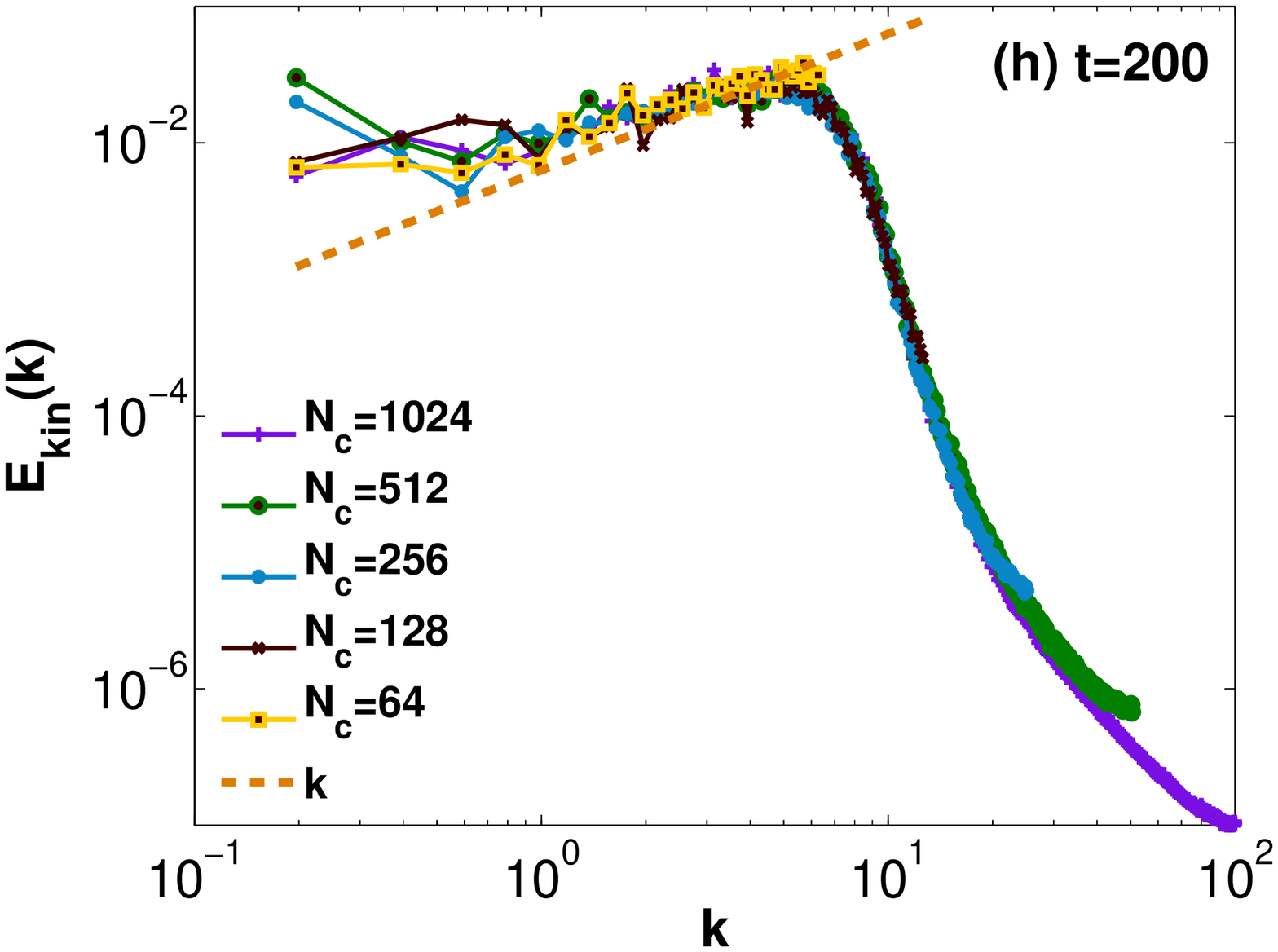}
\includegraphics[height=3.9cm]{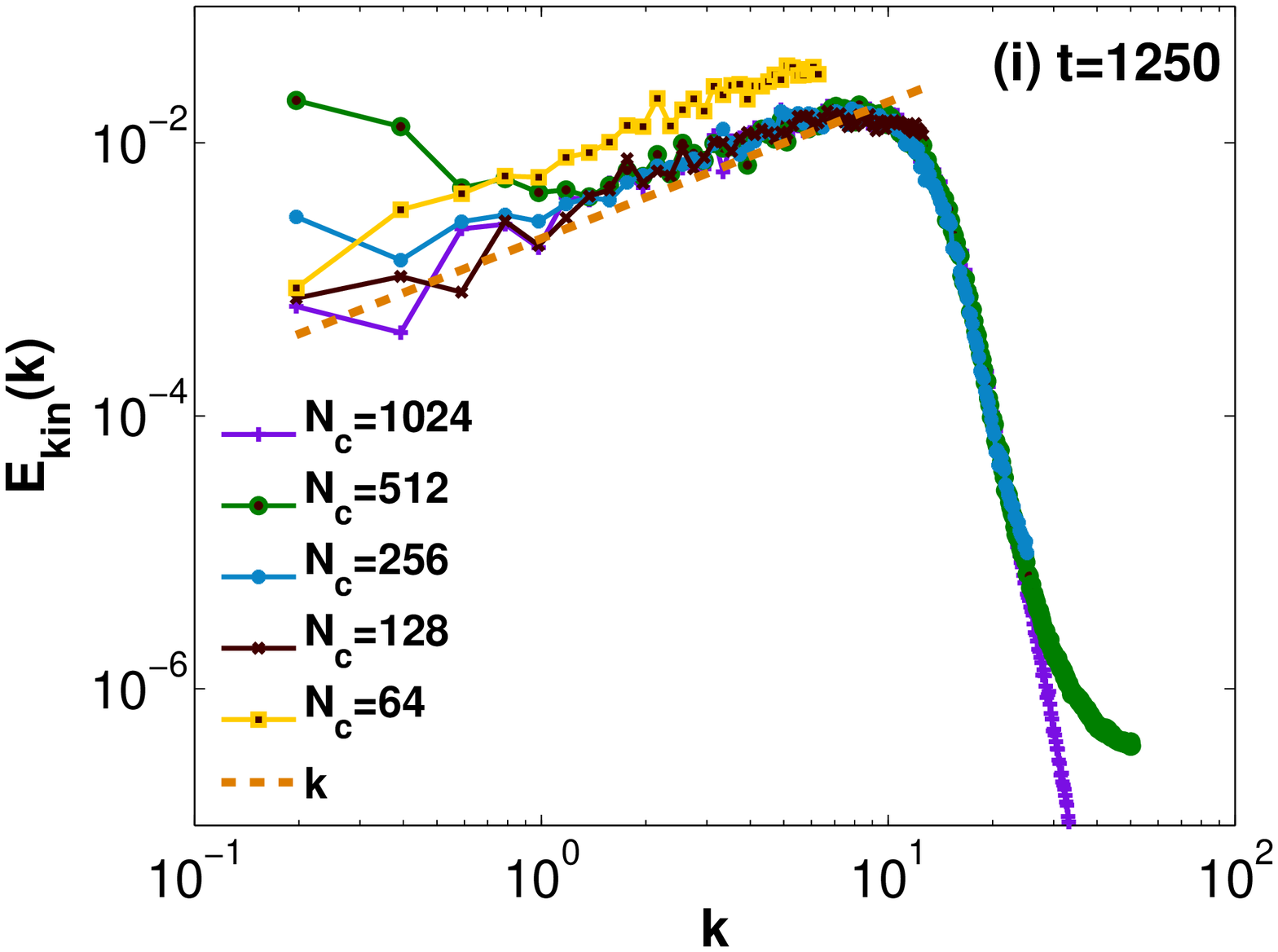}
\end{center}
\caption{\small Log-log (base 10) plots of the spectra $E_{kin}(k)$ from 
our DNS runs (a)-(c) $\tt A6$, $\tt A9$, and $\tt A10$ ($k_0=5\Delta k$ and 
$\sigma=2\Delta k$), (d)-(f) $\tt A11$, $\tt A12$ and $\tt A13$  
($k_0=15\Delta k$ and $\sigma=2\Delta$), and
(g)-(i) $\tt A1$, $\tt A5$-$\tt A8$ ($N^2_c=1024^2$, $512^2$, 
$256^2$, $128^2$, and $64^2$).
The complete time evolutions of the spectra in
(a)-(c), (d)-(f), and, (g)-(i) are illustrated in the panels 
V2, V3, and V4 of video S3.}
\label{fig:kediffgNc}
\end{figure*}

\begin{figure*}
\begin{center}
\includegraphics[height=4.cm]{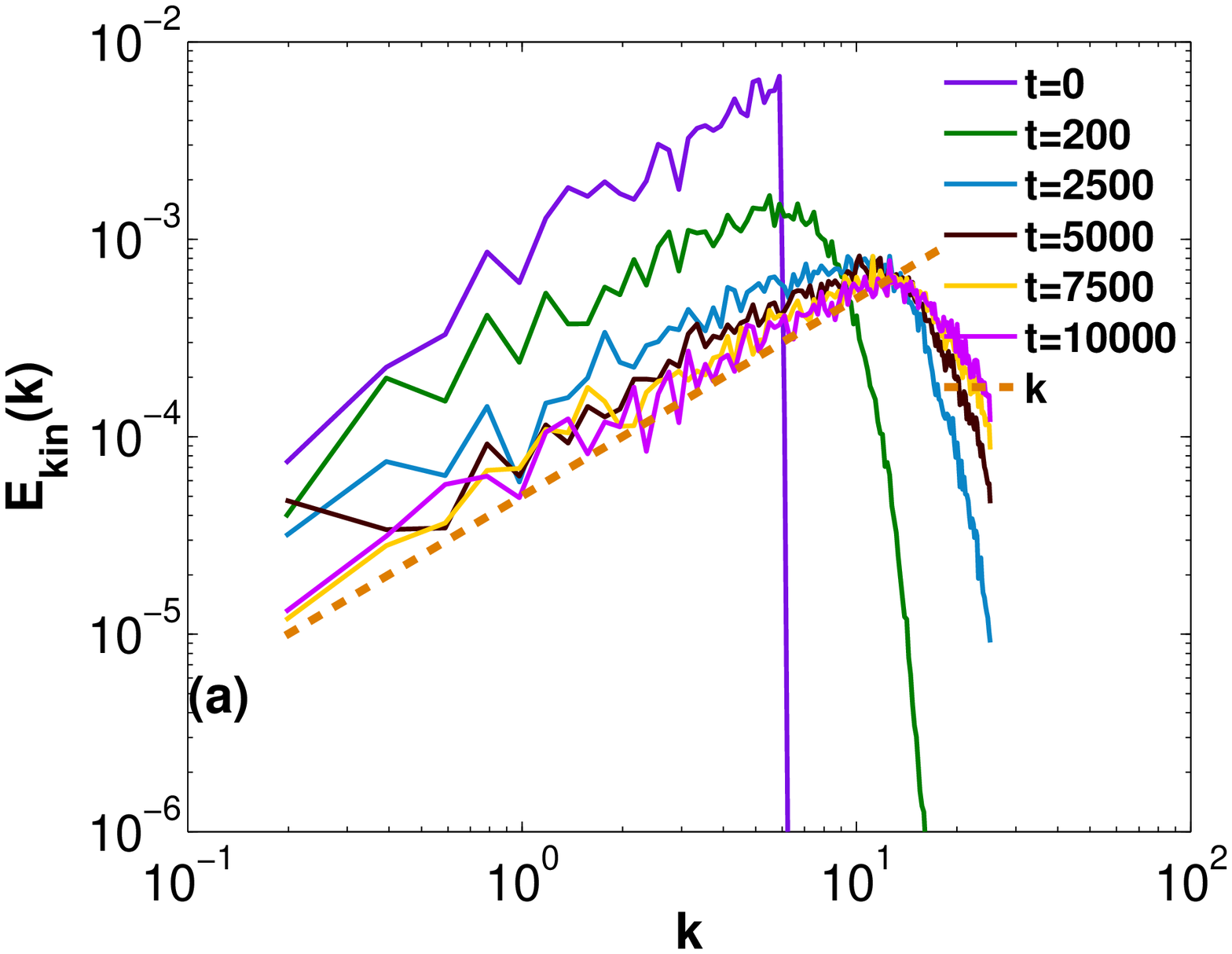}
\includegraphics[height=4.cm]{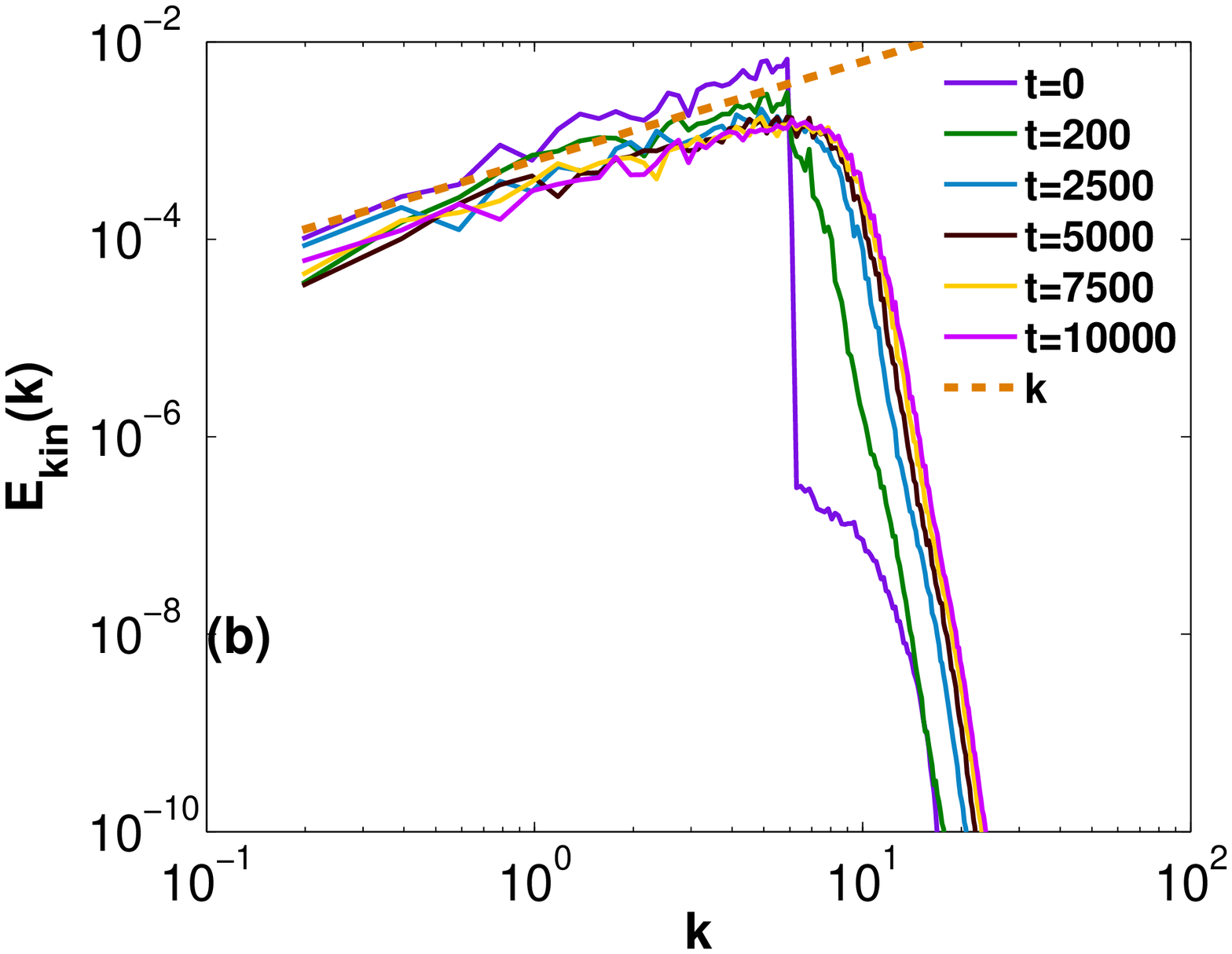}
\includegraphics[height=4.cm]{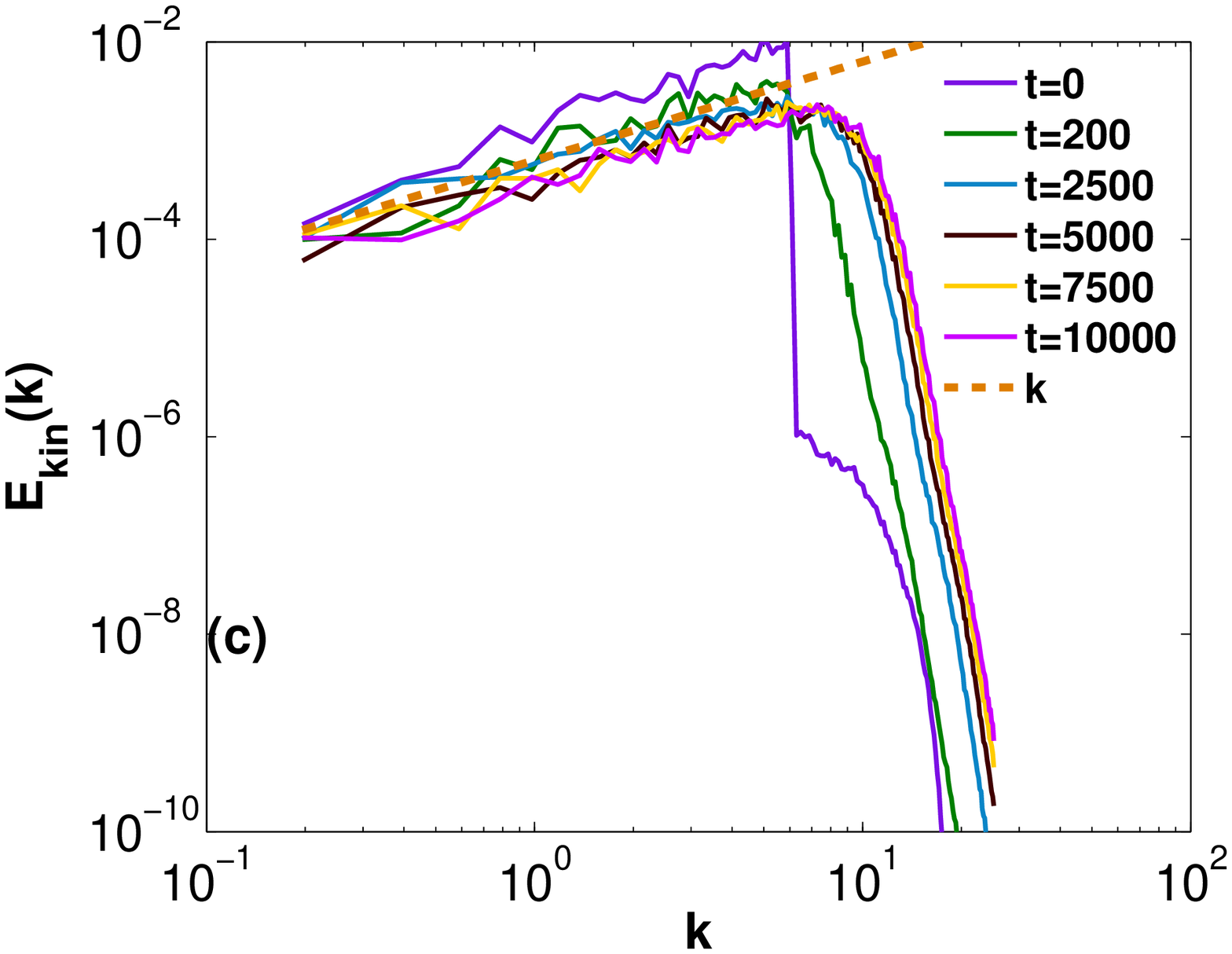}
\includegraphics[height=4.cm]{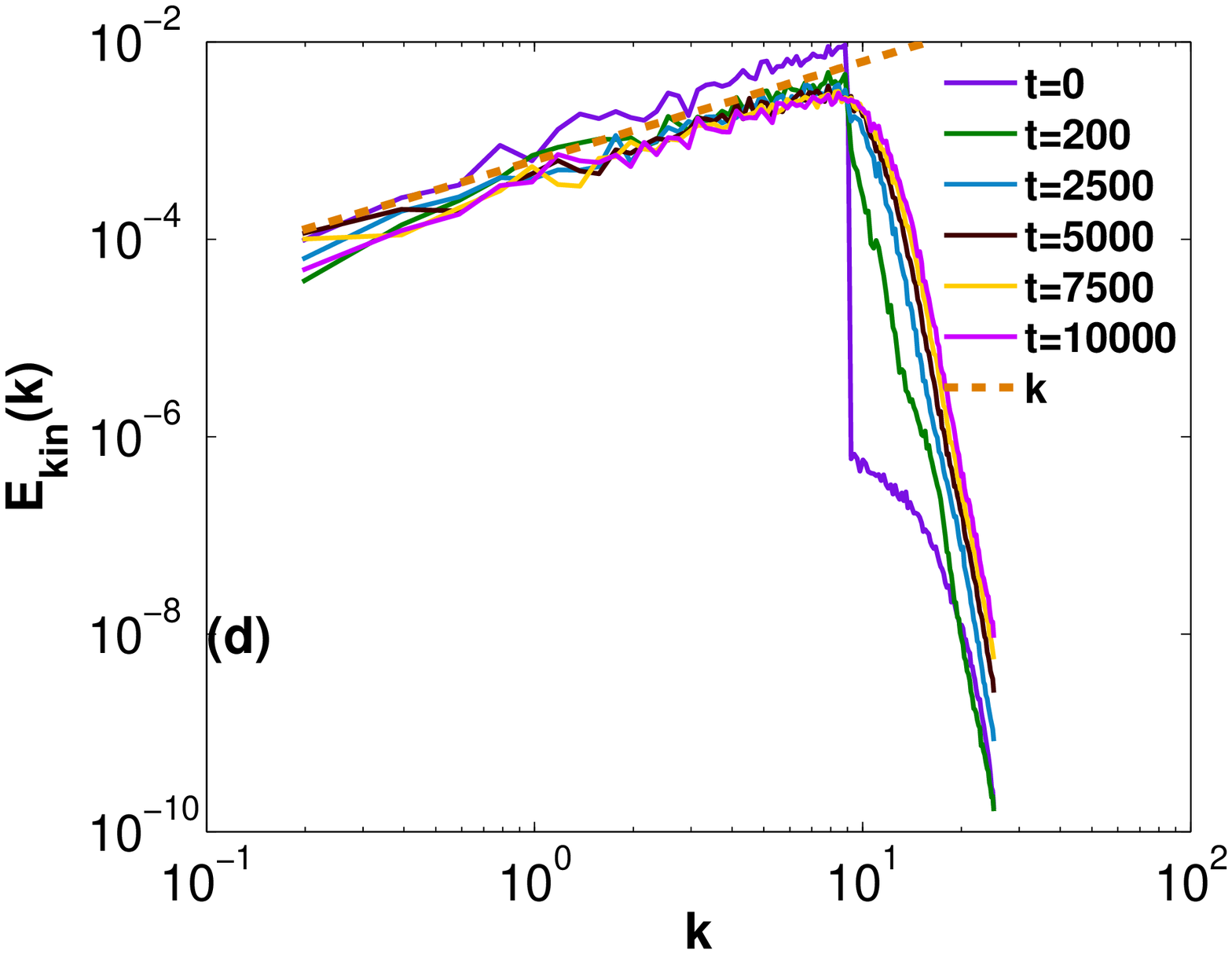}
\includegraphics[height=4.cm]{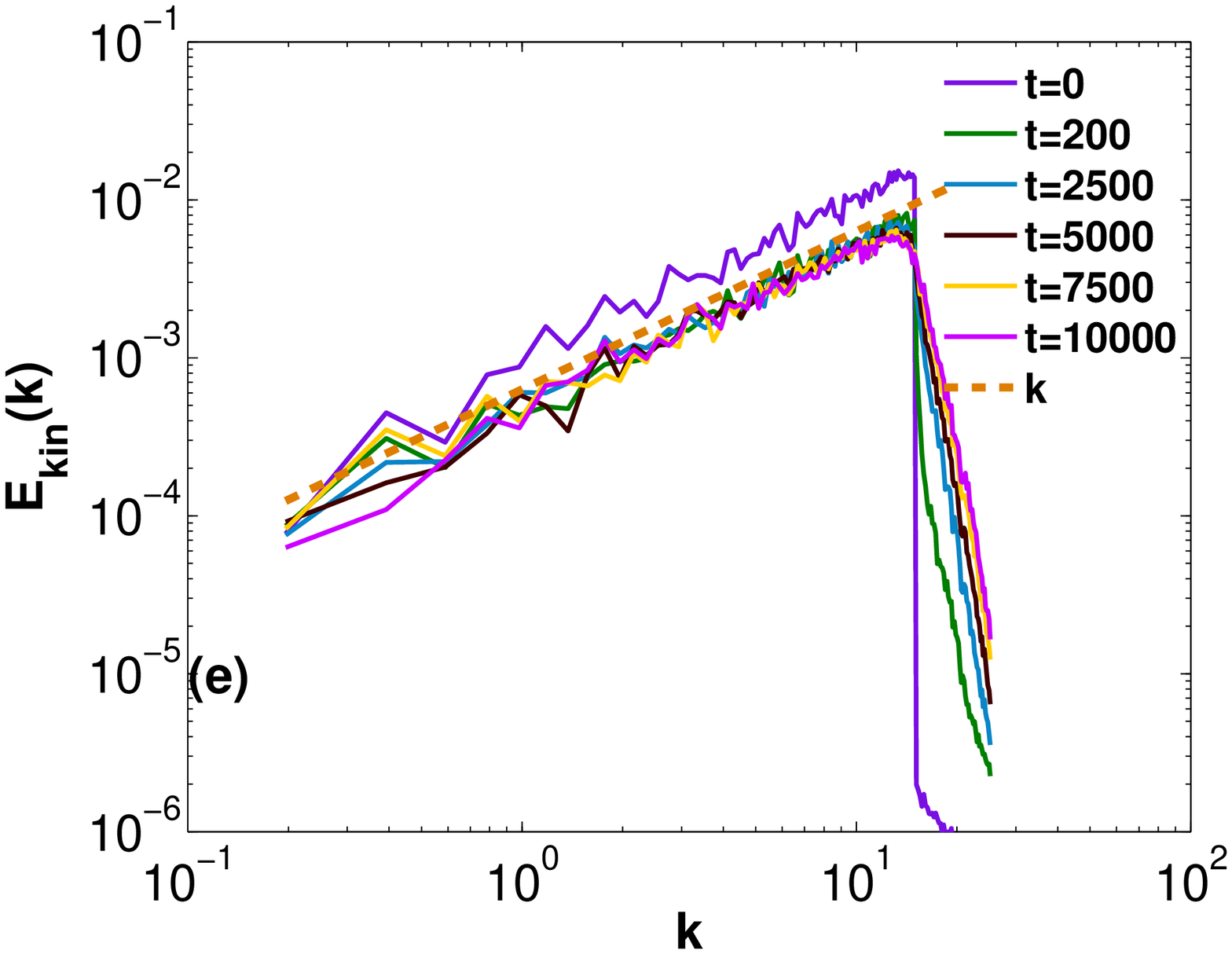}
\includegraphics[height=4.cm]{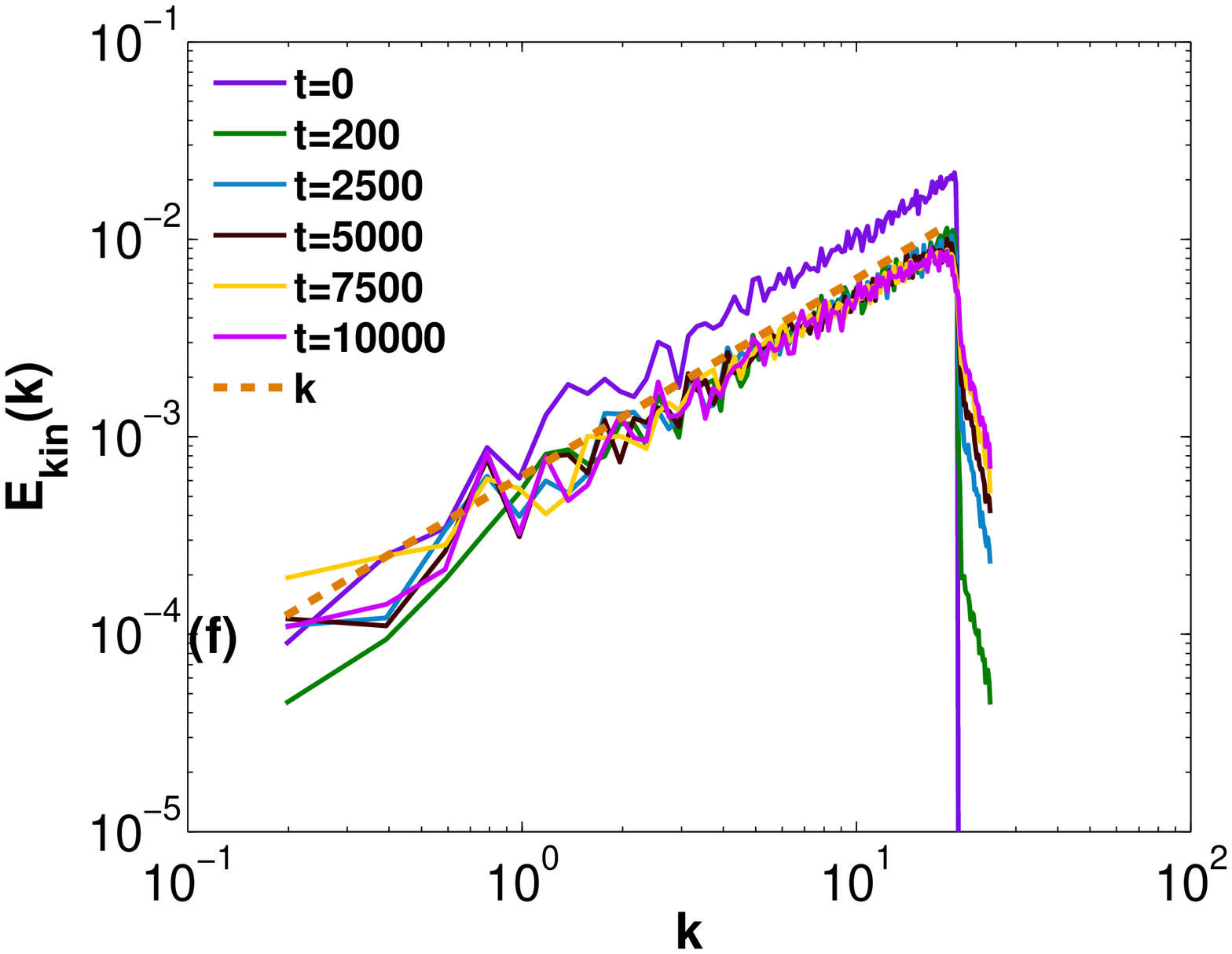}
\end{center}
\caption{\small Log-log (base 10) plots of the spectra $E_{kin}(k)$ from 
our DNS runs (initial conditions of type ${\tt IC3}$ (a) $\tt
C1$, (b) $\tt C2$, (c) $\tt C3$, (d) $\tt C4$, (e) $\tt C5$, and 
(f) $\tt C6$.}
\label{fig:ic3ckes}
\end{figure*}

With initial conditions of types $\tt IC1$ and $\tt IC2$, we
cannot control the initial value $k_c(t=0)\equiv k^{in}_c$
easily. However, initial conditions of type $\tt IC3$, which we
obtain from the SGLE, allow us to control $k^{in}_c$ and start,
therefore, with initial spectra that display partial
thermalization for $k < k^{in}_c$~\cite{Krstulovic2011pre} and a
sharp fall thereafter.  In figure \ref{fig:ic3ckes} we show the
time evolution of $E^c_{kin}(k)$ for such initial conditions from
runs $\tt C1$-$\tt C6$. For different representative values of
$k^{in}_c$, $g$, and $D$, we now study the time evolution of
$k_c(t)$, which characterizes the growth of the partially
thermalized scaling region. Here too, as with initial conditions
of types $\tt IC1$ and $\tt IC2$, if all other parameters like
$k^{in}_c=6.0$ and $D$ are held fixed, the speed of
thermalization increases with $g$ (cf. figure \ref{fig:ic3ckes}
(a) for the run $\tt C1$, with $g=5000$, and figure \ref{fig:ic3ckes}
(b) for the run $\tt C2$, with $g=1000$).  For these runs $\tt
C1$-$\tt C6$, the growth of the energy spectra, in the region $k>
k^{in}_c$, starts with the smoothening of the sharp cut-off at
$k^{in}_c$; the higher the value of $k^{in}_c$, the slower is
this growth (cf. figures \ref{fig:ic3ckes} (b), (d), (e), and (f)
for runs $\tt C2$, $\tt C4$, $\tt C5$, and $\tt C6$,
respectively). By contrast, an increase in $D$ (or $T$) in the
SGLE, accelerates this growth (cf. figures \ref{fig:ic3ckes} (b)
and (c) for runs $\tt C2$ and $\tt C3$, respectively).

\begin{figure*}
\begin{center}
\includegraphics[height=5.cm]{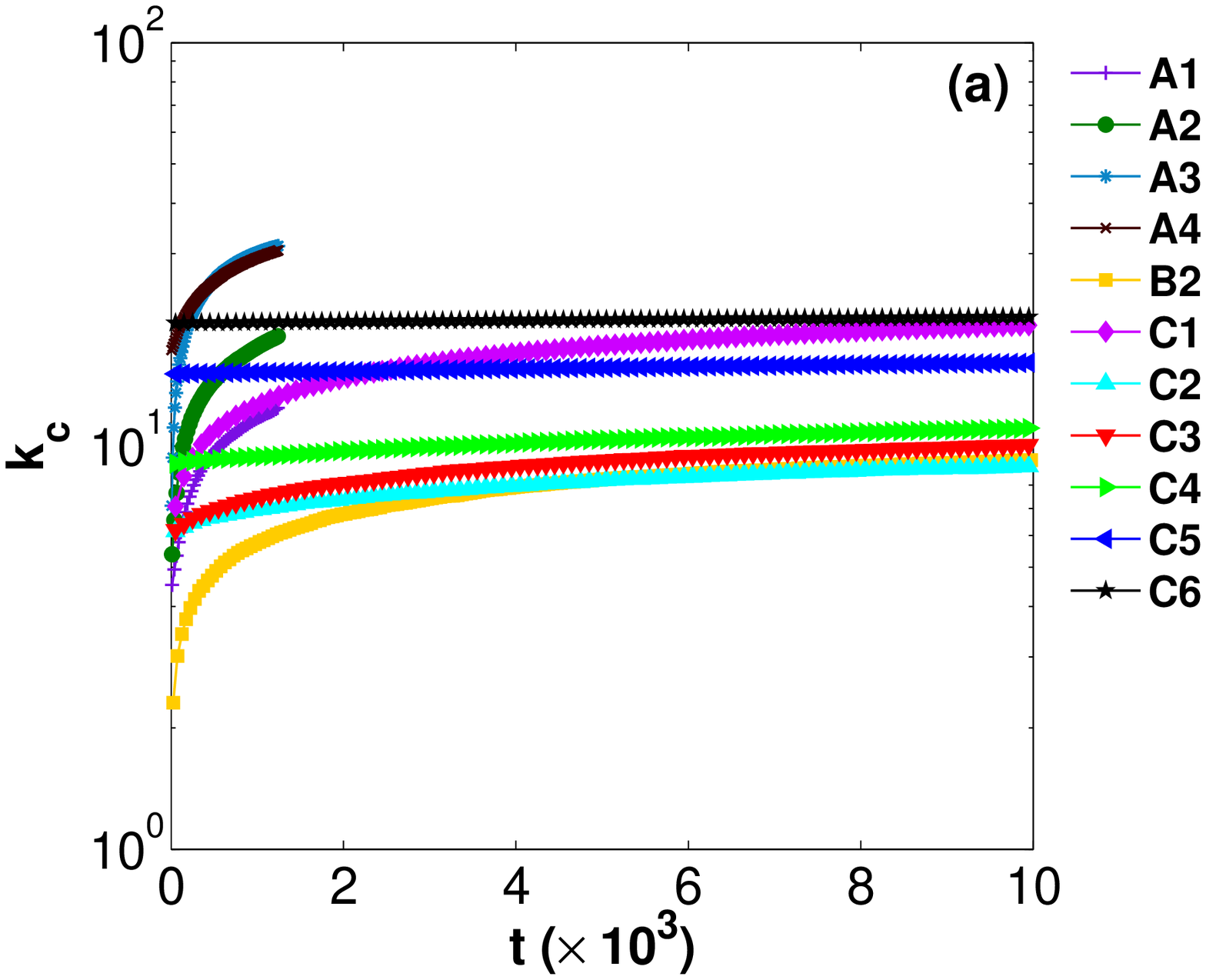}
\includegraphics[height=5.cm]{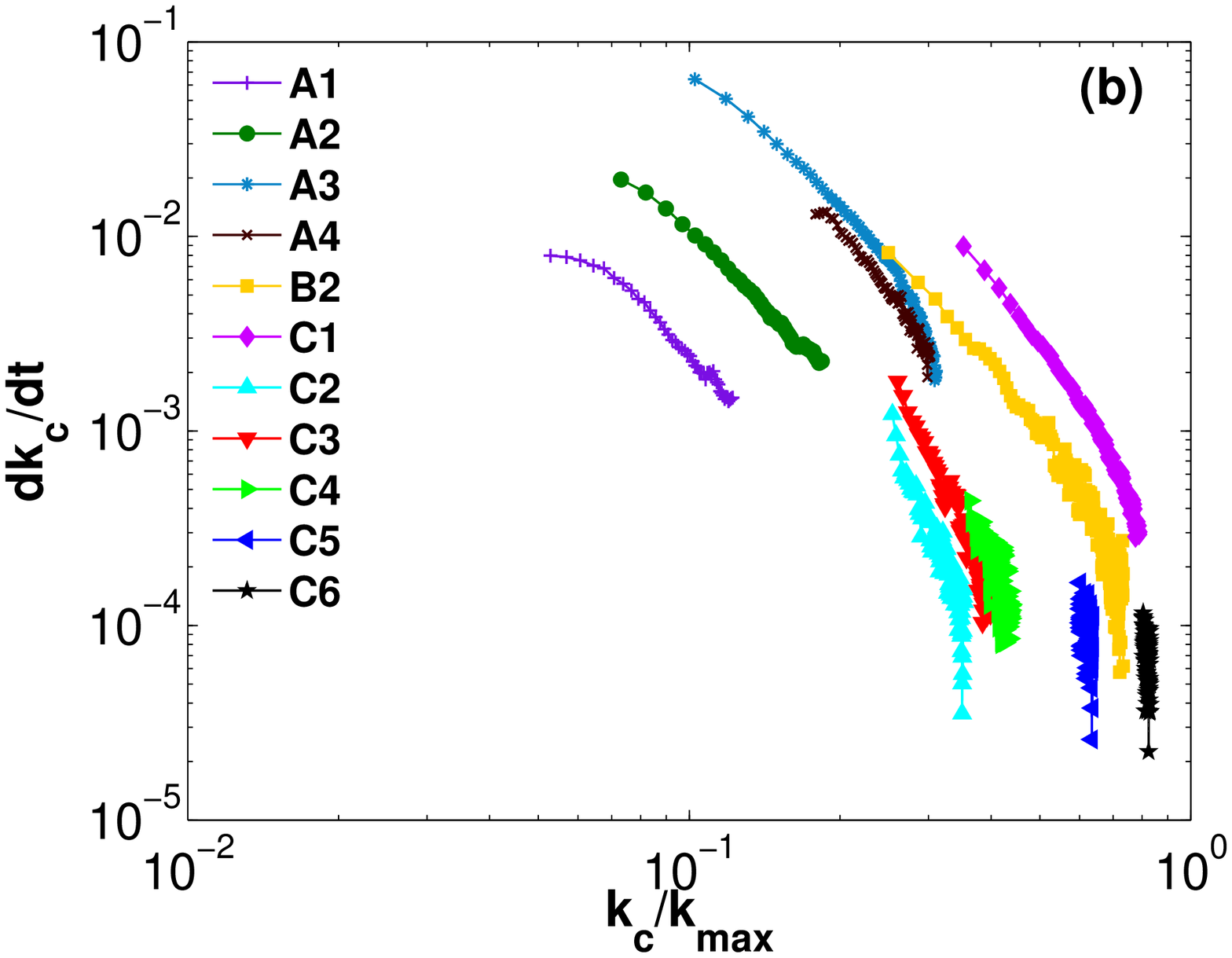}
\end{center}
\caption{\small Plots of (a) the self-truncation wave-number $k_c(t)$ 
versus time $t$ and (b) $dk_c/dt$ versus $k_c/k_{max}$ from our DNS runs 
$\tt A1$-$\tt A4$, $\tt B2$, and $\tt C1$-$\tt C6$. 
}
\label{fig:kctevolve}
\end{figure*}

The growth of $k_c(t)$ with $t$, illustrated in figure
\ref{fig:kctevolve} (a), can be fit to the form $k_c(t) \sim
t^\alpha$; however, as we show below, $\alpha$ depends on the
initial condition. We obtain the exponent $\alpha$ either from
slopes of log-log plots of (i) $k_c(t)$ versus $t$ or (ii)
$dk_c/dt$ versus $k_c/k_{max}$; we denote the values from
procedures (i) and (ii) as $\alpha_1$ and $\alpha_2$,
respectively. Note that in (ii) we have a parametric
plot~\cite{Krstulovic2011prl,Krstulovic2011pre}, shown in figure
\ref{fig:kctevolve} (b); this yields a straight-line scaling
regime with slope $\chi$ and $\alpha_2 = 1/(1-\chi)$. The values
of $\alpha_1$ and $\alpha_2$, listed in table~\ref{table:alpha},
show that $\alpha_1\simeq\alpha_2$; the discrepancy between these
two values for $\alpha$ is a convenient measure of the errors of
our estimates. For runs $\tt C4$, $\tt C5$, and $\tt C6$, we
cannot obtain $\alpha_2$ reliably; the small values of $\alpha_1$
for these runs indicate very slow growth of $k_c(t)$; indeed, in 
runs $\tt C5$ and $\tt C6$, a case can be made for a logarithmic
growth of $k_c(t)$ with $t$. 

\begin{table}
\small
\centering
   \begin{tabular}{@{\extracolsep{\fill}} c c c c c c c c c c c c }
    \hline
    $ $ & $E$ & $k_{max}$ &$\xi$ &$\xi k_{max}$  &$k^i_c$ &$k^f_c$ & $\alpha_1 $ & $\alpha_2$ \\ 
   \hline \hline
    {\tt A1} & $2.120$ & $100.53$ &$1.01$ &$101.73$ &$4.52$  &$12.42$ & $0.28$ & $0.26$ \\
    {\tt A2} & $3.045$ & $100.53$ &$0.72$ &$71.9$  &$5.39$ &$18.72$ & $0.28$  & $0.28$  \\
    {\tt A3} & $5.82$ & $100.53$ &$0.45$ &$45.49$ &$7.11$ &$31.3$ & $0.29$  & $0.27$  \\
    {\tt A4} & $49.69$ & $100.53$ &$1.01$ &$101.73$ &$17.31$ &$30.53$ & $0.2$  & $0.21$ \\
    {\tt B2} & $0.589$ & $12.57$ &$1.01$ &$12.72$ &$2.23$ &$9.23$ & $0.24$  & $0.25$  \\
    {\tt C1} & $2.536$ & $25.13$ &$0.45$ &$11.37$ &$7.08$ &$19.91$ & $0.22$  & $0.22$ \\
    {\tt C2} & $0.583$ & $25.13$ &$1.01$ &$25.43$ &$6.15$ &$8.90$ & $0.12$  & $0.14$  \\
    {\tt C3} & $0.637$ & $25.13$ &$1.01$ &$25.43$ &$6.18$ &$10.05$ & $0.14$  & $0.15$ \\
    {\tt C4} & $0.6999$ & $25.13$ &$1.01$ &$25.43$ &$9.05$ &$11.07$ & $0.09$  & $-$ \\
    {\tt C5} & $1.085$ & $25.13$ &$1.01$ &$25.43$ &$15.09$ &$16.08$ & $0.04$  & $-$  \\
    {\tt C6} & $1.557$ & $25.13$ &$1.01$ &$25.43$ &$20.17$ &$20.87$ & $0.02$  & $-$  \\
\hline
\end{tabular}
\caption{\small Summary of the self-truncation results from our 
DNS runs $\tt A1$-$\tt A4$, $\tt B2$, and $\tt C1$-$\tt C6$: 
$E$ is the total energy; $k_{max}=2\pi N_c/2L$; 
$\xi=L/\sqrt{g}$ is the healing length; $k^i_c$ and  $k^f_c$ are 
the initial and final values of $k_c$ (averaged over a few time
steps); $\alpha_1$ is the slope obtained from the log-log (base $10$) 
plot of $k_c$ versus $t$ and $\alpha_2=1/(1-\chi)$, where $\chi$ is 
the slope obtained from the log-log (base $10$) plot of 
$dk_c/dt$ versus $k_c/k_{max}$.}
\label{table:alpha}
\end{table}
\subsection{Complete thermalization}
\label{subsection:completethermalization}

The partially thermalized stage of the dynamical evolution of the
2D, Fourier-truncated, GP equation may either gradually become
completely thermalized, in which state a power-law scaling region is
present in the entire energy and the occupation number spectra,
or remain self-truncated with logarithmic growth.  In figures
\ref{fig:A1A4B1ckes} (g)-(i) and \ref{fig:A7B2ckes} (a)-(c), we
show the compressible kinetic energy spectra $E^c_{kin}$ for the
runs $\tt B1$ and $\tt A7$, where $E^c_{kin}$ shows power-law
scaling over the entire wave number range, from
$k=2\pi /L$ up to $k_{max}$, towards the end of the respective
simulations; a na\"ive fit is consistent with $E^c_{kin}(k) \sim k$ 
(but see below).

\subsubsection{Correlation functions and the BKT transition}
\label{subsub:corrfuncBKT}

A uniform, 2D, interacting Bose gas exhibits a BKT phase at low
energies (temperatures in the canonical ensemble). Thus, the
completely thermalized state of the 2D, Fourier-truncated, GP equation should yield a
BKT phase~\cite{Kogut1979rmp,Foster2010pra}, with the correlation
function $c(r) \sim r^{-\eta}$, at energies $E < E_{BKT}$; and
$c(r)$ should decay exponentially with $r$ if $E > E_{BKT}$. We
show this explicitly now by using initial conditions of type $\tt
IC1$ with $N_c=64$ and $N_c=128$ and $g=1000$; we obtain
different energies by changing $k_0$ and $\sigma$ (runs D1-D13 and
E1-E12 in table~\ref{table:paract128n64}).

\begin{table*}
\small
	\begin{tabular}{@{\extracolsep{\fill}} c c c c c | c c c c c}
\hline
	$N_c=128$ &$k_0$ &$\sigma$ &$E$ &$\eta$ &$N_c=64$ &$k_0$ &$\sigma$ &$E$ &$\eta$\\
	$$ &$(\times \Delta k)$ &$(\times \Delta k)$ &$$ &$$ 
&$$ &$(\times \Delta k)$ &$(\times \Delta k)$ &$$ &$$\\
\hline \hline
	{\tt D1} &$5$ &$2$ &$2.1$ &$0.008$ &{\tt E1} &$0$ &$2$ &$1.12$ &$0.012$ \\
	{\tt D2} &$10$ &$2$ &$5.05$ &$0.024$ &{\tt E2} &$3$ &$2$ &$1.64$ &$0.025$ \\
	{\tt D3} &$12$ &$2$ &$6.74$ &$0.034$ &{\tt E3} &$5$ &$2$ &$2.2$ &$0.040$ \\
	{\tt D4} &$14$ &$2$ &$8.74$ &$0.047$ &{\tt E4} &$8$ &$2$ &$3.68$ &$0.083$ \\
	{\tt D5} &$16$ &$2$ &$11.05$ &$0.080$ &{\tt E5} &$10$ &$2$ &$5.04$ &$0.164$ \\
	{\tt D6} &$18$ &$2$ &$13.68$ &$0.111$ &{\tt E6} &$11$ &$2$ &$5.84$ &$0.255$ \\
	{\tt D7} &$20$ &$2$ &$16.62$ &$0.181$ &{\tt E7} &$12$ &$2$ &$6.75$ &$ $ \\ 
	{\tt D8} &$21$ &$2.5$ &$18.34$ &$0.239$ &{\tt E8} &$13$ &$2$ &$7.74$ &$$ \\
	{\tt D9} &$24$ &$3$ &$23.75$ &$$ &{\tt E9} &$14$ &$2$ &$8.78$ &$$ \\
	{\tt D10} &$25$ &$2$ &$25.3$ &$$ &{\tt E10} &$15$ &$2$ &$9.88$ &$$ \\
	{\tt D11} &$26$ &$2$ &$27.27$ &$$ &{\tt E11} &$16$ &$2$ &$11.05$ &$$ \\
	{\tt D12} &$28$ &$2$ &$31.44$ &$$ &{\tt E12} &$17$ &$2$ &$12.32$ &$$ \\
	{\tt D13} &$30$ &$2$ &$35.9$ &$$ &{} &$$ &$$ &$$ &$$ \\
\hline
\end{tabular}
\caption{\small List of parameters for our complete-thermalization DNS runs 
$\tt D1$-$\tt D13$ ($N^2_c=128^2$) and $\tt E1$-$\tt E12$ ($N^2_c=64^2$):
$N^2_c$ is the number of collocation points; $k_0$ is the energy-injection 
scale; $\sigma$ is Fourier-space width of $\psi$ at $t=0$; $E$ is the total 
energy; and $\eta$ is the exponent of the correlation function 
$c(r) \sim r^{-\eta}$ for $E<E_{BKT}$. $g=1000$ for all the DNS runs and they
have been performed on a square simulation domain of area $\mathcal{A}=L^2$, 
with $L=32$. 
}
\label{table:paract128n64}
\end{table*}

\begin{figure*}
\begin{center}
\includegraphics[height=3.9cm]{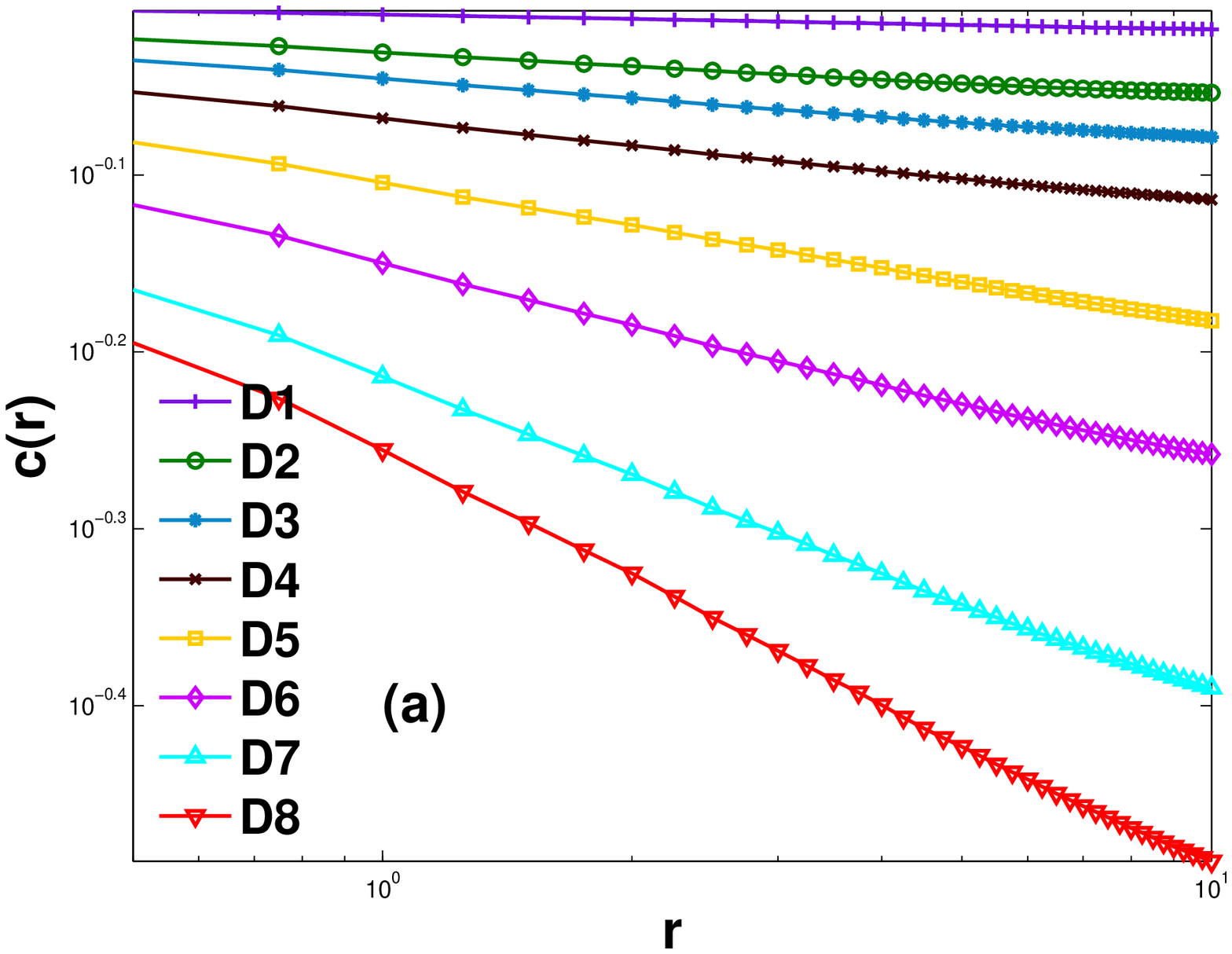}
\includegraphics[height=3.9cm]{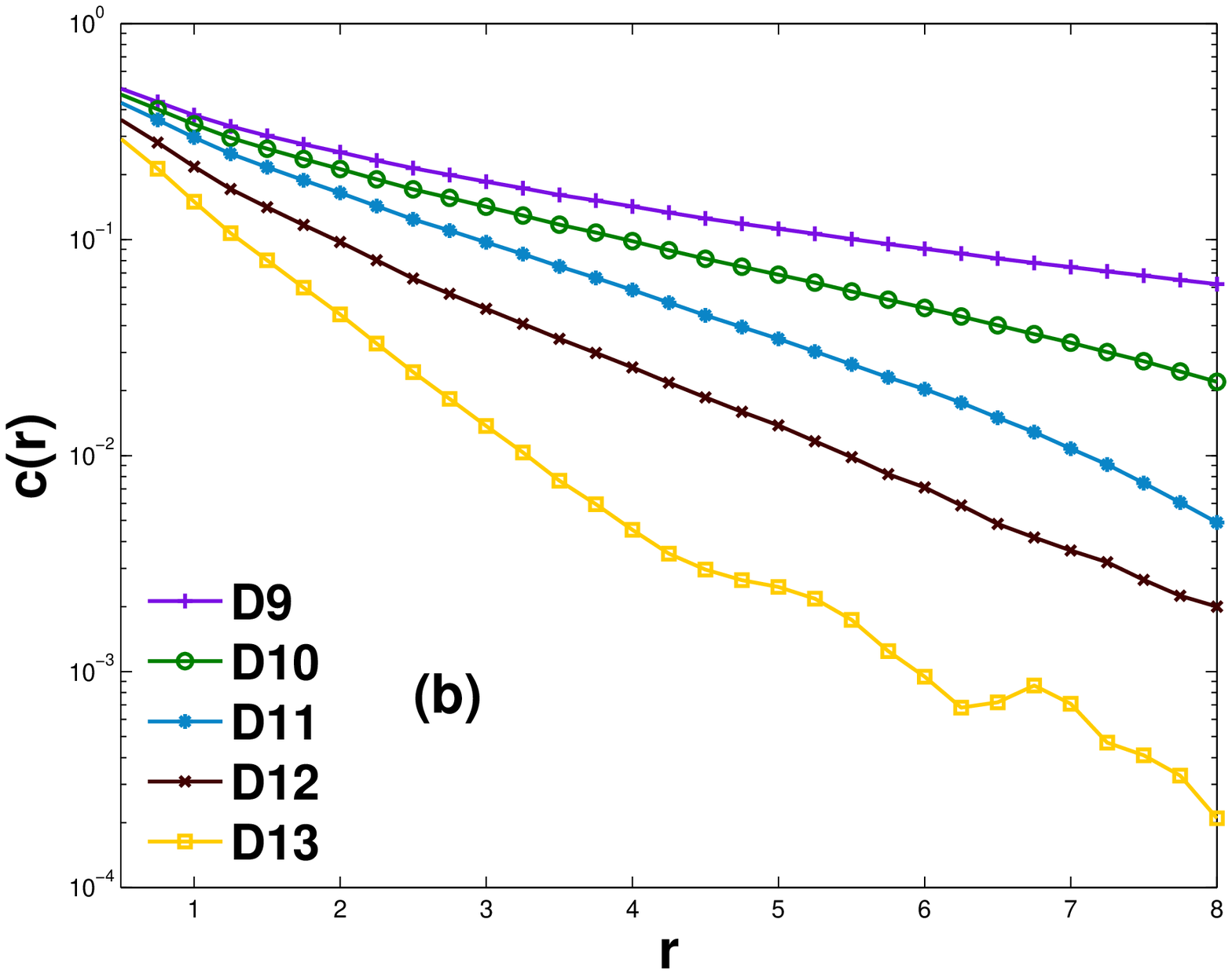}
\includegraphics[height=3.9cm]{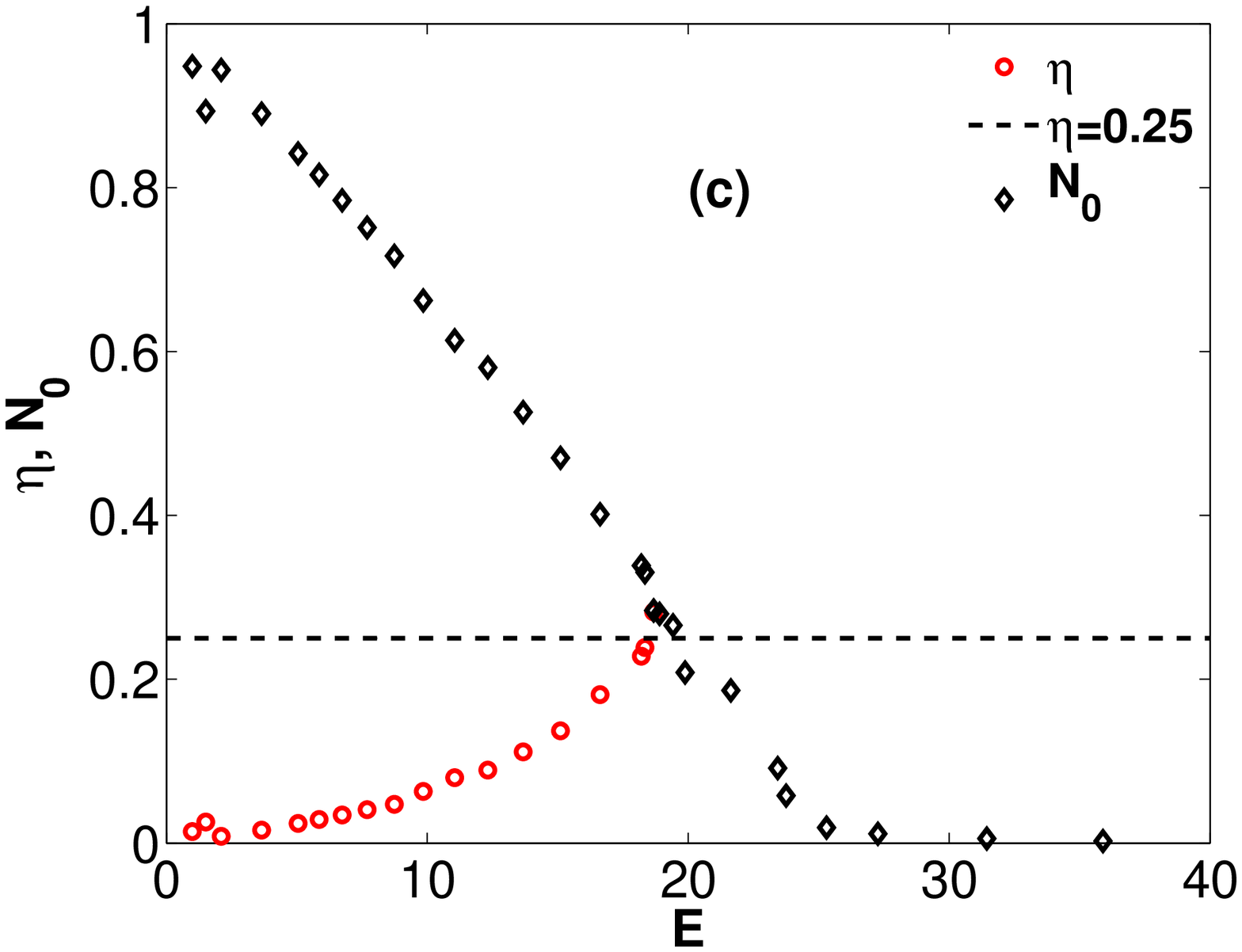}
\includegraphics[height=3.9cm]{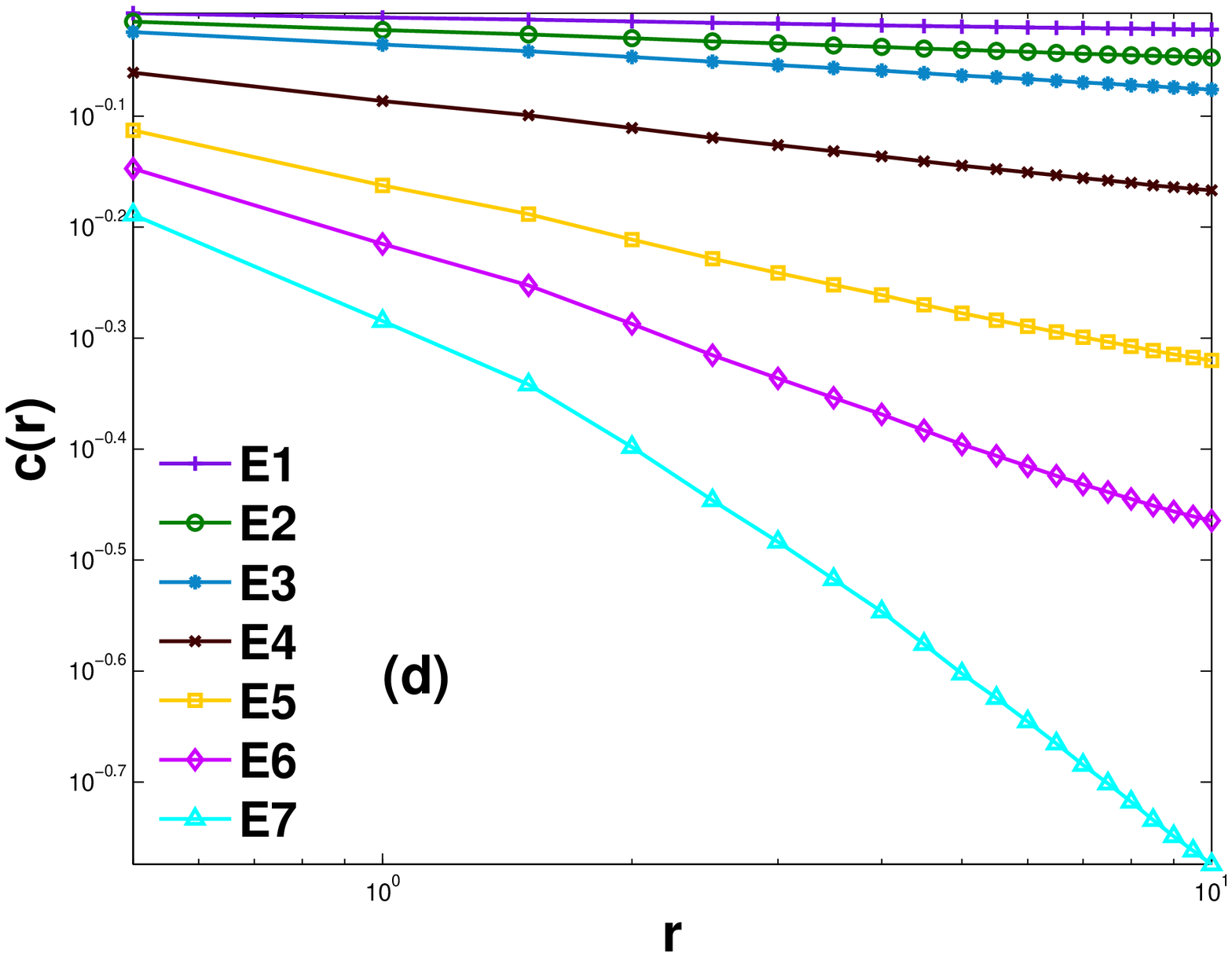}
\includegraphics[height=3.9cm]{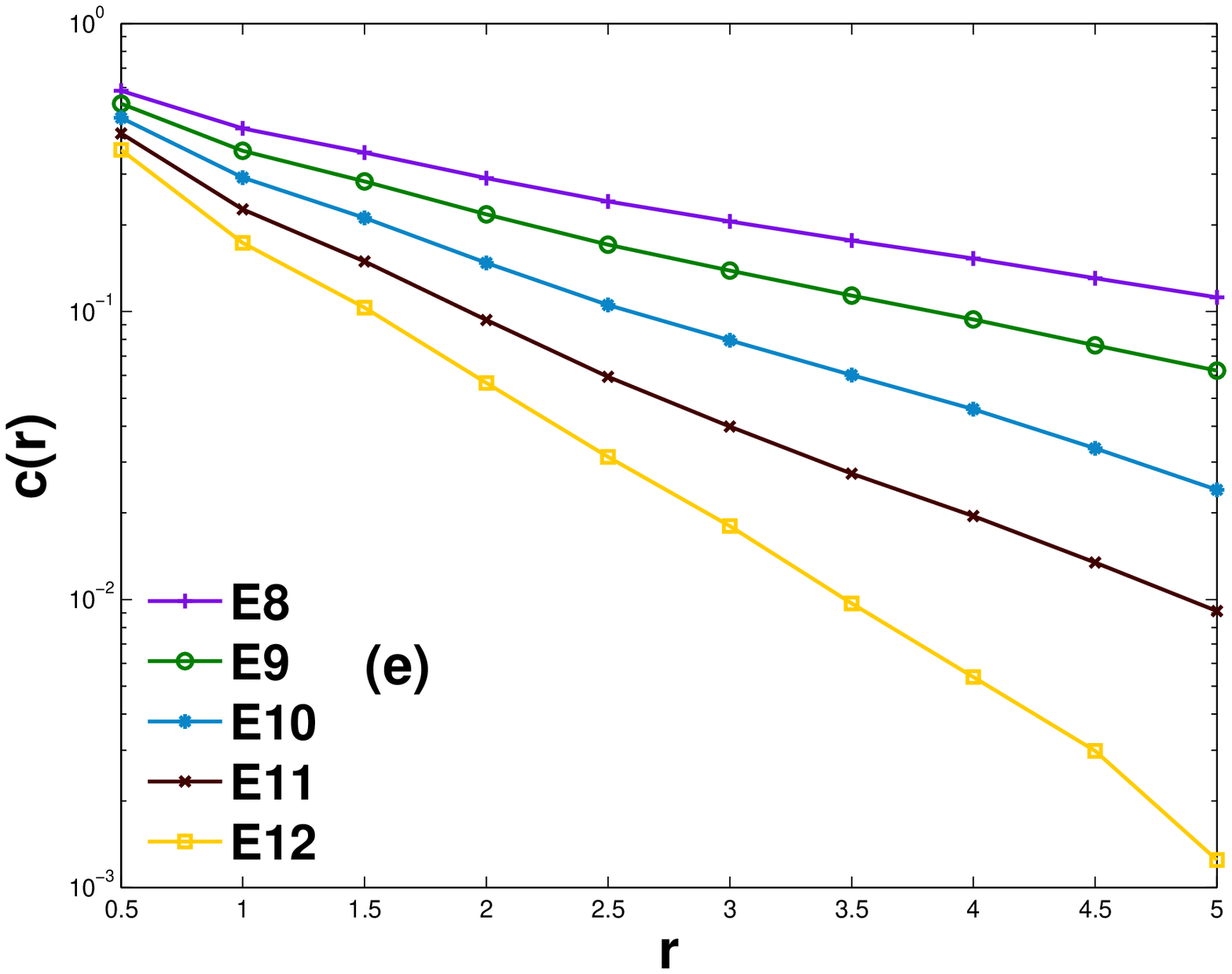}
\includegraphics[height=3.9cm]{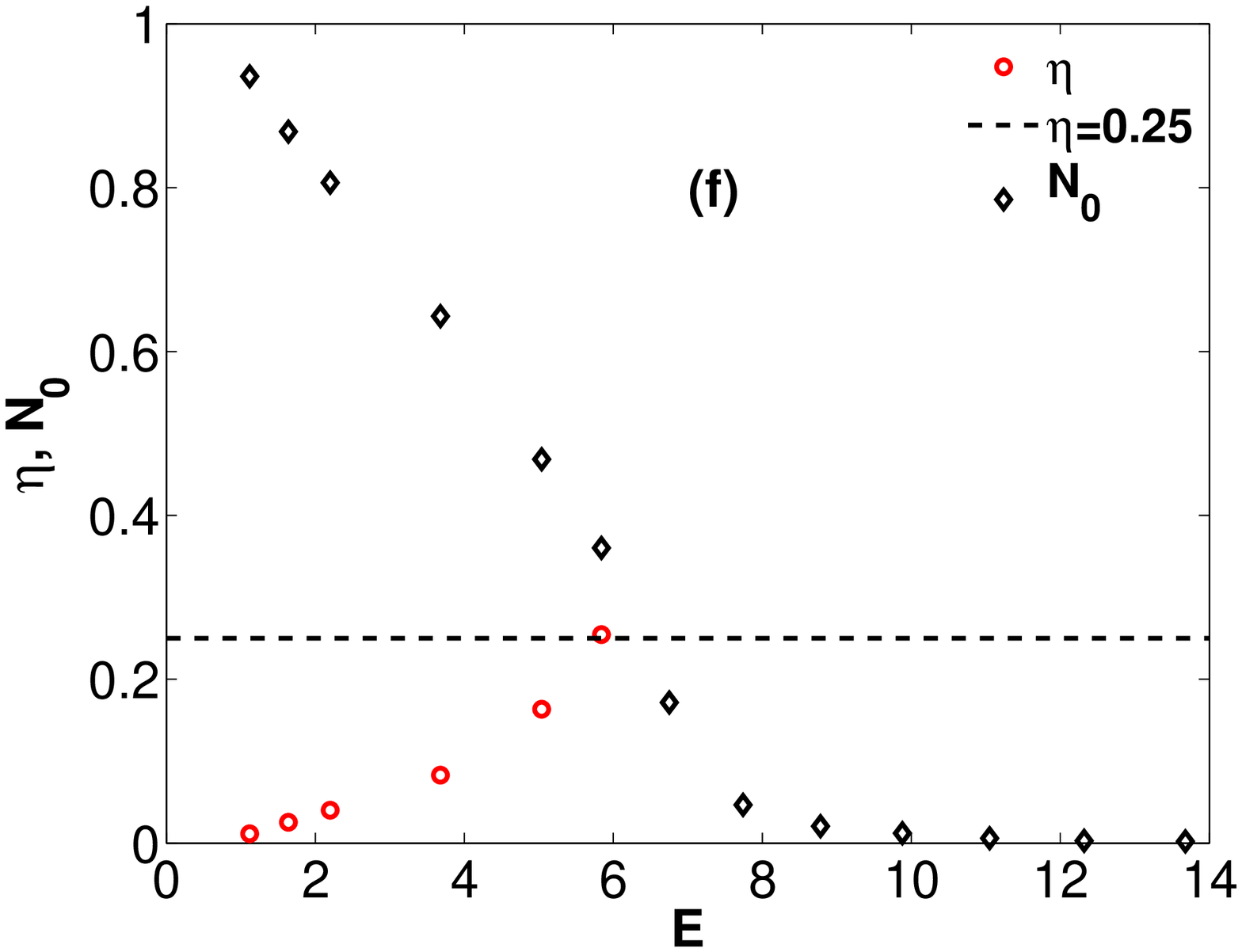}
\end{center}
\caption{\small Plots of $c(r)$ versus $r$ for different
energies in the complete-thermalization regime, for $N^2_c=128^2$
((a) and (b)) and $N^2_c=64^2$ ((d) and (e)).
(a) and (d) Log-log (base $10$) plots of $c(r)$ versus $r$ for 
different energies $E<E_{KT}$; the slopes of the linear parts of 
these plots yield the exponent $\eta$ (table~\ref{table:paract128n64});
(b) and (e) semilog (base(10) plots of $c(r)$ versus $r$ for
different energies $E>E_{KT}$;
(c) ($N^2_c=128^2$) and (f) ($N^2_c=64^2$) show plots of $\eta$ and 
$N_0$ versus $E$ (on the time scales of our runs $N_0$ is
nonzero; see the text for a detailed discussion).
}
\label{fig:corrkt}
\end{figure*}

In figure \ref{fig:corrkt}, we present plots of the correlation
functions $c(r)$. To illustrate the BKT transition clearly, we
present log-log plots of $c(r)$ versus $r$, for $E<E_{\rm BKT}$,
in figures \ref{fig:corrkt} (a) and (d), where the straight lines
indicate power-law regimes; and, for $E>E_{\rm BKT}$, we use
semi-log plots, as in figures \ref{fig:corrkt} (b) and (e), where
the straight lines signify an exponential decay of $c(r)$ with
$r$.  Given the resolution of our DNS runs, we find that, in a
small energy range in the vicinity of $E_{\rm BKT}$, we cannot
fit power-law or exponential forms satisfactorily; this leads to
an uncertainty in our estimate for $E_{\rm BKT}$.  Aside from
this uncertainty, the behavior of $c(r)$, in the regime of
complete thermalization, is in accord with our expectations
for the BKT phase; in particular, the exponent $\eta$ (see
equation~\eref{eq:corrpowerlaw}) depends on $E$ for $E < E_{\rm
BKT}$ as shown in figures~\ref{fig:corrkt} (c) and (f). Our
values for $\eta$, for the runs with $E <E_{\rm BKT}$ and with
$N_c=64$ and $N_c=128$, are listed in table
\ref{table:paract128n64}. Note that $E_{\rm BKT} \simeq 6
(N_c=64)$ and $E_{\rm BKT} \simeq 19 (N_c=128)$, i.e., $E_{\rm
BKT}$ depends on $N_c$, the number of collocation
points; we show analytically below how a low-temperature
analysis can be used to understand this dependence of $E_{\rm
BKT}$ on $N_c$.  In the completely thermalized state of the
Fourier-truncated, 2D, GP system, $N_0$ must vanish in the
thermodynamic limit by virtue of the Hohenberg-Mermin-Wagner
theorem~\cite{Mermin1966,Hohenberg1967} and $n(k) \sim k^{-1+\eta}$; 
it is not easy to realize this limit in
finite-size systems and with the limited run times that are
dictated by computational resources (see the plots of $N_0$ in
figure~\ref{fig:nk0}); however, finite-size scaling can be used to extract
the exponent $\eta$ from the $k=0$ part of $n(k)$ as shown in
reference~\cite{Damle1996}; similarly, $E^c_{kin}(k)$ should also show a 
power-law form with an exponent that depends on $\eta$, but this is 
difficult to realize in numerical calculations with limited  spatial 
resolutions and run lengths.

\subsubsection{Analytical estimation of the energy of the BKT transition}

The energy of a pure condensate of a uniform, weakly interacting, 2D 
Bose gas, which is described by the GP equation~\eqref{eq:2dgpe}, is
$E_0=g/(2\mathcal{A})$. We define the energy of our system 
to be $E=E_0(1+\delta \mathcal{E})$; this energy $E$ is fixed by the 
initial condition; and $\delta\mathcal{E}$ measures the relative
amount by which $E$ exceeds $E_0$. As we show in the~\ref{app:two}, 
the $N_c$ dependence of the energy $E_{\rm BKT}$, at which the 
BKT transition occurs, can be obtained approximately
as follows. We begin with
\begin{equation} 
 \delta \mathcal{E}_{\rm BKT}= \delta \tilde{\mathcal{E}}_{\rm BKT}  \frac{8}
  {\log ({\pi }^2\,{N_c}^2\,
      \left( 1 + \frac{{\pi }^2\,{N_c}^2}{2\,{g}} \right) )}, \label{eq:BKTanlyl}
 \end{equation}
where $\delta \tilde{\mathcal{E}}_{\rm BKT}$, the estimate for
the BKT transition energy that follows from an energy-entropy
argument (see \eqref{eq:xikmax} in the Appendix and~\cite{Kogut1979rmp}), is
\begin{equation} 
 \delta \tilde{\mathcal{E}}_{\rm BKT}=\frac{{\pi }^2\,{N_c}^2}{2\,{g}}
=\frac{\xi^2 \kmax^2}{2} \label{eq:BKTenergyentropy},
 \end{equation}
whence we obtain
\begin{equation}\label{eq:BKTanlylkmaxxi}
 \delta \mathcal{E}_{\rm BKT}= \frac{4\,\kmax^2\,{\xi }^2}
  {\log (\kmax^2\,\mathcal{A}\,\left( 1 + 
       \frac{\kmax^2\,{\xi }^2}{2} \right) )}.
\end{equation}
We can now write
\begin{equation}
E_{\rm BKT} = E_0 \left(1 + \frac{4\pi^2N^2_c}{g\log(\pi^2N^2_c(1+\frac{\pi^2N^2_c}{2g}))}
\right);
\label{eq:approxebkt}
\end{equation}
by using this expression we can determine the ratio $E_{\rm
BKT}(N_c^a)/E_{\rm BKT}(N_c^b)$ for runs with two different
values, $N_c^a$ and $N_c^b$, for the number of collocation
points; we can also obtain this ratio from our DNS, by
determining the value of $E$ at which the exponent $\eta$ becomes
$1/4$. In Table~\ref{table:ektnumtheory} we compare $E_{\rm
BKT}(N_c)$ for $N_c = 64$ and $N_c = 128$; our analytical
approximation~\eqref{eq:approxebkt} yields 
$E^{128}_{\rm BKT}/E^{64}_{\rm BKT}\simeq 3.15$; this is in 
excellent agreement with the value $\simeq 3.14$ that we obtain 
for this ratio from our DNS results.

\begin{table}
\small
\centering
   \begin{tabular}{@{\extracolsep{\fill}} c c c c c c}
	 \hline
		$N_c$ &$E_0$ &$\delta \tilde{\mathcal{E}}_{\rm BKT}$ 
&$\delta \mathcal{E}_{\rm BKT}$ &$E^{\rm A}_{\rm BKT}$ &$E^{\rm DNS}_{\rm BKT}$\\
\hline \hline
	{$64$} &$0.488$ &$20.21$ &$11.84$ &$6.27$ &$5.84$\\
	{$128$} &$0.488$ &$80.85$ &$39.44$ &$19.75$ &$18.34$\\
\hline
\end{tabular}
\caption{\small The values of $E_0$, 
$\delta \tilde{\mathcal{E}}_{\rm BKT}$ (see~\eqref{eq:BKTenergyentropy}), 
$\delta \mathcal{E}$ (see~\eqref{eq:BKTanlylkmaxxi}), 
$E^{\rm A}_{\rm BKT}$ (see~\eqref{eq:approxebkt}), and $E^{\rm DNS}_{\rm BKT}$ 
from our DNS runs $\tt D1$-$\tt D13$ ($N_c=64$) and $\tt E1$-$\tt E12$ ($N_c=64$).
$E_0$ is the ground state energy of a pure condensate of a uniform, interacting, 
2D Bose gas and $E^{\rm DNS}_{\rm BKT}$ is BKT-transition energy determined 
using our DNS runs.
}
\label{table:ektnumtheory}
\end{table} 


\section{Conclusions}
\label{section:conclusions}

We have carried out an extensive study of the statistical
properties of the dissipationless, unforced, 2D, Fourier-truncated,
GP equation. Our study has been designed specifically to
study and identify the universal features, if any, of the
turbulent evolution of the solutions of this equation, by 
undertaking a systematic DNS. In our study, we have used
statistical measures such as velocity-component PDFs and 
energy and occupation-number spectra, for a large number of
initial conditions. To the best of our knowledge, such a
comprehensive study of the Fourier-truncated, 2D, GP equation has
not been attempted hitherto. 

Our comprehensive study of the Fourier-truncated,
2D, GP equation, which makes use of the three types of initial
conditions (section~\ref{subsection:numericsandics}) and a wide
range of parameters (tables~\ref{table:param}
and~\ref{table:paract128n64}), allows us to systematize the
dynamical evolution of this system into four different regimes,
with qualitatively different statistical properties. This
demarkation of the evolution into different regimes has not been
systematized in earlier studies, which have concentrated only on
one or two of these regimes. For example, the study of
reference~\cite{White2010prlvpdf} has investigated states with a significant
number of vortex-anitvortex pairs and obtained for them PDFs of
velocity components that have power-law tails of the type shown
in figure~\ref{fig:velpdf}. References~\cite{Damle1996,Foster2010pra,Small2011bktphotonlat} 
have investigated the BKT nature of the thermalized state. Wave-turbulence
studies~\cite{Svistunov1991,Nazarenko2007freedecay2d,Nazarenko2Dforced2006} have focussed on power-law
regions in energy and occupation-number spectra of the type we
find in our third regime. The DNS studies in
~\cite{Nazarenko2007freedecay2d,Numasato2Dgp2010,Nowak2011prb,Nowak2012pra,
BradleyPhysRevX2012,ReevesinversearXiv2012} have
considered the time evolution of spectra and PDFs for the
Fourier-truncated, 2D, GP equation; in some cases, these studies
introduce dissipation or hyperviscosity and forcing; they have
also reported different power laws in
spectra~\cite{Numasato2Dgp2010,Nowak2011prb,Nowak2012pra}. Our work suggests that, at
least in the dissipationless, unforced, Fourier-truncated, 2D, GP
equation, the only robust power laws in spectra are the the ones
we have reported above; all other apparent power laws occur
either (a) for very special initial conditions~\cite{BradleyPhysRevX2012} or (b)
last for fleetingly small intervals of time and extend over very
small ranges of $k$.

To recapitulate, we find that, in the first dynamical-evolution
regime of the Fourier-truncated, 2D, GP equation, there are
initial-condition-dependent transients.  In the second regime the
energy and the occupation-number spectra start to develop
power-law scaling regions, but the power-law exponent and the
extent of the scaling region change with time and are influenced
by the initial conditions. In the third regime, of partial
thermalization, we find $E^c_{kin}(k)$ and $E_{int}(k)+E_q(k)$
$\sim k$, and $n(k) \sim 1/k$, for $k < k_c(t)$ and, for $k >
k_c$, we find an initial-condition-dependent self-truncation
regime, in which the spectra drop rapidly; the self-truncation
wave number  $k_c(t)$ grows either as $t^{\alpha}$ or
logarthimically for different intial conditions (table~\ref{table:alpha}). In the
fourth, complete-thermalization regime, power-law forms of
correlation functions and spectra, for $E < E_{\rm BKT}$, are
consistent with their nontrivial BKT forms; however, considerable
care must be exercised, as explained in section~\ref{subsub:corrfuncBKT} and
~\cite{Damle1996,Foster2010pra,Small2011bktphotonlat}, to distinguish these nontrivial
power laws from their wave-turbulence
analogs~\cite{Svistunov1991,Nazarenko2007freedecay2d,Nazarenko2Dforced2006}.

\ack

We thank CSIR, DST, and UGC(India) for financial support, and SERC 
(IISc) for computational resources.

\appendix
\section{}
\label{app:one}

The GP equation, which describes the dynamical evolution of the 
wave function $\psi(\mathbf{x},t)$ of a weakly interacting, 2D Bose 
gas at low temperatures, is
\begin{equation} \label{eq:std2dgpe}
i\hbar\frac{\partial\psi(\mathbf{x},t)}{\partial t} =
-\frac{\hbar^2}{2m}\nabla^2\psi(\mathbf{x},t) 
+ g_{\rm 2D}|\psi|^2\psi(\mathbf{x},t),
\end{equation}
where $g_{2D}$ is the effective interaction strength. As we 
have mentioned earlier (see ~\eqref{eq:totalenergy} and
~\eqref{eq:particleN}), the GP equation conserves the energy, 
given by the Hamiltonian 
\begin{equation}
H = \int_{\mathcal{A}} d^2x \left(\frac{\hbar^2}{2m}|\nabla \psi|^2 
+ \frac{g_{\rm 2d}}{2}|\psi|^4 \right),
\end{equation}
and the total number of particles $n=\int_{\mathcal{A}} |\psi|^2d^2x$. 
To obtain ~\eqref{eq:2dgpe}, we first divide~\eqref{eq:std2dgpe} by 
$\hbar$ and define $g=g_{\rm 2D}/\hbar$; we then set $\hbar/2m=1$, 
with $m=1$, so that $|\psi|^2$ is the same as $\rho$; this is
tantamount to using units with $\hbar=2$. 

\section{}
\label{app:two}

The Berezinskii-Kosterlitz-Thouless (BKT) transition is best studied 
by using the renormalization group~\cite{Kogut1979rmp}; here, we 
restrict ourselves to the heuristic, energy-entropy 
argument to obtain a rough estimate of the BKT transition temperature 
$T_{\rm BKT}$. In the $XY$ model, this transition is studied by using the 
Hamiltonian
\begin{equation} \label{eq:hamilxy}
H_{\rm{XY}}=-J \sum_{<i,j>} \cos(\theta_i-\theta_j),
\end{equation}
where $<i,j>$ denotes nearest-neighbour pairs of sites, on a 2D 
square lattice, $J$ is the nearest-neighbour exchange coupling, and
$(\theta_i - \theta_j)$ is the angle between the nearest-neighbour, 
$XY$ spins on sites $i$ and $j$. In the continuum limit, the above 
Hamiltonian becomes, to lowest order in spatial gradients, 
\begin{equation} \label{eq:conthamilxy}
H_{\rm{XY}}= \frac{J}{2} \int d^2x (\nabla \theta(x))^2.
\end{equation}
By comparing ~\eqref{eq:conthamilxy} with the kinetic-energy term 
in~\eqref{eq:std2dgpe}, we find that
\begin{equation} \label{eq:coupling}
J=\frac{|\langle \psi \rangle|^2 \hbar^2}{m} =\frac{\rho \Gamma^2}{(2
\pi)^2},
\end{equation} 
where $\Gamma$ denotes the Onsager-Feynman quantum of velocity 
circulation $\Gamma=4 \pi \hbar/2 m=h/m$. A rough estimate for the
BKT transition temperature $T_{\rm BKT}$ is given below:
\begin{equation}  \label{eq:roughTBKT}
\tilde{T}_{\rm BKT}=\frac{\pi J}{2k_B}=\frac{\pi \mid \langle \psi
\rangle \mid^2 \hbar^2}{2mk_B}
=\frac{\rho \Gamma^2}{8\pi k_B},
\end{equation}
here $\tilde{T}_{\rm BKT}$ denotes the estimate for  $T_{\rm BKT}$
that follows from an energy-entropy argument~\cite{Kogut1979rmp}.
For $T<T_{\rm BKT}$, the phase correlation function $c(r)$
(see~\eqref{eq:phasecorr}) and the angle-integrated spectrum $\hat{c}(k)$,
which follows from a Fourier tranform of $c(r)$, scale as
\begin{equation}
	c(r) \sim (a/r)^{\frac{T}{4 T_{\rm BKT}}}
\end{equation}
and 
\begin{equation}
	\hat{c}(k) \sim k^{-1+\frac{T}{4 T_{\rm BKT}}},
\end{equation}
respectively. Above $T_{\rm BKT}$ the correlation length
\begin{equation}
\ell = \frac{\int k^{-1} E(k) dk}{\int E(k) dk}
\end{equation}
is finite; and, as $T\to T_{\rm BKT}$, it displays the essential
singularity
\begin{equation}
\ell\sim\exp(b (T_{\rm BKT}/(T-T_{\rm BKT}))^{1/2}).
\end{equation}

\subsection{}

We now develop an analytical framework, which is valid at
low-temperatures $T \ll T_{\rm BKT}$, that can be used to test
some of the results of our DNS runs in the region of complete
thermalization. We first calculate equilibrium
thermodynamic functions for a weakly-interacting, 2D Bose gas,
in the grand-canonical ensemble; we then obtain their
analogues in the microcanonical ensemble. In the
grand-canonical ensemble the probability of a given
state is 
\begin{equation}
\mathbb{P} = \frac{1}{\Xi}e^{-\beta (H - \mu N)},
\end{equation}
where $\Xi$ is the grand partition function, $\beta$ the
inverse temperature, $\mu$ the chemical potential, and $N$
the number of bosons. The grand-canonical potential is 
\begin{equation}
\Omega = - \beta^{-1}\log(\Xi);
\end{equation}
and the mean energy $E$, entropy $S$, and $N$ are
\begin{subequations} \label{eq:thermrel}
\begin{align}
N &= -\frac{\partial \Omega}{\partial \mu}, \\
S &= \beta^2 \partial \Omega/\partial \beta,  \\
E &= \frac{\partial \Omega}{\partial \beta}+\mu N
= \frac{S}{\beta} + \mu N.
\end{align}
\end{subequations}
We adapt to 2D the 3D study of
Ref.~\cite{Krstulovic2011pre}, expand $\psi$ in terms of 
Fourier modes $A_{\bf k}$, and obtain $\Omega$
as the sum of the saddle-point part $\Omega_{sp}$ and
$\Omega_{Q}$, the deviations from the saddle point that are
quadratic in $A_{\bf k}$. We write
$\Omega=\Omega_{sp}+\Omega_{Q}$, where
$\Omega_{sp}=-\mathcal{A}\mu^2/2g$ and  
\begin{equation} \label{eq:grandpotph}
\Omega_{Q} = -\int^{p_{\rm max}}_{0}\frac{\left(p\mathcal{A} \log(\frac{2m}
{\beta \sqrt{p^4+4mp^2\mu}})\right)}{2\pi \beta \hbar^2}.
\end{equation}
We can also calculate the condensate depletion $\delta N$, where
the particle number $N=N_0+\delta N$ and $N_0$ is the number of particles 
in the $k=0$ mode, as follows:
\begin{equation} \label{eq:deltaNintg}
\delta N = \int^{p_{\rm max}}_{0} 
\frac{mp\mathcal{A}\left(p^{-2}+\frac{1}{p^2+4m\mu}\right)}
{2\pi \beta \hbar^2}.
\end{equation}
The integrals in the \eqref{eq:grandpotph} and \eqref{eq:deltaNintg} 
can be performed analytically, but, in contrast to the 
3D case where the primitives are zero at $p=0$, the $2D$ primitive 
for $\Omega_{\rm ph}$ is finite at  $p=0$ and for $\delta N$ 
it is infra-red (I.R.) divergent. By subtracting the I.R. finite and 
divergent terms from $\Omega_{Q}$ and $\delta N$, respectively,
we get the following expressions, in $2D$, in the thermodynamic limit
$\mathcal{A}\to\infty$: 
\begin{equation}
\begin{split}
\Omega & = -\frac{\mu^2\mathcal{A}}{2g}
- \frac{p^2_{\rm max}\mathcal{A}}{4\pi \beta \hbar^2} 
 + \frac{m\mu \mathcal{A} \log(1+\frac{p^2_{\rm max}}{4m\mu})}
{2\pi \beta \hbar^2} \\
& \quad - \frac{p^2_{\rm max}\mathcal{A} \log(\frac{2m}
{\beta \sqrt{p^4_{\rm max}+4m\mu p^2_{\rm max}}})}{4\pi \beta \hbar^2}
\end{split}
\end{equation}
and
\begin{equation} \label{eq:deltaN}
\delta N = \frac{m\mathcal{A}\left( \log(1+\frac{p^2_{\rm max}}{4m\mu}) +
\log(\frac{p^2_{\rm max}\mathcal{A}}{\hbar^2}) \right)}{4\pi \beta \hbar^2}.
\end{equation}
By using the thermodynamic relations~\eqref{eq:thermrel}, we get  
\begin{equation} \label{Eq:Ntot}
N=\frac{\mu \mathcal{A}}{g} - \frac{m\mathcal{A}\log(1+\frac{p^2_{\rm max}}{4m\mu})}
{2\pi \beta \hbar^2}
\end{equation}
and
\begin{equation}  \label{eq:Etot}
E=\frac{\mu^2\mathcal{A}}{2g} 
+ \frac{p^2_{\rm max}\mathcal{A}}{4\pi \beta \hbar^2} 
- \frac{m\mu \mathcal{A} \log (1 + \frac{p^2_{\rm max}}{4m\mu })}
{2\pi \beta \hbar^2}.
\end{equation}

\subsection{}

We next determine the chemical potential $\mu$, which fixes the 
total density $\rho=m N/\mathcal{A}$ at a given value, by 
solving the equation  
\begin{equation} \label{eq:solvemu}
\rho - \frac{m\mu}{g} + \frac{m^2
\log (1 + \frac{p^2_{\rm max}}{4m\mu})}{2\pi \beta \hbar^2} = 0;
\end{equation}
at $\beta=\infty$, i.e., zero temperature (subscript $0$) we obtain 
\begin{equation}
\mu_0=\frac{g\,\rho}{m};
\end{equation}
to order $\beta^{-1}$ we get 
\begin{equation} \label{eq:muSol}
\mu=\mu_0+\delta \mu, 
\end{equation}
where 
\begin{equation}
\delta \mu=\frac{mg \left( 4g\rho^2 + \rho p^2_{\rm max} \right)
    \log (1 + \frac{p^2_{\rm max}}{4g \rho})}{m^2 p^2_{\rm max} + 
    2\pi \beta \hbar^2 \rho p^2_{\rm max} + 
    8\pi \beta \hbar^2 g\rho^2}.
\end{equation}
We insert $\mu$ from \eqref{eq:muSol} into \eqref{eq:deltaN}, 
define the change in density $\delta \rho = m \delta
N/\mathcal{A}$, use the energy $E$ from \eqref{eq:Etot}, 
and then expand to order $\beta^{-1}$ to obtain
\begin{equation} \label{Eq:deltarhosol}
\delta \rho =\frac{m^2 \left( \log (1 + \frac{p^2_{\rm max}}{4g\rho}) 
+ \log (\frac{p^2_{\rm max}\mathcal{A}}{\hbar^2}) \right)}
{4\pi \beta \hbar^2}
\end{equation}
and
\begin{equation} \label{Eq:Etotsol}
E=\frac{g\rho^2\mathcal{A}}{2m^2} + \frac{p^2_{\rm max}\mathcal{A}}
   {4\pi \beta \hbar^2}. 
\end{equation}

By using \eqref{eq:roughTBKT} and $\rho=m \mid \langle \psi
\rangle\mid^2$, we obtain
\begin{equation} \label{Eq:betaBKTrough}
\tilde{\beta}_{\rm BKT}=\frac{1}{k_{\rm B}\tilde{T}_{\rm BKT}}
=\frac{2m^2}{\pi \rho \hbar^2}, 
\end{equation}
which we can use along with 
\eqref{Eq:deltarhosol} to relate the condensate relative 
depletion $\delta \rho /\rho$ to $\beta/\tilde{\beta}_{\rm BKT}$, 
where $\beta=1/(k_{\rm B}T)$ and $k_{\rm B}$ is the Boltzmann
constant, as given below:
\begin{equation} \label{eq:deltarhoisrhoi}
\frac{\delta \rho} {\rho} ={\frac {\tilde{\beta}_{\rm BKT}}{8\beta} } 
\log \left(\frac{p^2_{\rm max}\left(1 + \frac{p^2_{\rm max}}{4g\rho} \right) 
\mathcal{A}}{\hbar^2}\right).
\end{equation}
We use this low-temperature result \eqref{eq:deltarhoisrhoi}
to estimate the inverse-temperature scale $\beta_{\rm BKT}$, at which 
the depletion of the $k=0$ condensate mode becomes significant
for a finite-size system with $N_c^2$ collocation points (which fixes
the maximum momentum $p_{\rm max}$); in particular, 
we can solve \eqref{eq:deltarhoisrhoi}, for $\delta \rho/\rho=1$, 
to obtain
\begin{equation} \label{eq:betaBKT}
{\frac {\beta_{\rm BKT}}{\tilde{\beta}_{\rm BKT}} } ={\frac {1}{8} } 
\log \left(\frac{p^2_{\rm max}\left( 1 + 
       \frac{p^2_{\rm max}}{4g\rho} \right) \mathcal{A}}{\hbar^2}\right). 
\end{equation}
By making the replacements that correspond do defining $\hbar$, 
$m$,  and $g$ in terms of $c$ and $\xi$, as in  
\cite{Krstulovic2011pre}, $p_{\rm max} \to \hbar k_{\rm max}$, 
$\hbar  \to \sqrt{2} c m \xi$, and $g \to c^2m^2/\rho$, we can
rewrite \eqref{eq:betaBKT} as
\begin{equation} \label{eq:FacBKT}
{\frac {\beta_{\rm BKT}}{\tilde{\beta}_{\rm BKT}} } =
\frac{1}{8} \log \left(\kmax^2 \mathcal{A}\,
     ( 1 + \frac{{{k_{\rm max}}}^2\,{\xi }^2}{2} ) \right). 
\end{equation}

\subsection{}

Our DNS runs, which use initial conditions of type $\tt IC1$ and
$\tt IC2$, give the dynamical evolutions of the
Fourier-truncated, 2D GP equation, which is a Hamiltonian system.
The energy $E$, particle number $N$, and area $\mathcal{A}$
are conserved in this evolution, so our calculation
can be viewed as a simulation of this Hamiltonian system
in the microcanonical ensemble, which yields, eventually, the
fully thermalized state that we have described above. Therefore, 
we now transform the results, which we have obtained in the 
previous subsection, into their counterparts in the microcanonical 
ensemble. In the low-temperature limit,
\eqref{Eq:Etotsol} yields 
\begin{equation} \label{eq:betamicroc}
\beta=\frac{m^2\,p^2_{\rm max}\,\mathcal{A}}
  {2\pi \hbar^2 \left( 2m^2 E - g\rho^2 \mathcal{A} \right)}.
\end{equation}
The energy of a pure condensate is
\begin{equation} \label{eq:dimenergycond}
E_0 = \lim_{\beta \to \infty}
E=\frac{g\,{\rho}^2\,\mathcal{A}}{2\,m^2};
\end{equation}
and the energy and the inverse temperature $\beta$ \eqref{eq:betamicroc} 
can be related as follows:
\begin{equation}
E = E_0 (1+{\delta \mathcal{E}}),
\end{equation}
where $\delta \mathcal{E}$ is the relative increase of energy
above $E_0$, and 
\begin{equation} \label{Eq:betadeltae}
\beta = \frac{m^2\,p^2_{\rm max}}
  {2\pi \hbar^2 g\rho^2 \delta \mathcal{E}}.
\end{equation}
If we now substitute $\beta=\beta_{\rm BKT}$ by using
\eqref{eq:betaBKT}, we obtain,  
in terms of $c$, $\xi$ and $\rho$ (see text just below 
\eqref{eq:betaBKT})
\begin{equation}
E_0=\frac{c^2\,\rho\,\mathcal{A}}{2},
\end{equation}
\begin{equation} \label{Eq:xikmaxBKTrough}
\delta \tilde{\mathcal{E}}_{\rm BKT}=\frac{k^2_{\rm max} \xi^2}{2},
 \end{equation}
and
\begin{equation}
 \delta \mathcal{E}_{\rm BKT}= \frac{4k^2_{\rm max}\xi^2}
  {\log \left(k^2_{\rm max}\mathcal{A} ( 1 + 
       \frac{k^2_{\rm max} \xi^2}{2} ) \right)}.
\end{equation}
All the energies mentioned in the main paper are dimensionless;
thus, to convert the energies given in this Appendix to dimensionless forms, 
we divide them by $\hbar$. Hence, the energy of a pure condensate 
is obtained, in the dimensionless form, by dividing 
\eqref{eq:dimenergycond} by $\hbar$, which gives
\begin{equation}
E_0 = \frac{g}{2\mathcal{A}}=\frac{1}{2}\frac{g}{L^2}.
\end{equation}

\section*{References}

\bibliographystyle{unsrt}
\bibliography{reference}

\end{document}